\documentclass[aj,twocolumn]{aastex62}

\usepackage{placeins}
\usepackage{comment}

\newcommand{\hatstara}{HAT-P-69}
\newcommand{\hatstarab}{HAT-P-69\,b}
\newcommand{\hatstarb}{HAT-P-70}
\newcommand{\hatstarbb}{HAT-P-70\,b}

\newcommand{\hatstarateff}{$7394_{-600}^{+360}$}
\newcommand{\hatstarafeh}{$-0.069_{-0.075}^{+0.058}$}
\newcommand{\hatstaravsini}{$77.44_{-0.57}^{+0.55}$}
\newcommand{\hatstaravmac}{$5.76_{-0.24}^{+0.24}$}
\newcommand{\hatstaramass}{$1.648_{-0.026}^{+0.058}$}
\newcommand{\hatstararadius}{$1.926_{-0.031}^{+0.060}$}
\newcommand{\hatstaralogg}{$4.110_{-0.064}^{+0.034}$}
\newcommand{\hatstaralum}{$10.0_{-0.9}^{+1.8}$}
\newcommand{\hatstaraoblate}{$0.9678_{-0.0022}^{+0.0012}$}
\newcommand{\hatstarairot}{$58.2_{-1.2}^{+1.6}$}

\newcommand{\hatstaraperiod}{$4.7869491_{-0.0000021}^{+0.0000018}$}
\newcommand{\hatstaralam}{$21.2_{-3.6}^{+4.6}$}
\newcommand{\hatstaraplmass}{$3.58_{-0.58}^{+0.58}$}
\newcommand{\hatstaraplrad}{$1.676_{-0.033}^{+0.051}$}

\newcommand{\hatstarbteff}{$8450_{-690}^{+540}$}
\newcommand{\hatstarbfeh}{$-0.059_{-0.088}^{+0.075}$}
\newcommand{\hatstarbvsini}{$99.85_{-0.61}^{+0.64}$}
\newcommand{\hatstarbvmac}{$5.870_{-0.52}^{+0.58}$}
\newcommand{\hatstarbmass}{$1.890_{-0.013}^{+0.010}$}
\newcommand{\hatstarbradius}{$1.858_{-0.091}^{+0.119}$}
\newcommand{\hatstarblogg}{$4.181_{-0.063}^{+0.055}$}
\newcommand{\hatstarblum}{$16.7_{-4.6}^{+5.3}$}
\newcommand{\hatstarboblate}{$0.9574_{-0.0057}^{+0.0063}$}
\newcommand{\hatstarbirot}{$58.8_{-4.8}^{+7.5}$}

\newcommand{\hatstarbperiod}{$2.74432452_{-0.00000068}^{+0.00000079}$}
\newcommand{\hatstarblam}{$113.1_{-3.4}^{+5.1}$}
\newcommand{\hatstarbplmass}{$<6.78\, (3\sigma)$}
\newcommand{\hatstarbplrad}{$1.87_{-0.10}^{+0.15}$}

\newcommand{\ortotal}{$0.41\pm0.10$\,\%}

\newcommand{\orG}{$0.71\pm0.31$\,\%}
\newcommand{\orF}{$0.43\pm0.15$\,\%}
\newcommand{\orA}{$0.26\pm0.11$\,\%}

\newcommand{\nPlanet}{18}
\newcommand{\nCandidate}{3}
\newcommand{\nFalsePositive}{10}
\newcommand{\nTotal}{31}

\newcommand{\ctbd}[1]{}

\newcommand{\lc}{light curve}

\newcommand{\Lc}{Light curve}

\newcommand{\kms}{\ensuremath{\rm km\,s^{-1}}}
\newcommand{\ms}{\ensuremath{\rm m\,s^{-1}}}

\newcommand{\gcmc}{\ensuremath{\rm g\,cm^{-3}}}

\newcommand{\logg}{\ensuremath{\log{g}}}
\newcommand{\vsini}{\ensuremath{v \sin{I_\star}}}
\newcommand{\feh}{\ensuremath{\rm [Fe/H]}}

\newcommand{\rsun}{\ensuremath{R_\sun}}
\newcommand{\msun}{\ensuremath{M_\sun}}
\newcommand{\lsun}{\ensuremath{L_\sun}}

\newcommand{\rstar}{\ensuremath{R_\star}}
\newcommand{\mstar}{\ensuremath{M_\star}}
\newcommand{\lstar}{\ensuremath{L_\star}}

\newcommand{\teffstar}{\ensuremath{T_{\rm eff\star}}}

\newcommand{\loggstar}{\ensuremath{\log{g_{\star}}}}

\newcommand{\rpl}{\ensuremath{R_{p}}}
\newcommand{\mpl}{\ensuremath{M_{p}}}

\newcommand{\rhopl}{\ensuremath{\rho_{p}}}

\newcommand{\arstar}{\ensuremath{a/\rstar}}

\newcommand{\rjup}{\ensuremath{R_{\rm J}}}
\newcommand{\mjup}{\ensuremath{M_{\rm J}}}

\graphicspath{{./}{figures/}}

\received{\today}
\submitjournal{AJ}

\shorttitle{Planets around A stars from HATNet and \emph{TESS}}
\shortauthors{Zhou et al.}

\begin{document}

\title{Two new HATNet hot Jupiters around A stars,\\
and the first glimpse at the occurrence rate of hot Jupiters from \emph{TESS}
\footnote{\scriptsize{
Based on observations obtained with the Hungarian-made Automated
Telescope Network. Based in part on
observations obtained with the Tillinghast Reflector 1.5 m telescope and the
1.2 m telescope, both operated by the Smithsonian Astrophysical Observatory
at the Fred Lawrence Whipple Observatory in Arizona. This work makes use of the Smithsonian Institution High Performance Cluster (SI/HPC). Based in part on observations made with the Southern African Large Telescope (SALT) }}}

\correspondingauthor{George~Zhou}
\email{george.zhou@cfa.harvard.edu}

\author{G.~Zhou}
\affiliation{Center for Astrophysics \textbar{} Harvard \& Smithsonian, 60 Garden St., Cambridge, MA 02138, USA.}
\affiliation{Hubble Fellow}

\author{C.X.~Huang}
\affiliation{Department of Physics, and Kavli Institute for Astrophysics and Space Research, Massachusetts Institute of Technology, Cambridge, MA 02139, USA.}

\author{G.\'A.~Bakos}
\affiliation{Department of Astrophysical Sciences, Princeton University, NJ 08544, USA.}
\affiliation{Packard Fellow}
\affiliation{MTA Distinguished Guest Fellow, Konkoly Observatory, Hungary}

\author{J.D.~Hartman}
\affiliation{Department of Astrophysical Sciences, Princeton University, NJ 08544, USA.}

\author{David W.~Latham}
\affiliation{Center for Astrophysics \textbar{} Harvard \& Smithsonian, 60 Garden St., Cambridge, MA 02138, USA.}

\author{S.N.~Quinn}
\affiliation{Center for Astrophysics \textbar{} Harvard \& Smithsonian, 60 Garden St., Cambridge, MA 02138, USA.}

\author[0000-0001-6588-9574]{K.A.~Collins}
\affiliation{Center for Astrophysics \textbar{} Harvard \& Smithsonian, 60 Garden St., Cambridge, MA 02138, USA.}

\author{J.N.~Winn}
\affiliation{Department of Astrophysical Sciences, Princeton University, NJ 08544, USA.}

\author{I. Wong}
\affiliation{Department of Earth, Atmospheric and Planetary Sciences, Massachusetts Institute of Technology, Cambridge, MA 02139, USA}
\affiliation{51 Pegasi b Fellow}

\author{G. Kov\'acs}
\affiliation{Konkoly Observatory of the Hungarian Academy of Sciences, Budapest, 1121 Konkoly Thege ut. 15-17, Hungary}

\author{Z.~Csubry}
\affiliation{Department of Astrophysical Sciences, Princeton University, NJ 08544, USA.}

\author{W.~Bhatti}
\affiliation{Department of Astrophysical Sciences, Princeton University, NJ 08544, USA.}

\author{K.~Penev}
\affiliation{Physics Department, University of Texas at Dallas, 800 W Campbell Rd. MS WT15, Richardson, TX 75080, USA}

\author{A. Bieryla}
\affiliation{Center for Astrophysics \textbar{} Harvard \& Smithsonian, 60 Garden St., Cambridge, MA 02138, USA.}

\author{G.A. Esquerdo}
\affiliation{Center for Astrophysics \textbar{} Harvard \& Smithsonian, 60 Garden St., Cambridge, MA 02138, USA.}

\author{P. Berlind}
\affiliation{Center for Astrophysics \textbar{} Harvard \& Smithsonian, 60 Garden St., Cambridge, MA 02138, USA.}

\author{M.L. Calkins}
\affiliation{Center for Astrophysics \textbar{} Harvard \& Smithsonian, 60 Garden St., Cambridge, MA 02138, USA.}

\author{M.~de Val-Borro}
\affiliation{Astrochemistry Laboratory, Goddard Space Flight Center, NASA, 8800 Greenbelt Rd, Greenbelt, MD 20771, USA}

\author{R.~W.~Noyes}
\affiliation{Center for Astrophysics \textbar{} Harvard \& Smithsonian, 60 Garden St., Cambridge, MA 02138, USA.}

\author{J. L\'az\'ar}
\affiliation{Hungarian Astronomical Association, 1451 Budapest, Hungary}

\author{I. Papp}
\affiliation{Hungarian Astronomical Association, 1451 Budapest, Hungary}

\author{P. S\'ari}
\affiliation{Hungarian Astronomical Association, 1451 Budapest, Hungary}

\author{T. Kov\'acs}
\affiliation{Konkoly Observatory of the Hungarian Academy of Sciences, Budapest, 1121 Konkoly Thege ut. 15-17, Hungary}

\author[0000-0003-1605-5666]{Lars A. Buchhave}
\affiliation{DTU Space, National Space Institute, Technical University of Denmark, Elektrovej 328, DK-2800 Kgs. Lyngby, Denmark}

\author{T. Szklenar}
\affiliation{Hungarian Astronomical Association, 1451 Budapest, Hungary}

\author{B.~B\'eky}
\affiliation{Google Inc.}

\author{M.C. Johnson}
\affiliation{Department of Astronomy, The Ohio State University, 140 West 18th Ave., Columbus, OH 43210, USA}

\author{W.D. Cochran}
\affiliation{McDonald Observatory, University of Texas at Austin, 2515 Speedway, Stop C1400, Austin, TX 78712, USA}

\author{A.Y. Kniazev}
\affiliation{South African Astronomical Observatory, PO Box 9, 7935 Observatory, Cape Town, South Africa}
\affiliation{Southern African Large Telescope Foundation, PO Box 9, 7935 Observatory, Cape Town, South Africa}

\author{K.G. Stassun}
\affiliation{Department of Physics and Astronomy, Vanderbilt University, 6301 Stevenson Center, Nashville, TN 37235, USA}
\affiliation{Department of Physics, Fisk University, 1000 17th Avenue North, Nashville, TN 37208, USA}

\author{B.J. Fulton}
\affiliation{Caltech/IPAC-NExScI, 1200 East California Boulevard, Pasadena, CA 91125, USA}

\author{A. Shporer}
\affiliation{Department of Physics, and Kavli Institute for Astrophysics and Space Research, Massachusetts Institute of Technology, Cambridge, MA 02139, USA.}

\author{N. Espinoza}
\affiliation{Max-Planck-Institut f\"ur Astronomie, K\"onigstuhl 17, 69117 Heidelberg, Germany}
\affiliation{Instituto de Astrof\'isica, Facultad de F\'isica, Pontificia Universidad Cat\'olica de Chile, Av. Vicu\~na Mackenna 4860, 782-0436 Macul, Santiago,
Chile}
\affiliation{Millennium Institute of Astrophysics (MAS), Av. Vicu\~na Mackenna 4860, 782-0436 Macul, Santiago, Chile}

\author{D. Bayliss}
\affiliation{Department of Physics, University of Warwick, Gibbet Hill Road, Coventry CV4 7AL, UK}


\author{M. Everett}
\affiliation{National Optical Astronomy Observatory, Tucson, AZ, USA}



\author{S.~B.~Howell}
\affiliation{NASA Ames Research Center, Moffett Field, CA 94035}


\author{C. Hellier}
\affiliation{Astrophysics Group, Keele University, Staffordshire, ST5 5BG, UK}

\author{D.R. Anderson}
\affiliation{Astrophysics Group, Keele University, Staffordshire, ST5 5BG, UK}
\affiliation{Department of Physics, University of Warwick, Gibbet Hill Road, Coventry CV4 7AL, UK}

\author{A. Collier Cameron}
\affiliation{SUPA, School of Physics and Astronomy, University of St.~Andrews, North Haugh,  Fife, KY16 9SS, UK}


\author{R.G. West}
\affiliation{Department of Physics, University of Warwick, Gibbet Hill Road, Coventry CV4 7AL, UK}

\author{D.J.A. Brown}
\affiliation{Department of Physics, University of Warwick, Gibbet Hill Road, Coventry CV4 7AL, UK}

\author{N. Schanche}
\affiliation{SUPA, School of Physics and Astronomy, University of St.~Andrews, North Haugh,  Fife, KY16 9SS, UK}

\author{K. Barkaoui}
\affiliation{Astrobiology Research Unit, University of Li\'{e}ge, Belgium}
\affiliation{Oukaimeden Observatory, High Energy Physics and Astrophysics Laboratory, Cadi Ayyad University, Marrakech, Morocco}

\author{F. Pozuelos}
\affiliation{Space Sciences, Technologies and Astrophysics Research (STAR) Institute, University of Li\'{e}ge, Belgium}

\author{M. Gillon}
\affiliation{Astrobiology Research Unit, University of Li\'{e}ge, Belgium}

\author{E. Jehin}
\affiliation{Space Sciences, Technologies and Astrophysics Research (STAR) Institute, University of Li\'{e}ge, Belgium}

\author{Z. Benkhaldoun}
\affiliation{Oukaimeden Observatory, High Energy Physics and Astrophysics Laboratory, Cadi Ayyad University, Marrakech, Morocco}

\author{A. Daassou}
\affiliation{Oukaimeden Observatory, High Energy Physics and Astrophysics Laboratory, Cadi Ayyad University, Marrakech, Morocco}


\author{G. Ricker}
\affiliation{Department of Physics, and Kavli Institute for Astrophysics and Space Research, Massachusetts Institute of Technology, Cambridge, MA 02139, USA.}

\author{R. Vanderspek}
\affiliation{Department of Physics, and Kavli Institute for Astrophysics and Space Research, Massachusetts Institute of Technology, Cambridge, MA 02139, USA.}

\author[0000-0002-6892-6948]{S.~Seager}
\affiliation{Department of Physics, and Kavli Institute for Astrophysics and Space Research, Massachusetts Institute of Technology, Cambridge, MA 02139, USA.}
\affiliation{Department of Earth, Atmospheric and Planetary Sciences, Massachusetts Institute of Technology, Cambridge, MA 02139, USA}
\affiliation{Department of Aeronautics and Astronautics, MIT, 77 Massachusetts Avenue, Cambridge, MA 02139, USA}

\author{J.M. Jenkins}
\affiliation{NASA Ames Research Center, Moffett Field, CA 94035, USA}

\author{Jack J. Lissauer}
\affiliation{NASA Ames Research Center, Moffett Field, CA 94035, USA}


\author{J.D. Armstrong}
\affiliation{Institute for Astronomy, University of Hawaii, 34 Ohia Ku St., Pukalani, Maui, HI 96768, USA}


\author[0000-0003-2781-3207]{K. I.\ Collins}
\affiliation{Department of Physics and Astronomy, Vanderbilt University, 6301 Stevenson Center, Nashville, TN 37235, USA}

\author{T. Gan}
\affiliation{Physics Department and Tsinghua Centre for Astrophysics, Tsinghua University, Beijing 100084, China}

\author{R. Hart}
\affiliation{Centre for Astrophysics, University of Southern Queensland, Toowoomba, QLD, 4350, Australia}

\author{K. Horne}
\affiliation{SUPA Physics and Astronomy, University of St Andrews, North Haugh, St Andrews KY16 9SS, UK}


\author[0000-0003-0497-2651]{J. F.\ Kielkopf} 
\affiliation{Department of Physics and Astronomy, University of Louisville, Louisville, KY 40292, USA}



\author{L.D. Nielsen}
\affiliation{Observatoire de Gen\'{e}ve, Universit\'{e} de Gen\'{e}ve, 51 Chemin des Maillettes, 1290 Sauverny, Switzerland}

\author{T. Nishiumi}
\affiliation{Kyoto Sangyo University, Motoyama, Kamigamo, Kita-Ku, Kyoto-City, 603-8555, Japan}

\author[0000-0001-8511-2981]{N. Narita}
\affiliation{Department of Astronomy, The University of Tokyo, 7-3-1 Hongo, Bunkyo-ku, Tokyo 113-0033, Japan}
\affiliation{Astrobiology Center, 2-21-1 Osawa, Mitaka, Tokyo 181-8588, Japan}
\affiliation{JST, PRESTO, 7-3-1 Hongo, Bunkyo-ku, Tokyo 113-0033, Japan}
\affiliation{National Astronomical Observatory of Japan, 2-21-1 Osawa, Mitaka, Tokyo 181-8588, Japan}
\affiliation{Instituto de Astrof\'\i sica de Canarias (IAC), 38205 La Laguna, Tenerife, Spain}

\author{E. Palle}
\affiliation{Instituto de Astrof\'\i sica de Canarias (IAC), 38205 La Laguna, Tenerife, Spain}
\affiliation{Departamento de Astrof\'\i sica, Universidad de La Laguna (ULL), 38206, La Laguna, Tenerife, Spain}

\author{H.M. Relles}
\affiliation{Center for Astrophysics \textbar{} Harvard \& Smithsonian, 60 Garden St., Cambridge, MA 02138, USA.}

\author{R. Sefako}
\affiliation{South African Astronomical Observatory, P.O. Box 9, Observatory, Cape Town 7935, South Africa}


\author[0000-0001-5603-6895]{T.G. Tan}
\affiliation{Perth Exoplanet Survey Telescope, Perth, Western Australia}




\author{M. Davies}
\affiliation{NASA Ames Research Center, Moffett Field, CA 94035, USA}


\author{Robert F. Goeke}
\affiliation{Department of Physics, and Kavli Institute for Astrophysics and Space Research, Massachusetts Institute of Technology, Cambridge, MA 02139, USA.}

\author[0000-0002-5169-9427]{N. Guerrero}
\affiliation{Department of Physics, and Kavli Institute for Astrophysics and Space Research, Massachusetts Institute of Technology, Cambridge, MA 02139, USA.}

\author{K. Haworth}
\affiliation{Department of Physics, and Kavli Institute for Astrophysics and Space Research, Massachusetts Institute of Technology, Cambridge, MA 02139, USA.}




\author{S. Villanueva}
\affiliation{Department of Physics, and Kavli Institute for Astrophysics and Space Research, Massachusetts Institute of Technology, Cambridge, MA 02139, USA.}
\affiliation{Pappalardo Fellow}




\begin{abstract}
Wide field surveys for transiting planets are well suited to searching diverse stellar populations, enabling a better understanding of the link between the properties of planets and their parent stars. We report the discovery of \hatstarab{} (TOI 625.01) and \hatstarbb{} (TOI 624.01), two new hot Jupiters around A stars from the HATNet survey which have also been observed by the \emph{Transiting Exoplanet Survey Satellite} (\emph{TESS}). \hatstarab{} has a mass of \hatstaraplmass{}$\,M_\mathrm{Jup}$ and a radius of \hatstaraplrad{}$\,R_\mathrm{Jup}$, and resides in a prograde 4.79-day orbit. \hatstarbb{} has a radius of \hatstarbplrad{}$\,R_\mathrm{Jup}$ and a mass constraint of \hatstarbplmass{}$\,M_\mathrm{Jup}$, and resides in a retrograde 2.74-day orbit. We use the confirmation of these planets around relatively massive stars as an opportunity to explore the occurrence rate of hot Jupiters as a function
of stellar mass. We define a sample of 47{,}126 main-sequence stars brighter than $T_\mathrm{mag}=10$ that yields \nTotal{} giant planet candidates, including \nPlanet{} confirmed planets, \nCandidate{} candidates, and \nFalsePositive{} false positives. We find a net hot Jupiter occurrence rate of \ortotal{} within this sample, consistent with the rate measured by \emph{Kepler} for FGK stars. When divided into stellar mass bins, we find the occurrence rate to be \orG{} for G stars, \orF{} for F stars, and \orA{} for A stars. Thus, at this point, we cannot discern any statistically significant trend in the occurrence of hot Jupiters with stellar mass. 
\end{abstract}



\section{Introduction}
\label{sec:introduction}

Radial velocity and transit surveys have been responsible for the discovery of about 400 close-in giant planets with periods less than 10 days\footnote{NASA Exoplanet Archive, 2019 April}. These ``hot Jupiters'' are the best characterized exoplanets, and are testbeds for nearly all the techniques to measure the densities, composition, atmospheres, orbital, and dynamical properties of exoplanetary systems. Hot Jupiters are also extreme examples of planetary migration, thought to have formed beyond the ice line, and migrated to their present-day locations via interactions with the protoplanetary gas disk, or via dynamical interactions with nearby planets or stars followed by tidal migration \citep[as recently reviewed by ][]{2018ARA&A..56..175D}.   

About three-quarters of the known hot Jupiters have emerged
from ground-based, wide-field transit surveys. These surveys have been successful not only in detecting a large number of planets,
but also in searching a wide range of stellar types, thanks
to their wide-field sky coverage. Transiting Jovian planets have been confirmed around stars ranging from M dwarfs (HATS-6 \citealt{2015AJ....149..166H}; NGTS-1 \citealt{2018MNRAS.475.4467B}; HATS-71 \citealt{2018arXiv181209406B}) to A stars (e.g. WASP-33 \citealt{2010MNRAS.407..507C}; KELT-9 \citealt{2017Natur.546..514G}). 

The properties of planets are thought to be dependent on the properties of the host stars. In particular, more massive stars may host more massive protoplanetary disks \citep[e.g.][]{2006A&A...452..245N}. Radial velocity surveys of intermediate-mass subgiants (``retired A stars'') reported that giant planets are more abundant around more massive stars,
but tend to have wider and more circular orbits
than their lower-mass main-sequence counterparts \citep{2010PASP..122..905J,2014A&A...566A.113J,2015A&A...574A.116R,2018ApJ...860..109G}. Data from the \emph{Kepler} primary mission allowed for
the determination of occurrence rates for planets as small as $1\,R_\oplus$ around FGK stars \citep[e.g.][]{2012ApJS..201...15H, 2013ApJ...766...81F, 2013ApJ...778...53D, 2013PNAS..11019273P, 2015ApJ...809....8B,2018AJ....155...89P}. In particular, occurrence rates from \emph{Kepler} indicate that small planets with orbital periods less than
a year are more common around less massive stars \citep{2013ApJ...767...95D,2015ApJ...814..130M}.

Despite this progress, many questions remain unanswered. Planets around main-sequence A stars are still poorly explored. A stars have radii as large as $4\,R_\odot$ on the main sequence, causing the transit depth of a Jovian planet to be 16 times smaller than it would be for a solar-type star. As such, ground-based transit surveys have poor completeness in this regime.
The \emph{Kepler} mission could have performed a sensitive search
for giant planets around A stars, but in fact very little data
from main-sequence A stars were obtained, because the mission was geared toward the detection of smaller planets for which FGK stars are more
favorable. For these reasons, there has been no robust determination
of the frequency of giant planets around main-sequence A stars.

There has also been tension between the occurrence rates of hot Jupiters measured by \emph{Kepler} ($0.43 \pm 0.05\%$ from \citealt{2013ApJ...766...81F}, $0.57_{-0.12}^{+0.14}\%$ from \citealt{2018AJ....155...89P}, $0.43^{+0.07}_{-0.06}$ from
\citealt{2017AJ....153..187M}) and those from radial velocity surveys ($1.5\pm0.6\%$ from \citealt{2008PASP..120..531C},  $1.2 \pm 0.4\%$ from \citealt{2012ApJ...753..160W}). These differences have been attributed to metallicity \citep[e.g.][]{2012ApJ...753..160W}, stellar age, or multiplicity (\citealt{2015ApJ...799..229W}, although see also \citealt{2018AJ....155..244B}). Surveying different populations with a diverse set of host stars may help resolve these tensions. 

The launch of the {\it Transiting Exoplanet Survey Satellite} \citep[\emph{TESS},][]{2016SPIE.9904E..2BR} heralds a new era of exoplanet characterization. In particular, the 30-minute cadence Full Frame Images (FFI) are providing us with an opportunity to search a wide range of
stellar types. Unlike {\it Kepler}, with {\it TESS} there is no
need to pre-select the target stars to
be within a certain range of masses or sizes.
Based on observations of 7 sky sectors between late 2018-07  and 2019-02, \emph{TESS} has delivered space-based photometry for 126{,}950 stars brighter than $T_\mathrm{mag} = 10$. The promise of near-complete sensitivity from space-based photometry to hot Jupiters across the main-sequence, and the availability of follow-up results from the tremendous efforts of the \emph{TESS} follow-up program motivates another look into the occurrence rates of hot Jupiters. 

In this paper, we describe the confirmation of two planets discovered by the HATNet survey around A stars, members of a relatively unexplored planet demographic.
\emph{TESS} data for these objects became available
during our confirmation process, and were independently identified as planet candidates based on FFI photometry. The follow-up observations, modeling of the systems, and derived system parameters are described in Sections~\ref{sec:obs} and \ref{sec:analysis}. In Section~\ref{sec:occurate}, we describe our estimates of the occurrence rates of hot Jupiters around main sequence A, F, and G stars. The estimate makes use of a magnitude-limited sample of main-sequence stars $(T_\mathrm{mag} < 10)$ surveyed by \emph{TESS} during its first seven sectors, planets catalogued in the \emph{TESS} Objects of Interest (TOI) list, existing planets from literature recovered by \emph{TESS}, and false-positive rates estimated via vetting observations of the \emph{TESS} follow-up program.

\section{Observations}
\label{sec:obs}

\hatstara{} and \hatstarb{} were identified as transiting planet candidates by the HATNet survey \citep{Bakos:2004}. \hatstara{} was observed by HATNet between 2010-11 and 2011-06, resulting in approximately 24{,}000 photometric data points. Subsequently, it received photometric and spectroscopic follow-up observations over 2011-2019 that confirmed its planetary nature. It was then observed during Sector 7 of the \emph{TESS} mission, flagged as a transiting planet candidate by the MIT quicklook pipeline (Huang et al., in preparation), and assigned {\it TESS} Object of Interest (TOI) number 625. These highly precise space-based photometric observations are subsequently incorporated in the analyses below. \hatstara{} was also independently identified as a planet candidate (1SWASPJ084201.35+034238.0) by the WASP survey \citep{2006PASP..118.1407P}, and was the subject
of extensive photometric follow-up via the WASP survey team.
These observations are described in Section~\ref{sec:photometry}, and
included in the global analyses.

\hatstarb{} was identified as a planet candidate based on nearly 10{,}000 HATNet observations spanning the interval from 2009-09 to 2010-03. Subsequent ground-based photometric follow-up observations were attempted during the 2016-2017 time frame, but these observations failed to recover the transit event due to the accumulation of uncertainty in the transit ephemerides. \hatstarb{} was also independently identified as a hot Jupiter candidate by the MNIT quicklook pipeline, and given the designation
TOI-624. The revised ephemeris from \emph{TESS} allowed us to successfully perform photometric and spectroscopic follow-up observations that confirmed the planetary nature of the system. \hatstarb{} was also identified by the WASP survey independently as a planet candidate (1SWASPJ045812.56+095952.7), receiving substantial ground-based photometric follow-up prior to the \emph{TESS} observations. 


\subsection{Photometry}
\label{sec:photometry}

\subsubsection{Candidate identification by HATNet}
\label{sec:hatnet}

\begin{figure*}
\begin{tabular}{cc}
\hatstara{} & \hatstarb{} \\
\includegraphics[width=0.5\textwidth]{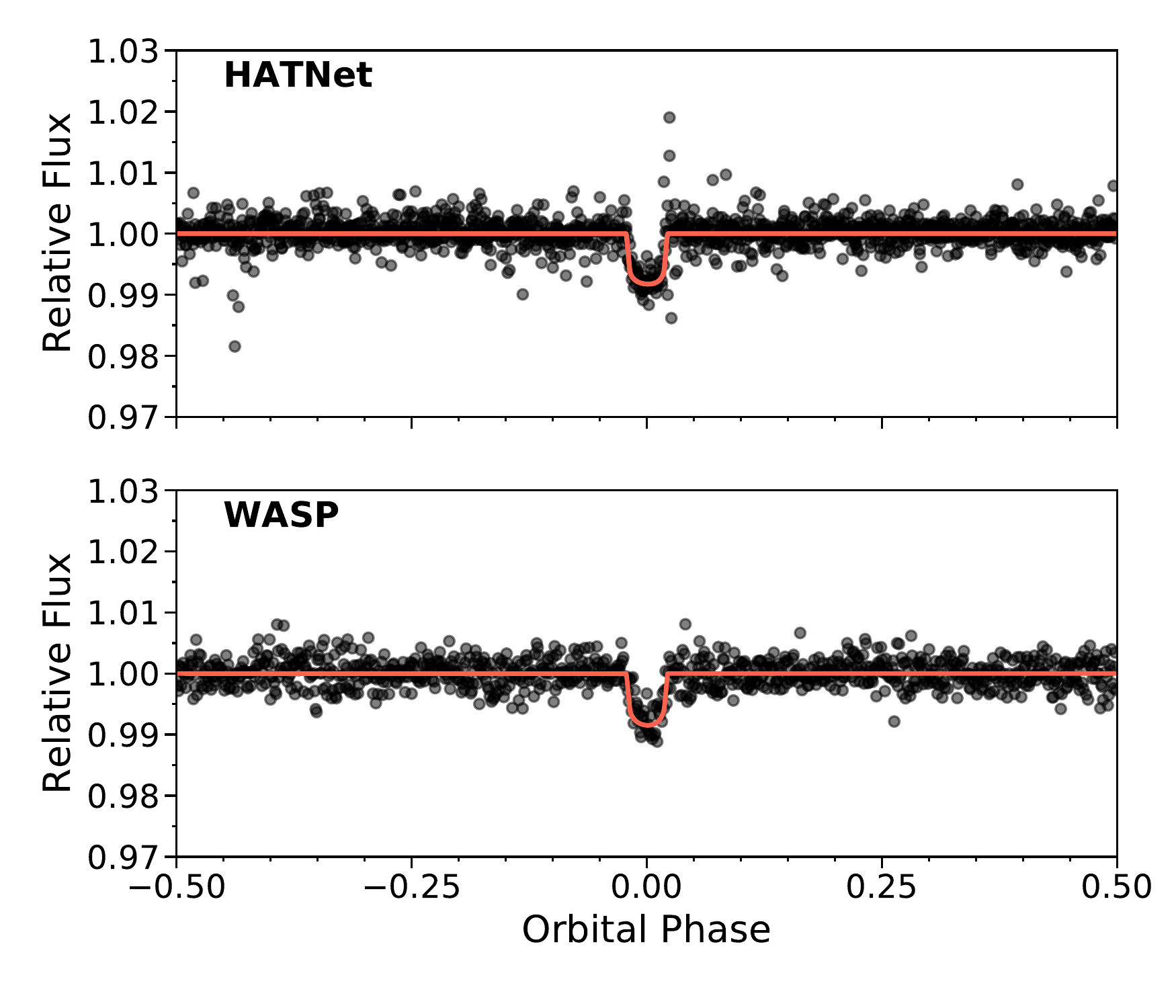}     &  
\includegraphics[width=0.5\textwidth]{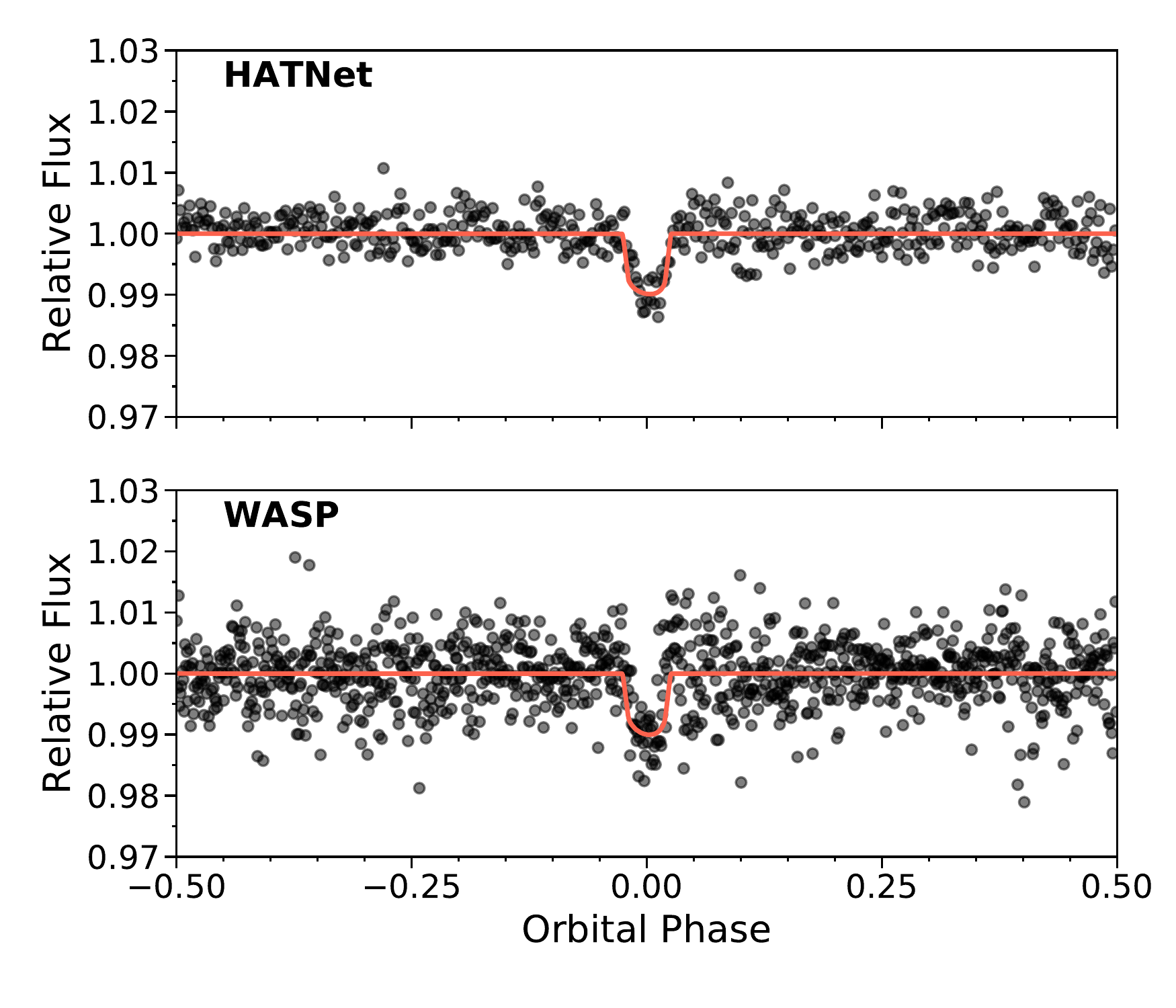}\\
\end{tabular}
\caption{
Discovery light curves of \hatstara{} \textbf{(left)} and \hatstarb{} \textbf{(right)}. The light curves have been averaged in phase
with bins of width 0.002. The top panels show the HATNet light curves,
and the bottom panels show the WASP light curves.
\label{fig:hnlc}}
\end{figure*}

The HATNet survey \citep{Bakos:2004} is one of the longest running wide-field photometric surveys for transiting planets. It employs a network of small robotic telescopes at the Fred Lawrence Whipple Observatory (FLWO) in Arizona, and at Mauna Kea Observatory in Hawaii. Each survey field is $8^\circ\times8^\circ$, and observations are obtained with the Sloan $r'$ filter. Observations are reduced following the process laid out by
\citet{2010ApJ...710.1724B}. Light curves were extracted via aperture photometry. Systematic effects were mitigated using External Parameter Decorrelation \citep[EPD,][]{2007ApJ...670..826B}, and the Trend Filtering Algorithm \citep[TFA,][]{2005MNRAS.356..557K}. Periodic transit signals were identified via the Box-fitting Least Squares analysis \citep[BLS,][]{2002A&A...391..369K}. The HATNet observations are summarized in Table~\ref{tab:photobs}, and the discovery light curves are shown in Figure~\ref{fig:hnlc}.

\subsubsection{\emph{TESS} observations}
\label{sec:tess}

\hatstara{} and \hatstarb{} were observed by \emph{TESS} during Year 1 of its primary mission. \hatstara{} is present in the Camera 1 FFIs obtained
during the Sector 7 campaign, between 2019-01-07 and 2019-02-02. \hatstarb{} is present on the Camera 1 FFIs in Sector 5, between 2018-11-15 and 2018-12-11. \emph{TESS} FFIs provide approximately 27\,days of nearly
continuous monitoring for all stars within its field of view. 

\begin{figure}
\begin{tabular}{cc}
    \hatstara{} & \hatstarb{} \\
    \includegraphics[width=0.2\textwidth]{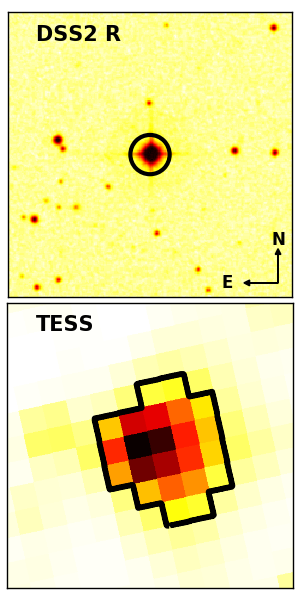} &
    \includegraphics[width=0.2\textwidth]{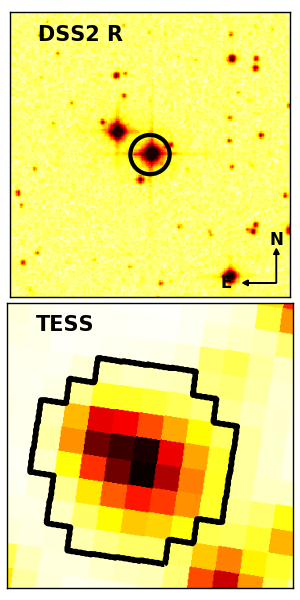}  \\
\end{tabular}
\caption{
    Fields surrounding each of the planet-hosting stars. \textbf{Top} $4\arcmin \times 4\arcmin$ Digitized Sky Survey $R$ cutouts of \hatstara{} and \hatstarb{}. \textbf{Bottom} \emph{TESS} Full Frame Image cutouts of \hatstara{} and \hatstarb{}. The DSS and \emph{TESS} cutouts are plotted at the same scale and orientation. The photometric apertures used to extract the \emph{TESS} light curves are marked. 
\label{fig:tesspixel}}
\end{figure}

We extracted the FFI light curves of the two systems with the \emph{lightkurve} package \citep{lightkurve} using the public FFI images on MAST archive produced from the Science Processing Operations Center (SPOC) pipeline \citep{2016SPIE.9913E..3EJ}. The raw aperture photometry light curves are diluted due to the presence of nearby bright stars. In particular, \hatstarb{} is located within $33\arcsec$ (1.6 pixels) of a fainter star with a magnitude difference of $\Delta T_{\rm mag} = 0.75$. We extracted $10\times10$ pixel subrasters
surrounding each star, and defined photometric apertures to include
all pixels with fluxes higher than 68\% of the fluxes of nearby pixels. For \hatstarb{}, this aperture includes both the target star and the nearby neighbor. For \hatstara{}, the photometric aperture does not contain
any other stars within 6 magnitudes of the target star.
Nearby pixels of apparently blank sky were used to estimate the background flux surrounding the target star. Figure~\ref{fig:tesspixel} shows each star as observed by \emph{TESS}, along with the photometric aperture. An $R$ band image of the star field from the Digitized Sky Survey 2 \citep{2000ASPC..216..145M} is also shown for reference. The extracted light curve of \hatstarb{} was then deblended, based on the magnitudes of nearby stars from version 6 of the \emph{TESS} Input Catalog \citep{2018AJ....156..102S}.

\begin{figure*}
\centering
\includegraphics[width=0.9\textwidth]{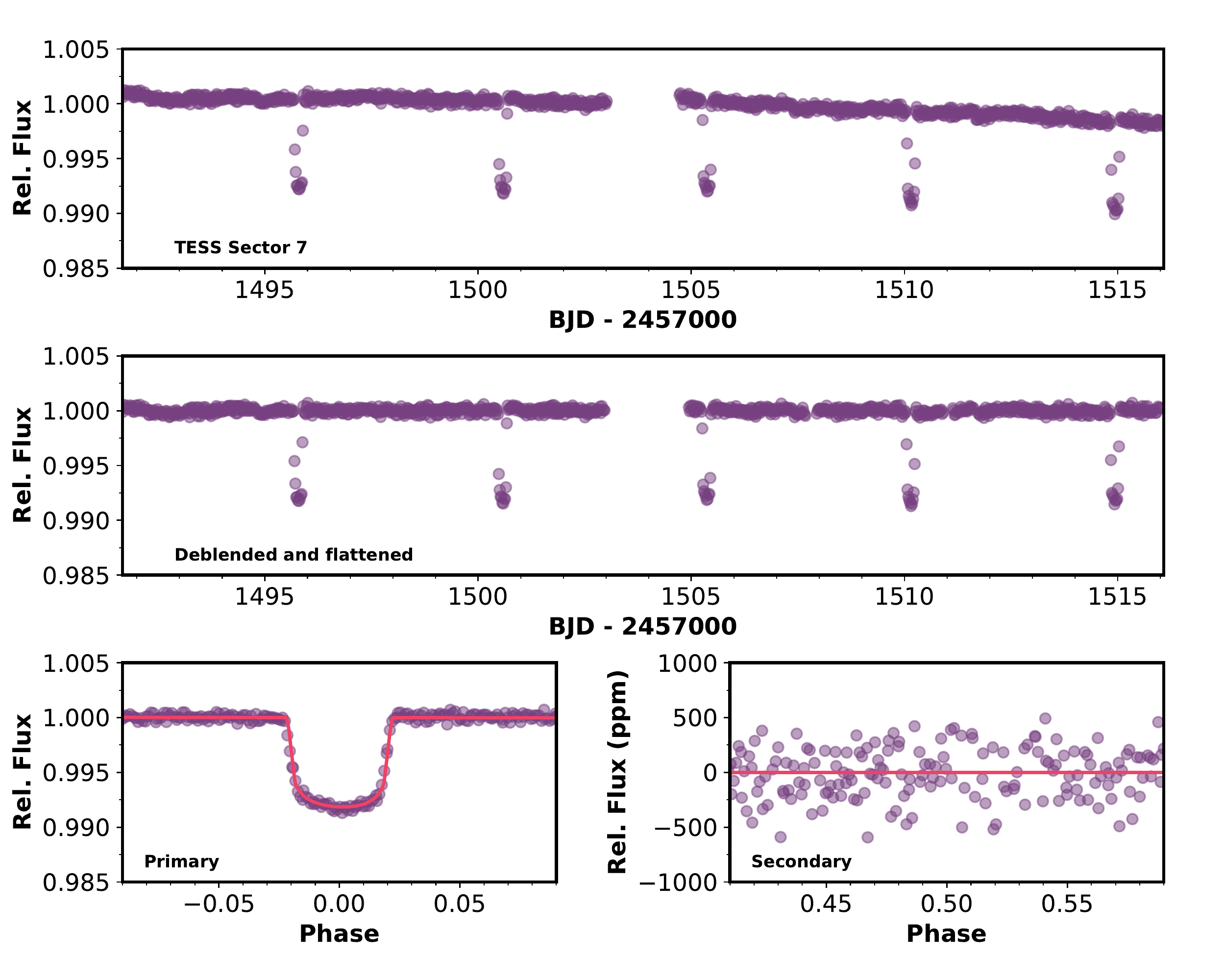}
\caption{
    \emph{TESS} light curve of \hatstara{}. \textbf{Top} Raw \emph{TESS} light curve. \textbf{Center} Detrended light curve. \textbf{Lower left} Detrended light curve phase folded to the transit ephemeris, showing the transit and associated best fit model (plotted in red). \textbf{Lower right} Detrended light curve in the region of the secondary eclipse, assuming circular orbit. 
\label{fig:HTR413001_tess}}
\end{figure*}

\begin{figure*}
\centering
\includegraphics[width=0.9\textwidth]{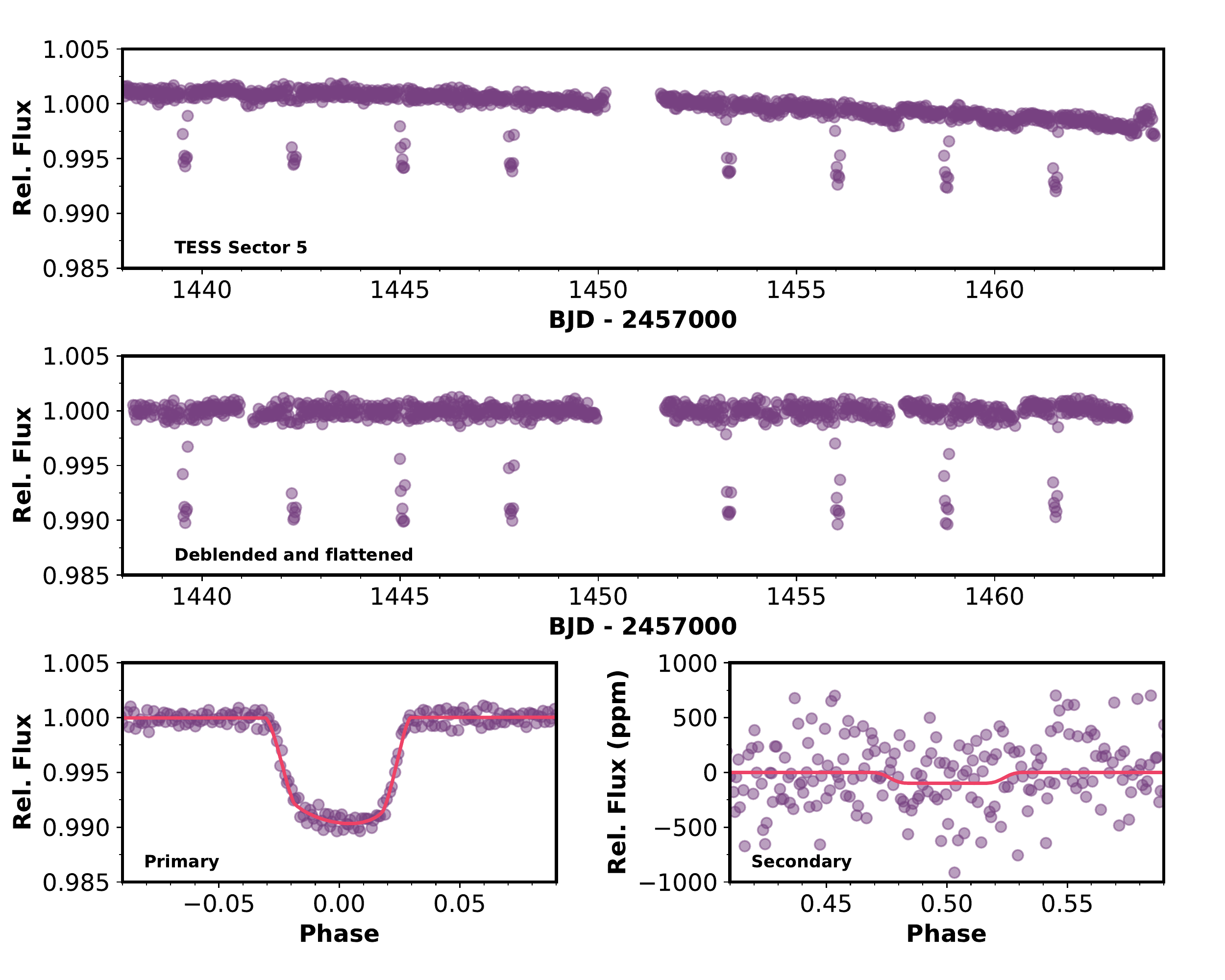}
\caption{
    TESS light curve of \hatstarb{}. Panel contents as per described in Figure~\ref{fig:HTR413001_tess}. The tentative detection of a secondary eclipse, with a depth of $159 \pm 65$ ppm is shown in the lower right panel. The best fit model is shown in red. 
\label{fig:HTR358007_tess}}
\end{figure*}

Figures~\ref{fig:HTR413001_tess} and \ref{fig:HTR358007_tess} present the \emph{TESS} light curves of the target stars. The \emph{TESS} light curves of \hatstara{} and \hatstarb{} show no large systematic variation, nor signs of pulsations or additional eclipsing companions. The \emph{TESS} transit signals agree in depth with the
depths that are measured from ground-based observations. 


\paragraph{Phase modulation and secondary eclipses}

Hot Jupiters on circular orbits are expected to be tidally locked \citep[e.g.][]{2008EAS....29....1M}, with a fixed dayside atmosphere facing the star at all times. As a result, there can be large temperature differences between the dayside and non-illuminated nightside. During secondary eclipse, when the planet passes behind the star, the total flux from the dayside is occulted. In addition, as the planet orbits the host star, the flux from the planet's sky-projected hemisphere changes periodically, producing an atmospheric brightness modulation. 

To search for these signals in the \emph{TESS} data, we fit a simple phase curve model to the full light curve (transits, secondary eclipses, and out-of-eclipse flux modulation), following the methods described in detail in \citet{2019AJ....157..178S}. Given the geometry of the system, the extrema of the atmospheric brightness modulation occur during conjunction, i.e., a cosine of the orbital phase. The out-of-eclipse flux is therefore given by $F(t) = 1+ B_1\cos(\phi)$, where $\phi=2\pi(t-T_c)$ is the orbital phase, and $B_1$ is the semi-amplitude of the phase curve signal. We include secondary eclipse signals halfway between transits, with a depth parametrized by $f_p$, i.e., the relative brightness of the planet's dayside hemisphere. 

Since we are interested in temporal signals in the out-of-eclipse light curve, we do not use the detrended time series and instead multiply the phase curve model by generalized polynomials in time to capture all non-astrophysical time-dependent signals in the raw light curve, which are likely attributable to instrumental systematics. The raw light curves shown in Figures 3 and 4 display clear long-term temporal trends, as well as discontinuities in flux that occur during momentum dumps. 

Given these discontinuities, we split each light curve into small segments separated by momentum dumps and fit a separate polynomial systematics model to each segment. The orders of the polynomials used in the final fit are determined by first fitting each segment individually and minimizing the Bayesian Information Criterion (BIC), defined as $BIC = \chi^2 + k \ln(n)$, where $k$ is the number of fitted parameters, and $n$ is the number of data points. After optimizing the polynomial orders, we carry out a joint fit of the full light curve.

For \hatstarb{}, we find that the non-astrophysical systematics in the segments are well-described by polynomials of second to third order. In the joint fit, we report a marginal $2.4\sigma$ secondary eclipse detection of $159 \pm 65$ ppm, while the atmospheric brightness modulation amplitude is consistent with zero. Figure~\ref{fig:HTR358007_tess} shows the systematics-corrected and phase-folded light curve in the vicinity of the secondary eclipse, along with the best-fit model. 

To evaluate the statistical significance of this \hatstarbb{} secondary eclipse detection, we compare the BIC of a joint fit that includes only transits and secondary eclipses (fixing $B_1$ to zero) with the BIC of a fit that assumes a flat out-of-transit light curve (fixing $B_1$ and $f_p$ to zero). The difference in BIC is less than 0.1, indicating that the secondary eclipse detection is not formally statistically robust. From an analogous analysis of the \hatstara{} phase curve, we do not detect any significant secondary eclipse depth or phase curve signal.

\subsubsection{Independent identification by WASP}
\label{sec:wasp}

\hatstara{} and \hatstarb{} were both independently identified as planet candidates by the WASP survey \citep{2019MNRAS.483.5534S}. The northern facility (SuperWASP-North) and the southern facility (WASP-South) both consist of arrays of eight 200\,mm f/1.8 Canon telephoto lenses on a common mount. Each camera is coupled with $2\mathrm{K} \times 2\mathrm{K}$ detectors, yielding a field of view of $7.8\times7.8^\circ$ per camera \citep{2006PASP..118.1407P}. \hatstara{} was observed by both WASP-South and SuperWASP-North, producing 25{,}200 photometric points spanning from 2009-01-14 to 2012-04-23. \hatstarb{} was observed by SuperWASP-North, producing 19{,}200 observations spanning 2008 October 13 to 2011 February 04. These long baseline observations are plotted in Figure~\ref{fig:hnlc}, and were included in the global modeling (Section~\ref{sec:globalmodel}) to help refine the transit ephemeris.

\subsubsection{Ground-based follow-up observations}
\label{sec:fulc}

A series of facilities provided follow-up photometry of \hatstara{} and \hatstarb{} to confirm the transit signal, improve the
determination of the the planet radius, and increase the precision
of the transit ephemeris. A number of transit observations were obtained
with the FLWO 1.2\,m telescope and KeplerCam, a $4\,\mathrm{K}\times4\mathrm{K}$ CCD camera operated with $2\times2$ binning, giving a plate scale of $0\farcs672\,\mathrm{pixel}^{-1}$. Photometry were extracted as per \citet{2010ApJ...710.1724B}. Follow-up photometry were also obtained using the Las Cumbres Observatory \citep[LCO,][]{2013PASP..125.1031B} network. These observations included transits obtained via the 0.8\,m LCO telescope located at the Byrne Observatory at Sedgwick, California, using the SBIG STX-16803 $4\mathrm{K}\times4\mathrm{K}$ camera with a field of view of $16\arcmin \times16\arcmin$. Observations were also obtained using the 1\,m LCO telescope at Siding Spring Observatory, Australia, using the Sinistro Fairchild CCD, with a field of view of $27\arcmin \times 27 \arcmin$ over the $4\mathrm{K} \times 4\mathrm{K}$ detector. Additional photometric follow-up were obtained using the TRAPPIST (TRAnsiting Planets and PlanetesImals Small Telescope) North facility \citep{2011Msngr.145....2J,2013A&A...552A..82G,2019AJ....157...43B} at Oukaimeden Observatory in Morocco. TRAPPIST-North is a $0.6$\,m robotic photometer employing a $2\mathrm{K}\times2\mathrm{K}$ CCD with a field of view of $19\farcm8 \times19\farcm8$ at a plate scale of $0\farcs6$ per pixel.

\begin{figure}
    \includegraphics[width=0.5\textwidth]{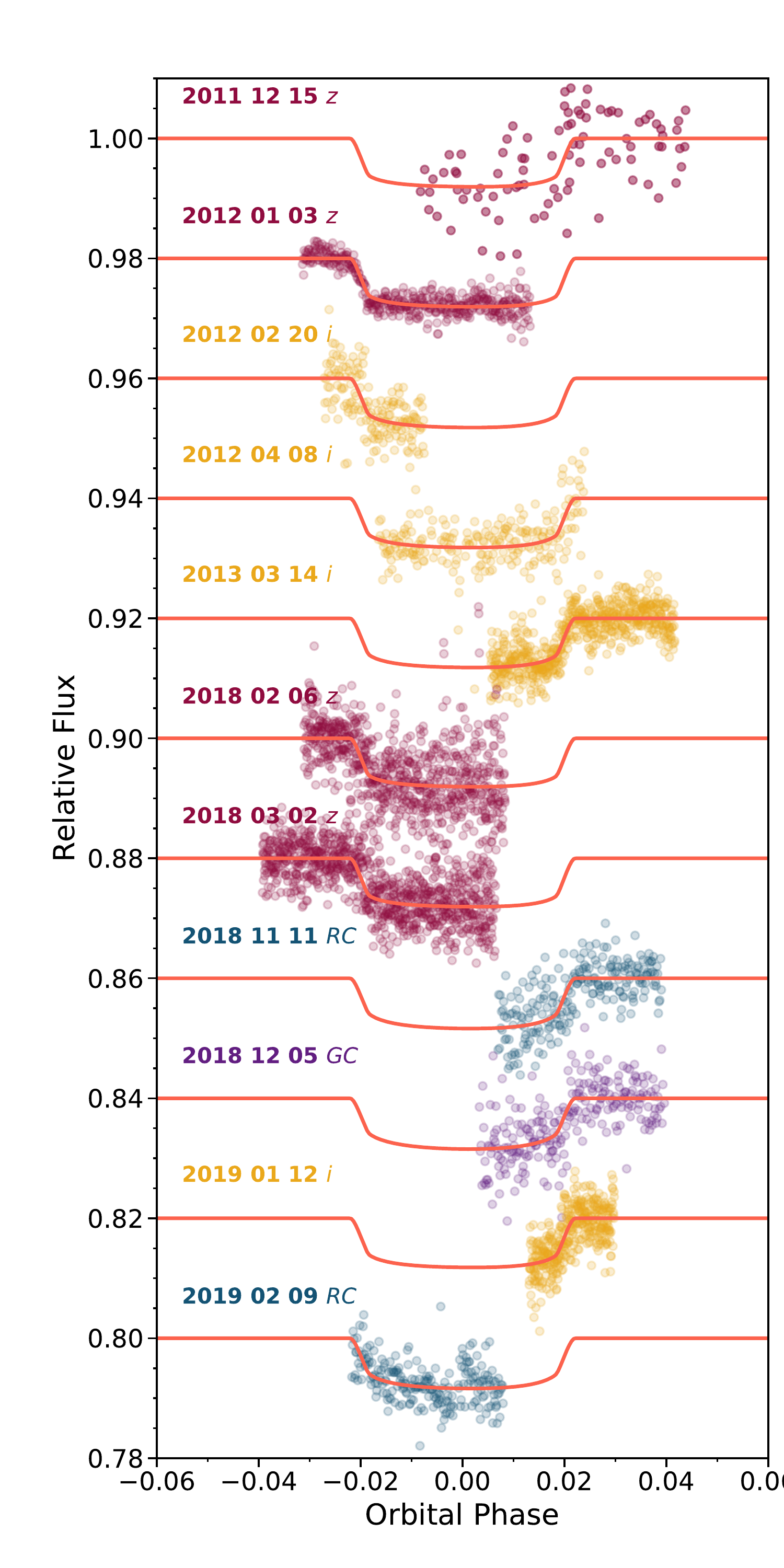}
\caption{
Ground based follow-up light curves for \hatstara{}, vertically separated for clarity. The photometric bandpass and date of the observations are labeled. The facilities contributing to each light curve are presented in Table~\ref{tab:photobs}. 
\label{fig:HTR413001_fulc}}
\end{figure}

\begin{figure}
    \includegraphics[width=0.5\textwidth]{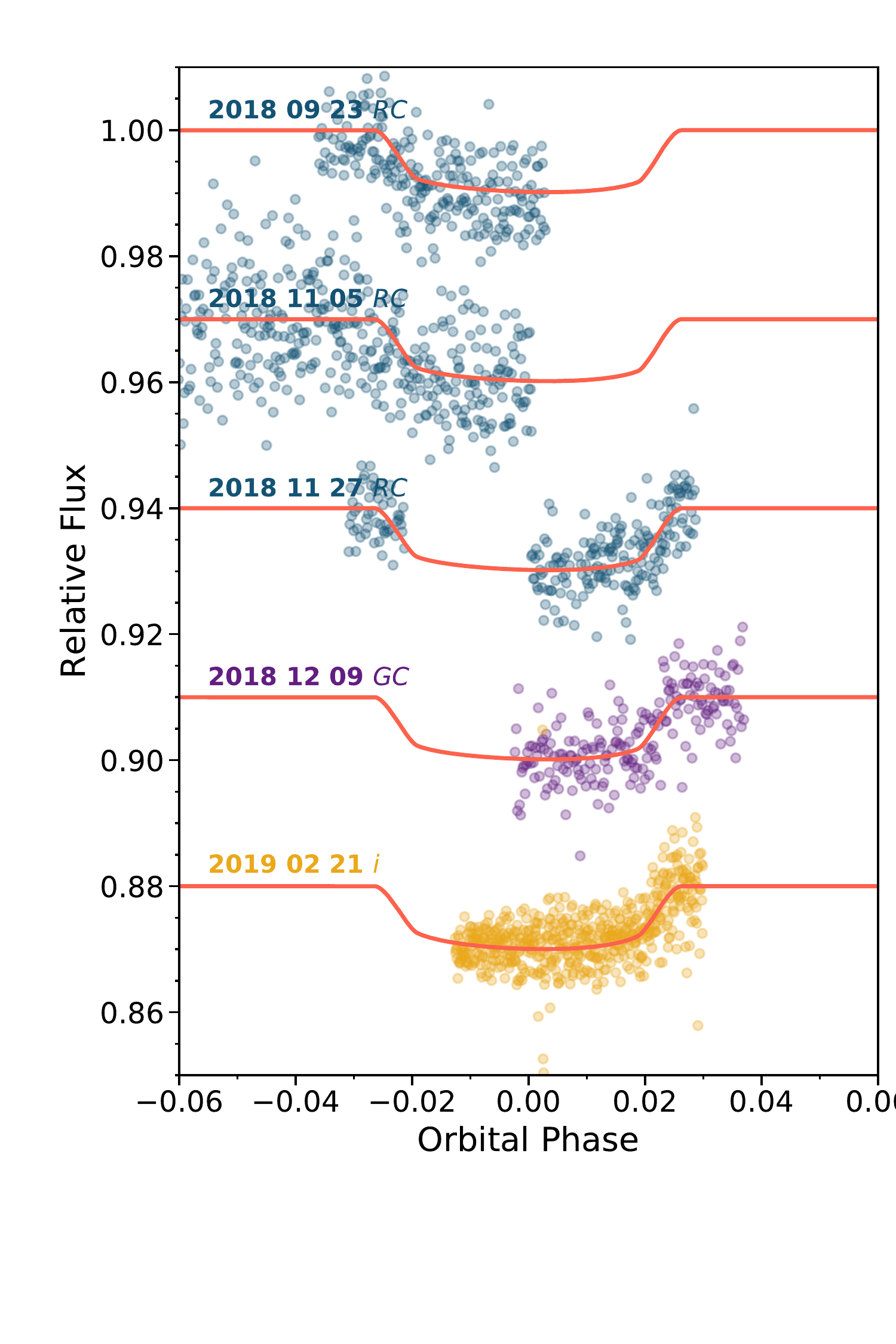}
\caption{
Ground based follow-up light curves for \hatstarb{}, description as per Figure~\ref{fig:HTR413001_fulc}. 
\label{fig:HTR358-007_fulc}}
\end{figure}

The dates, cadences, and filters used in these observations are summarized in Table~\ref{tab:photobs}. The light curves are made available in Tables~\ref{tab:lc_table_starb} and \ref{tab:lc_table_stara}, and shown in Figures~\ref{fig:HTR413001_fulc} and \ref{fig:HTR358-007_fulc}. 

\begin{deluxetable*}{lllrrr}
\tablewidth{0pc}
\tabletypesize{\scriptsize}
\tablecaption{
    Summary of photometric observations
    \label{tab:photobs}
}
\tablehead{
    &
    &
    &
    &
    &\\
    \multicolumn{1}{c}{Target}          &
    \multicolumn{1}{c}{Facility}          &
    \multicolumn{1}{c}{Date(s)}             &
    \multicolumn{1}{c}{Number of Images}\tablenotemark{a}      &
    \multicolumn{1}{c}{Cadence (s)}\tablenotemark{b}         &
    \multicolumn{1}{c}{Filter}            
}
\startdata
\hatstara{} & WASP-South/North & 2009-01-14 -- 2012-04-23 & 25282 & 432 & WASP Broadband\\
\hatstara{} & HAT-6 & 2010-11-02 -- 2011-04-21 & 10384 & 229 & $r$ \\
\hatstara{} & HAT-7 & 2010-11-02 -- 2011-05-25 & 8707 & 233 & $r$ \\
\hatstara{} & HAT-7 & 2011-02-14 -- 2011-06-03 & 4539 & 215 & $r$ \\
\hatstara{} & KeplerCam 1.2\,m & 2011-12-15 & 93 & 170 & $z$ \\
\hatstara{} & KeplerCam 1.2\,m & 2012-01-03 & 417 & 44 & $z$ \\
\hatstara{} & LCO BOS 1.0\,m & 2012-02-20 & 170 & 48 & $i$ \\
\hatstara{} & LCO BOS 1.0\,m & 2012-04-08 & 223 & 68 & $i$ \\
\hatstara{} & KeplerCam 1.2\,m & 2013-03-14 & 617 & 24 & $i$ \\
\hatstara{} & KeplerCam 1.2\,m & 2018-02-06 & 759 & 22 & $z$ \\
\hatstara{} & KeplerCam 1.2\,m & 2018-03-02 & 886 & 22 & $z$ \\
\hatstara{} & TRAPPIST 0.6\,m & 2018-11-11 & 234 & 60 & $RC^c$ \\
\hatstara{} & TRAPPIST 0.6\,m & 2018-12-05 & 251 & 60 & $GC^d$ \\
\hatstara{} & KeplerCam 1.2\,m & 2019-01-12 & 381 & 18 & $i$ \\
\hatstara{} & TRAPPIST 0.6\,m & 2019-02-09 & 223 & 52 & $RC$ \\
\hatstara{} & \emph{TESS} & 2019-01-08 -- 2019-02-01 & 1087 & 1800 & $TESS$ \\
[1.5ex]
\hatstarb{} & WASP-North & 2008-10-13 -- 2011-02-04 & 19266 & 351 & WASP Broadband \\
\hatstarb{} & HAT-9 & 2009-09-19 -- 2010-03-30 & 9987 & 224 & $r$ \\
\hatstarb{} & TRAPPIST 0.6\,m & 2018-09-23 & 238 & 40 & $RC$ \\
\hatstarb{} & TRAPPIST 0.6\,m & 2018-11-05 & 376 & 40 & $RC$ \\
\hatstarb{} & TRAPPIST 0.6\,m & 2018-11-27 & 231 & 35 & $RC$ \\
\hatstarb{} & TRAPPIST 0.6\,m & 2018-12-09 & 209 & 42 & $GC$ \\
\hatstarb{} & \emph{TESS} & 2018-11-15 -- 2018-12-10 & 1024 & 1800 & $TESS$ \\
\hatstarb{} & KeplerCam 1.2\,m & 2019-02-21 & 563 & 18 & $i$ \\
\enddata 
\tablenotetext{a}{
  Outlying exposures have been discarded.
}
\tablenotetext{b}{
  Median time difference between points in the \lc. Uniform sampling was not possible due to visibility, weather, pauses.
}
\tablenotetext{c}{
  RC: Red continuum filter centered at $7128\,\AA$ with width of $58\,\AA$
}
\tablenotetext{d}{
  GC: Green continuum filter centered at $5260\,\AA$ with width of $65\,\AA$
}
\end{deluxetable*}

\begin{deluxetable*}{rrrrrrr}

\tablewidth{0pc}
\tabletypesize{\scriptsize}
\tablecaption{
        Differential photometry of \hatstara{}
    \label{tab:lc_table_starb}
}
\tablehead{
    &
    &
    &
    & \\
    \multicolumn{1}{c}{BJD}          &
    \multicolumn{1}{c}{Mag (Raw)}\tablenotemark{a}             &
    \multicolumn{1}{c}{Mag (EPD)}            &
    \multicolumn{1}{c}{Mag (TFA)}            &
    \multicolumn{1}{c}{$\sigma$ Mag}      &
    \multicolumn{1}{c}{Instrument}         &
    \multicolumn{1}{c}{Filter}            
}
\startdata  
2455502.9688207 & 9.12413 & 10.01248 & 10.00835 & 0.00161& HATNet & r'\\
2455502.9733846 & 9.11519 & 10.0089 & 10.00331 & 0.0016& HATNet & r'\\
2455502.9776452 & 9.11541 & 10.01343 & 10.00835 & 0.0016& HATNet & r'\\
2455502.9819047 & 9.12139 & 10.01393 & 10.01161 & 0.0016& HATNet & r'\\
2455502.9862569 & 9.10128 & 10.00651 & 9.99933 & 0.00159& HATNet & r'\\
\enddata 
\tablenotetext{a}{
This table is available in a machine-readable ascii file. A portion is shown here for guidance regarding its form and content.\\
Raw, EPD, and TFA magnitudes are presented for HATNet light curves. The detrending and potential blending may cause the HATNet transit to be shallower than the true transit in the EPD and TFA light curves. This is accounted for in the global modeling by the inclusion of a dilution factor. Follow-up light curves have been treated with EPD simultaneous to the transit fitting. Pre-EPD magnitudes are presented for the follow-up light curves.
}
\end{deluxetable*}

\begin{deluxetable*}{rrrrrrr}

\tablewidth{0pc}
\tabletypesize{\scriptsize}
\tablecaption{
        Differential photometry of \hatstarb{}
    \label{tab:lc_table_stara}
}
\tablehead{
    &
    &
    &
    & \\
    \multicolumn{1}{c}{BJD}          &
    \multicolumn{1}{c}{Mag (Raw)}\tablenotemark{a}             &
    \multicolumn{1}{c}{Mag (EPD)}            &
    \multicolumn{1}{c}{Mag (TFA)}            &
    \multicolumn{1}{c}{$\sigma$ Mag}      &
    \multicolumn{1}{c}{Instrument}         &
    \multicolumn{1}{c}{Filter}            
}
\startdata  
2455093.9914136 & 8.8238 & 9.69444 & 9.69370 & 0.00177& HATNet & r'\\
2455093.9939800 & 8.83789 & 9.69611 & 9.70024 & 0.00179& HATNet & r'\\
2455093.9966693 & 8.84903 & 9.67121 & 9.67967 & 0.0018& HATNet & r'\\
2455093.9993076 & 8.80637 & 9.70271 & 9.69896 & 0.00177& HATNet & r'\\
2455094.0019585 & 8.84992 & 9.69871 & 9.69148 & 0.00181& HATNet & r'\\
\enddata 
\tablenotetext{a}{
This table is available in a machine-readable form in the online journal. A portion is shown here for guidance regarding its form and content.\\
Raw, EPD, and TFA magnitudes are presented for HATNet light curves. The detrending and potential blending may cause the HATNet transit to be shallower than the true transit in the EPD and TFA light curves. This is accounted for in the global modeling by the inclusion of a dilution factor. Follow-up light curves have been treated with EPD simultaneous to the transit fitting. Pre-EPD magnitudes are presented for the follow-up light curves.
}
\end{deluxetable*}

\subsection{Spectroscopy}
\label{sec:spectroscopy}

We carried out a series of spectroscopic follow-up observations to confirm the nature of the transiting candidates, constrain the masses, and measure the orbital obliquities of the companions. The observations are listed in Table~\ref{tab:specobssummary}, and summarized below. 

\begin{figure}
\centering
    \includegraphics[width=0.4\textwidth]{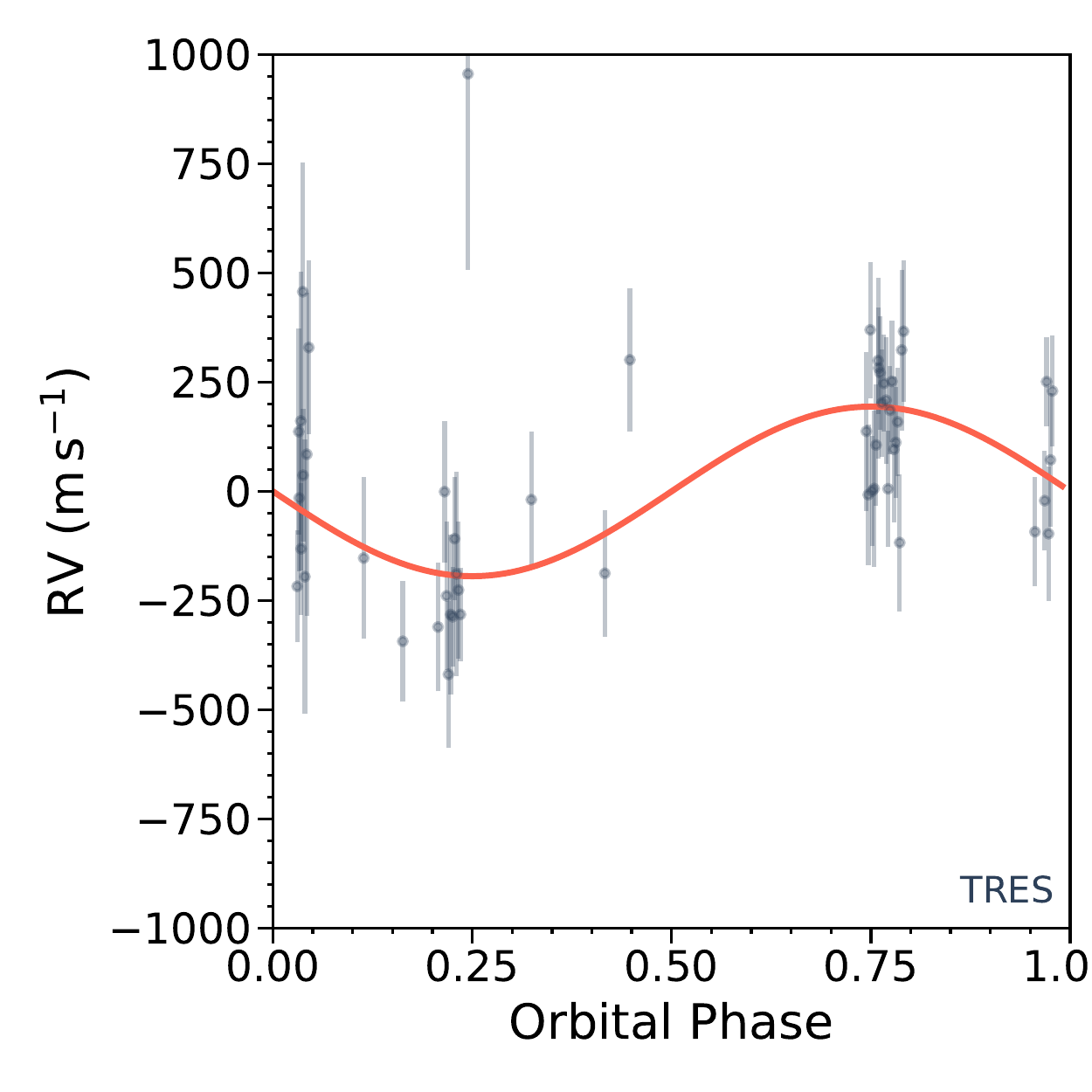}
\caption{
TRES radial velocities for \hatstara{}. The best fit orbit from the global model is plotted in red. The fitted radial velocity jitter has been added to the per-point uncertainties in quadrature. 
\label{fig:HTR413001_rv}}
\end{figure}

\begin{figure}
\centering
    \includegraphics[width=0.4\textwidth]{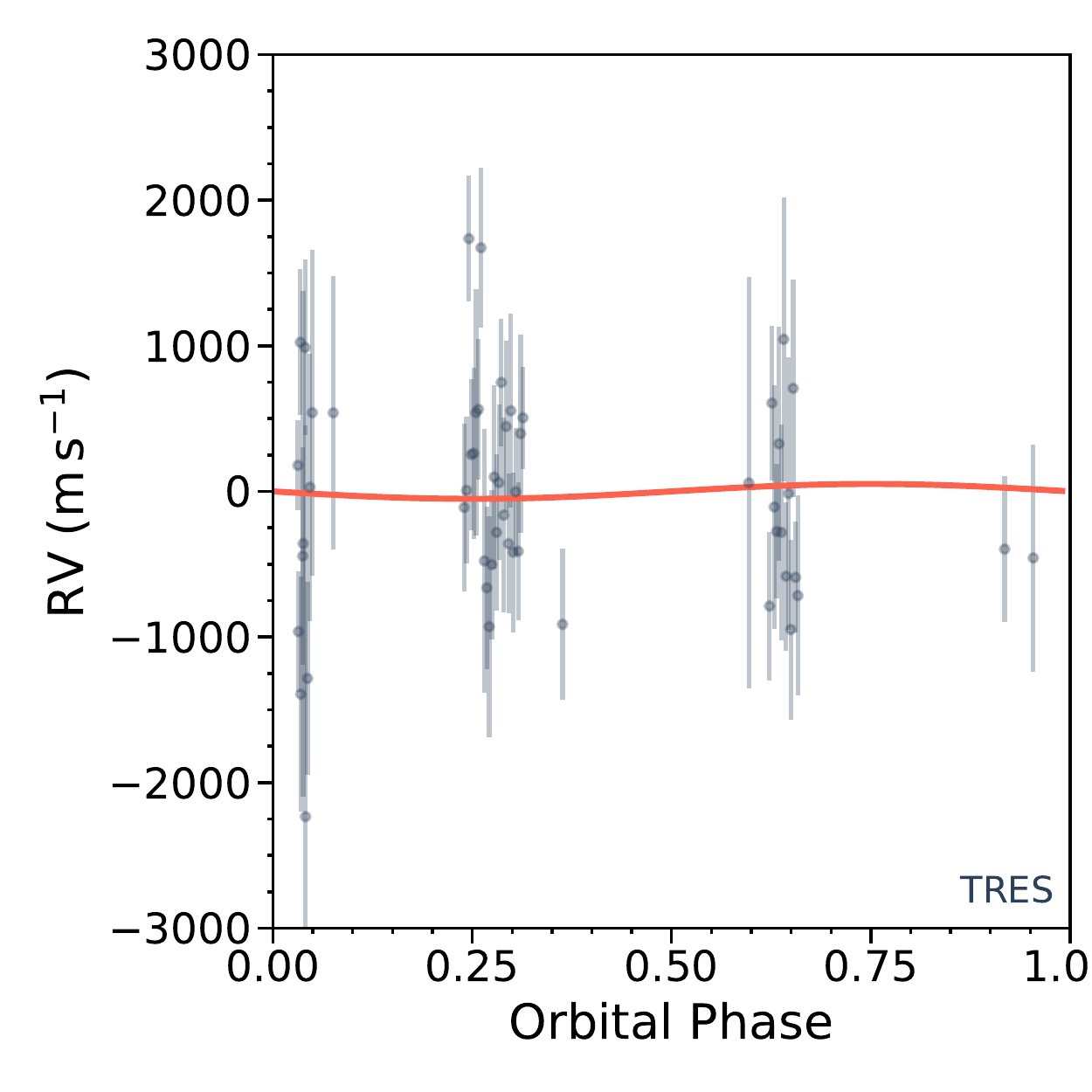}
\caption{
TRES radial velocities for \hatstarb{}, caption as per Figure~\ref{fig:HTR413001_rv}. 
\label{fig:HTR358007_rv}}
\end{figure}

The Tillinghast Reflector Echelle Spectrograph \citep[TRES,][]{Furesz:2008} on the 1.5\,m telescope at FLWO, Arizona, was used to obtain dozens of spectra for each system. TRES is a fiber fed echelle spectrograph, with a spectral resolution of $R = 44000$ over the wavelength region of $3850-9100$\,\AA. The observing strategy and data reduction process are described by \citet{2012Natur.486..375B}. Each spectrum is measured from the combination of three consecutive observations for optimal cosmic ray rejection, and the wavelength solution is provided by bracketing ThAr hollow cathode lamp exposures. A series of TRES spectra were obtained at phase quadratures to most efficiently constrain the mass of the planets. For \hatstara{}, relative radial velocities were obtained using a multi-order analysis \citep{2012ApJ...745...80Q} of the TRES spectra. For \hatstarb{}, we modeled the stellar line profiles derived from a least-squares deconvolution \citep[LSD,][]{1997MNRAS.291..658D} to derive the absolute radial velocities of each spectrum. In our
experience with rapidly rotating stars, the best radial velocities are
obtained by modeling of the LSD-derived line profiles. The TRES velocities for \hatstara{} and \hatstarb{} are listed in Tables~\ref{tab:rv_table_stara} and \ref{tab:rv_table_starb}, and plotted in Figures~\ref{fig:HTR413001_rv} and \ref{fig:HTR358007_rv}, respectively. 

\begin{figure*}
\centering
\begin{tabular}{ll}
    \multicolumn{2}{c}{\textbf{\large\hatstara{}}} \\ 
    \multicolumn{2}{c}{\includegraphics[width=0.3\textwidth]{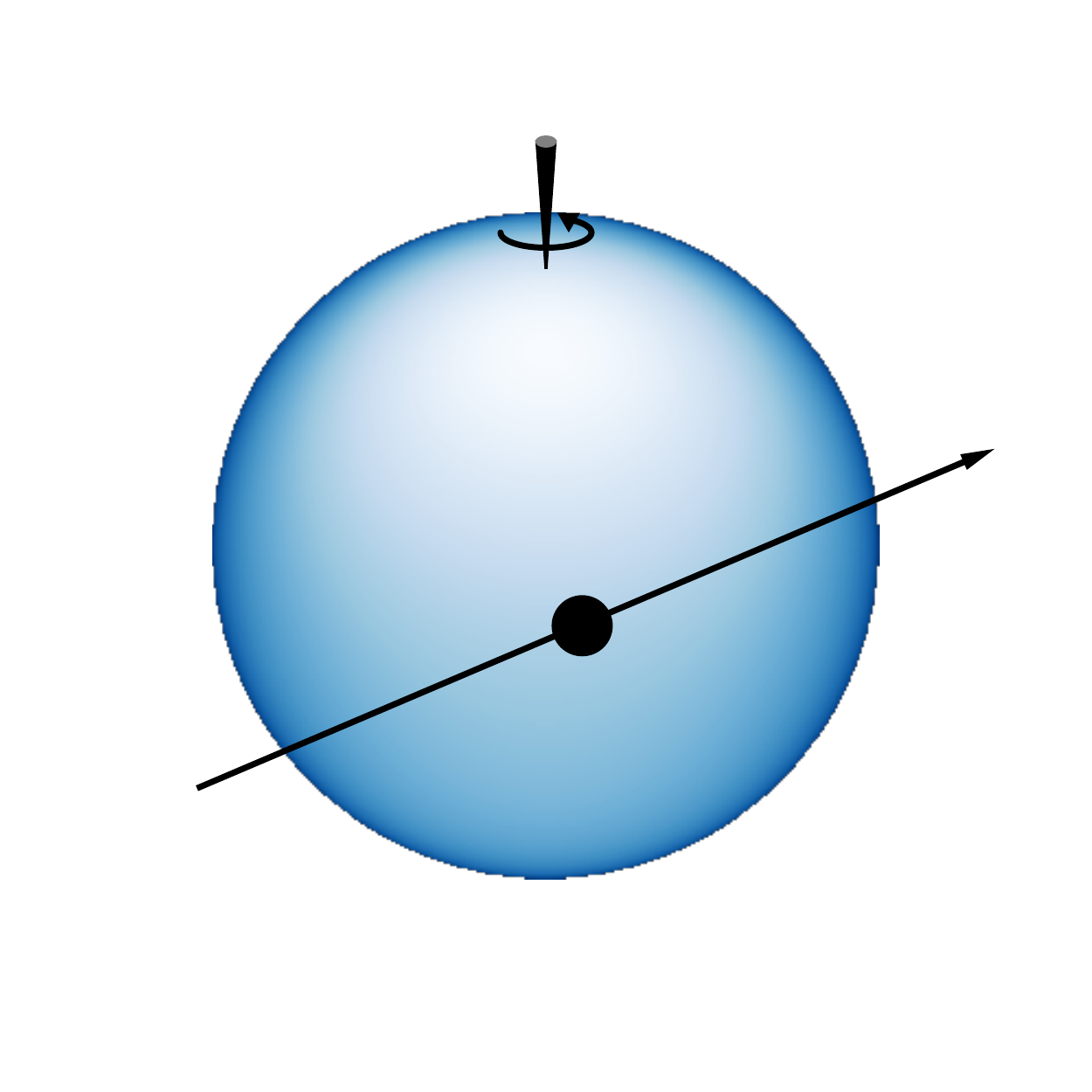}} \\
    \textbf{TRES} & \textbf{SALT} \\ 
    \includegraphics[width=0.3\textwidth]{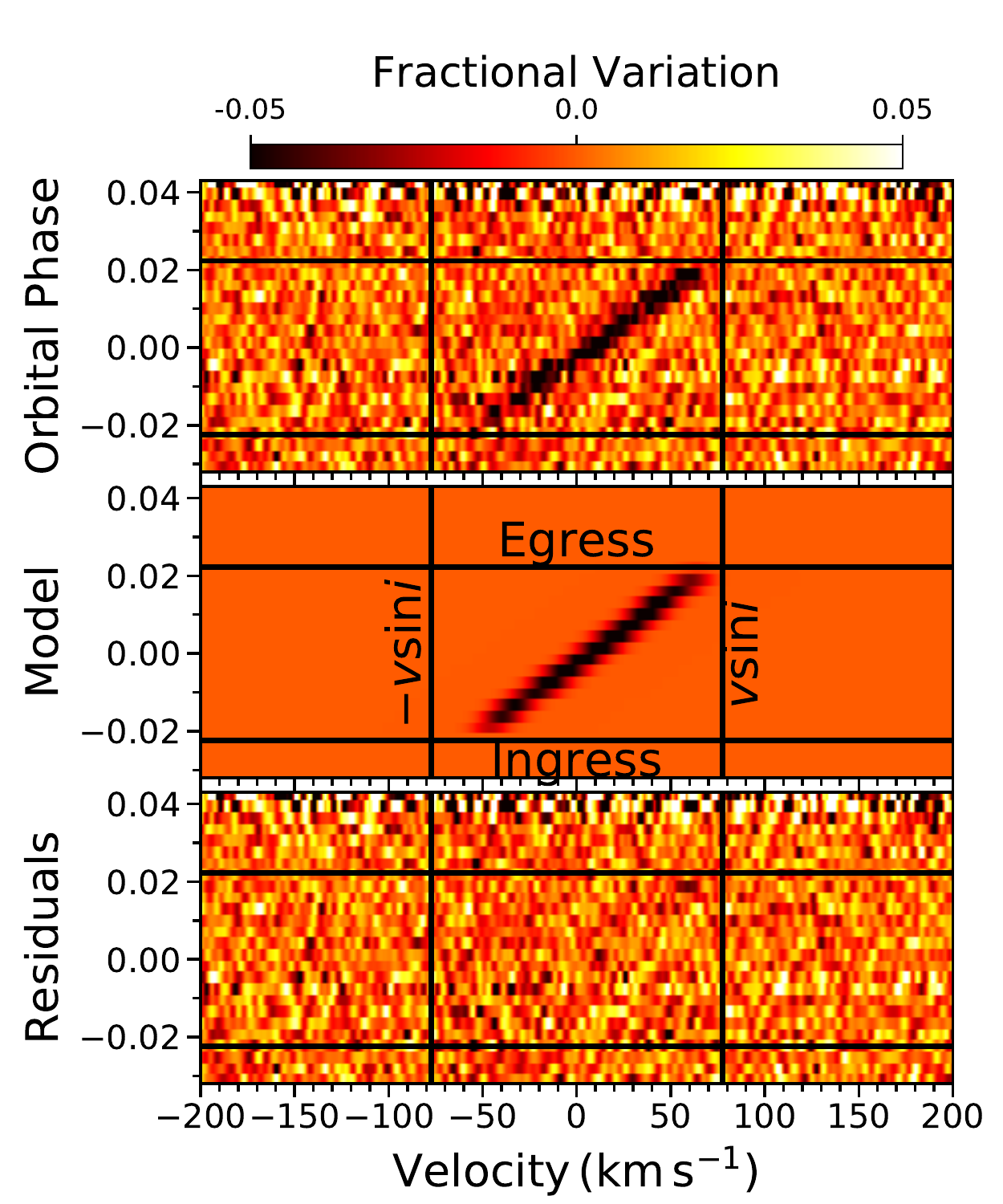} & 
    \includegraphics[width=0.3\textwidth]{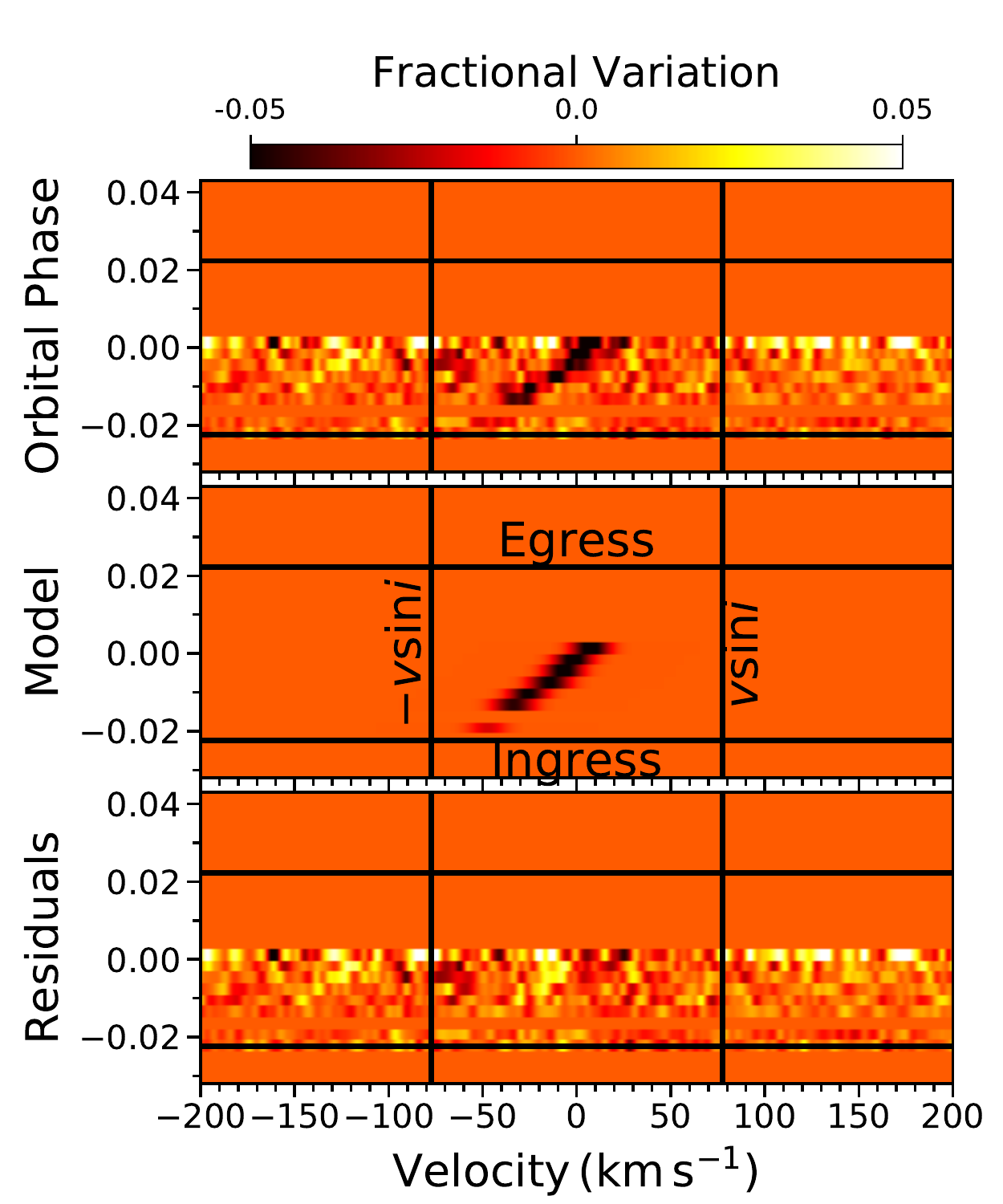} \\
\end{tabular}
\caption{
The Doppler transits of \hatstarab{}. Each Doppler map (top panel)
shows the intensity of the line profile as a function of
both velocity (relative to the line center) and orbital phase.
The ingress and egress phases are marked with horizontal lines.
The top segment shows the data from all the observed transits,
averaged into phase bins of size 0.003.
The middle panel shows the best-fitting model, and the lower
panel shows the residuals. A diagrammatic representation of the transit geometry of each system is shown at the top of the figure, with the relative sizes of the star and planet plotted to scale. The gravity darkening effect is exaggerated to allow it to be easily seen. The left panel shows the Doppler transit signal for \hatstarab{}, combined from 3 partial TRES transit observations. The right panels shows the partial transit of \hatstarab{} via SALT-HRS. Phases at which no data was obtained are colored in plain orange. \label{fig:dt_hatp69}}
\end{figure*}

\begin{figure}
\centering
\begin{tabular}{l}
    \multicolumn{1}{c}{\textbf{\large\hatstarb{}}}\\ 
    \multicolumn{1}{c}{\includegraphics[width=0.3\textwidth]{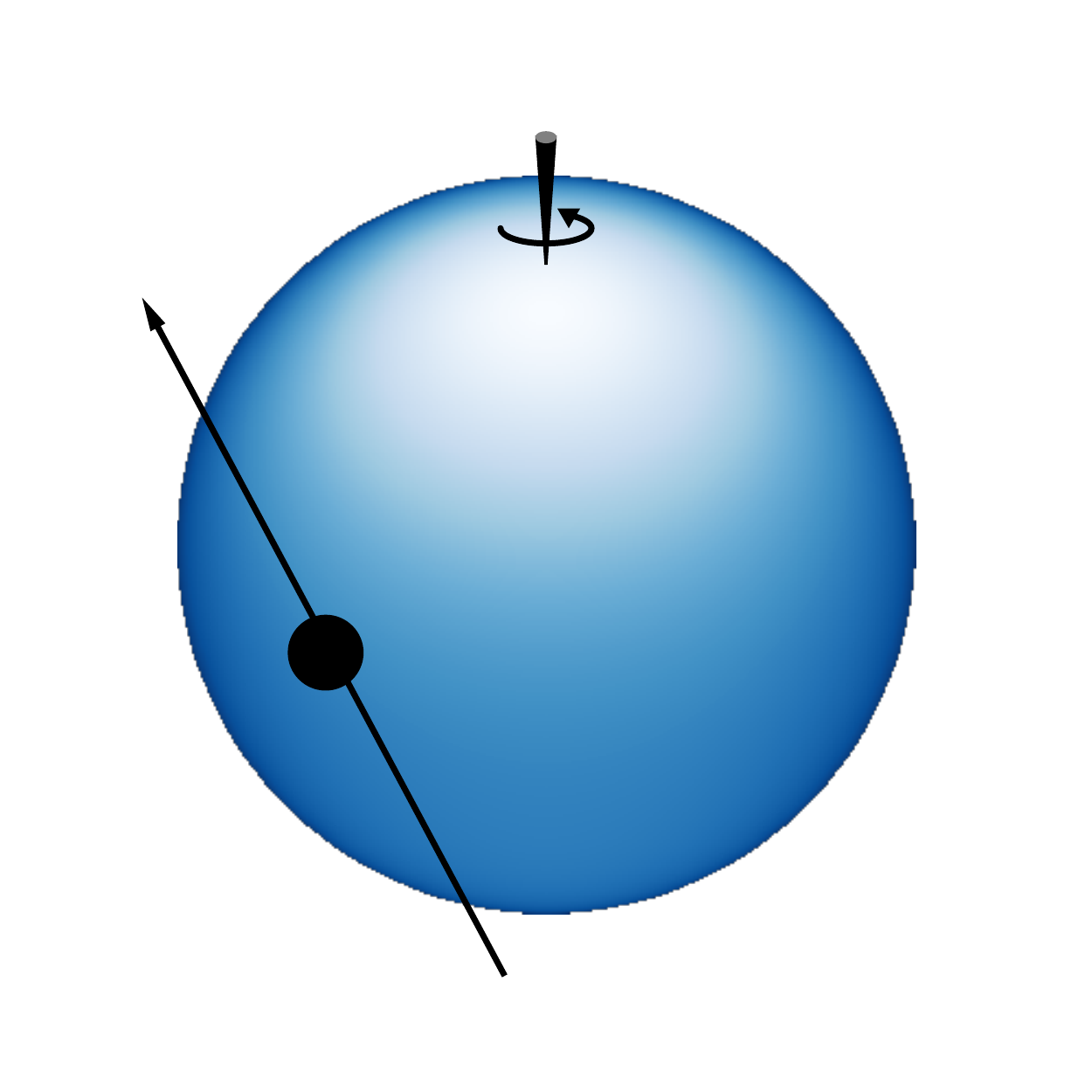}} \\
    \textbf{TRES} \\ 
    \includegraphics[width=0.3\textwidth]{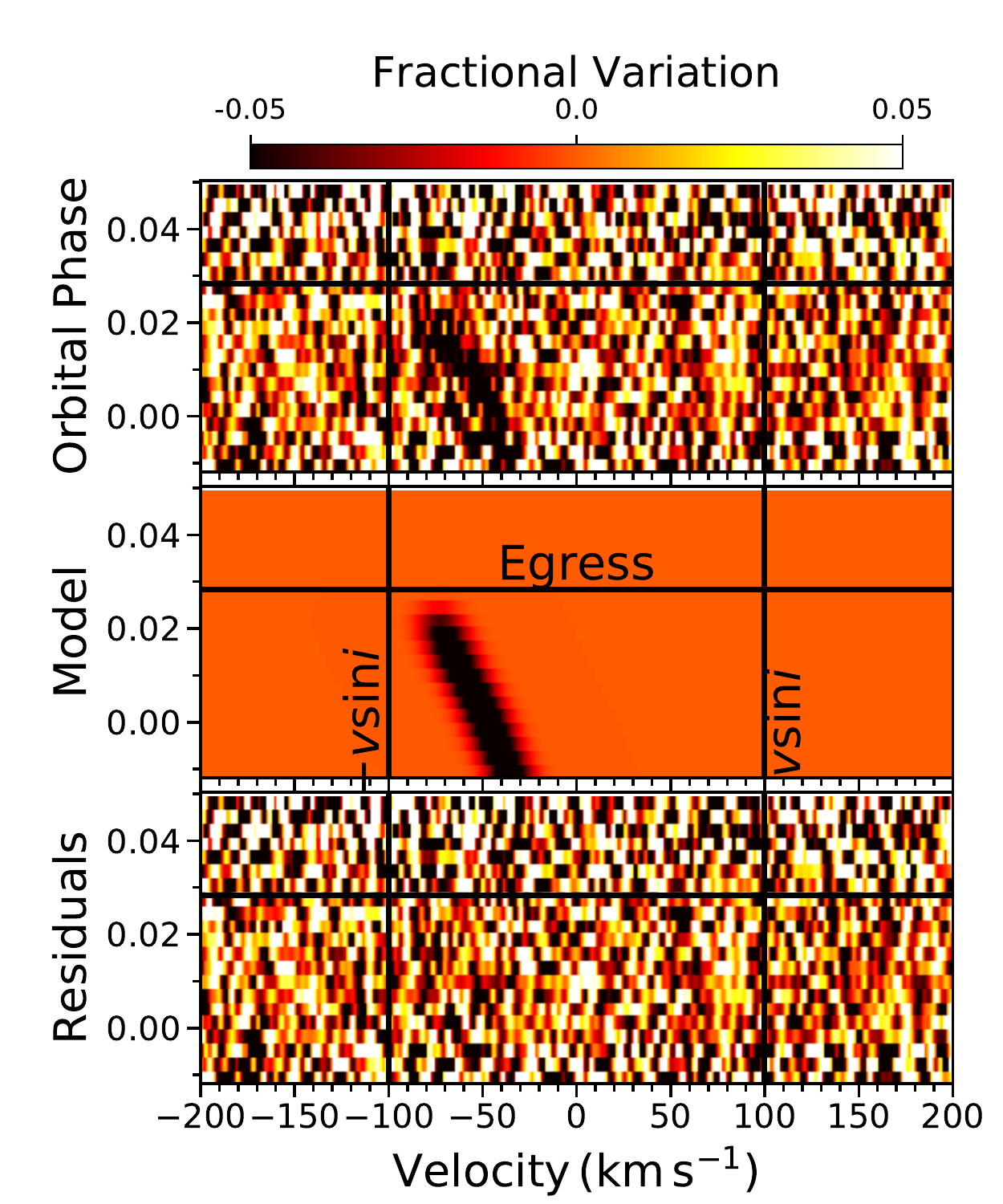} \\
\end{tabular}
\caption{
The Doppler transit \hatstarbb{} as measured via two partial TRES transits. The figure follows the format specified in Figure~\ref{fig:dt_hatp69}.
\label{fig:dt_hatp70}}
\end{figure}

Spectroscopic observations were also obtained with TRES throughout the transits of each planet. These observations allow us to measure variations in the stellar line profile due to the partial obscuration of the
photosphere of the rapidly rotating star \citep{2010MNRAS.407..507C}. By measuring the planetary ``shadow'' on the line profile of the star, we confirm that the photometric transit signal is indeed caused by a small body that is
transiting the bright rapidly rotating target star, as opposed to
being the diluted signal of a much fainter eclipsing binary that is
spatially blended with the target star in the photometric aperture.
The observing strategy and analysis largely follow the procedure laid out by \citet{2016MNRAS.460.3376Z}. We observed three partial transits of \hatstara{} on 2017-03-08, 2017-03-13 and 2019-01-12, with the Doppler shadow of the planet clearly detected in each individual transit (Figure~\ref{fig:dt_hatp69}). Two partial transits of \hatstarb{} were obtained on 2019-02-21 and 2019-03-04. Observations on 2019-02-21 were hampered by poor weather, but the subsequent transit on 2019-03-04 clearly revealed the planet shadow (Figure~\ref{fig:dt_hatp70}).
These observations are used in the global analysis (Section~\ref{sec:globalmodel}) to derive the projected spin-orbit angle of the systems.

One additional partial transit of \hatstarab{} was obtained via the High Resolution Spectrograph \citep[HRS][]{2014SPIE.9147E..6TC} on the Southern African Large Telescope (SALT). HRS is a fiber fed echelle spectrograph, used in the medium resolution mode yielding a spectral resolution of $R=40000$ over the wavelength region of $3700-5500$\,\AA{} over the blue arm of the spectrograph. Observations from the red arm of the spectrograph were not used due to the fewer line-count over its spectral coverage. The observations were obtained covering the ingress of \hatstarab{} on 2015-03-06, covering 11 spectra with integration times of 700\,s each. The target star remained at an altitude of $47-53^\circ$ throughout the transit observations.  The spectra were extracted and calibrated using the \emph{MIDAS} pipeline \citep{2016MNRAS.459.3068K,2017ASPC..510..480K}. The spectral line profiles were extracted via a similar process as that described above. The average line profile is subtracted, leaving a significant detection of the planetary transit over ingress (Figure~\ref{fig:dt_hatp69}).

In addition, a number of spectroscopic resources contributed to the initial spectroscopic vetting of the targets. Observations of \hatstara{} were obtained using the High Resolution Echelle Spectrometer (HIRES) on the 10\,m Keck-I at Mauna Kea Observatory. Observations were also obtained using the High Dispersion Spectrograph (HDS) on the 8.2\,m Subaru telescope on Mauna Kea Observatory. In both cases observations were made using the Iodine cell, but did not yield high precision velocities due to the rapid rotation of the star. They were not included in the analysis. We also made use of the CHIRON instrument on the SMARTS 1.5\,m telescope at Cerro Tololo Inter-American Observatory (CTIO), Chile \citep{2013PASP..125.1336T}, obtaining 4 observations of \hatstarb{}. Similarly, reconnaissance observations were obtained with the SOPHIE echelle facility on the 1.93\,m Haute-Provence Observatory, France, as well as the CORALIE spectrograph on the 1.2\,m Euler telescope at the ESO La Silla Observatory, Chile. Given that the TRES observations vastly outnumber these reconnaissance observations, we incorporate only the TRES data in our global modeling. 

\begin{deluxetable*}{llcrrrrr}
\tablewidth{0pc}
\tabletypesize{\scriptsize}
\tablecaption{
    Summary of spectroscopic observations\label{tab:specobssummary}
}
\tablehead{
\\
    \multicolumn{1}{c}{Target} &
    \multicolumn{1}{c}{Telescope/Instrument} &
    \multicolumn{1}{c}{Date Range}          &
    \multicolumn{1}{c}{Number of Observations} &
    \multicolumn{1}{c}{Resolution}          &
    \multicolumn{1}{c}{Observing Mode}          
}
\startdata
\hatstara{} & FLWO 1.5\,m TRES & 2011-10-10 -- 2017-03-14 & 45 & 44000 & RV \\
\hatstara{} & SALT HRS & 2015-03-06 & 11 & 40000 & Transit \\
\hatstara{} & FLWO 1.5\,m TRES & 2017-03-08 & 18 & 44000 & Transit\\
\hatstara{} & FLWO 1.5\,m TRES & 2017-03-13 & 17 & 44000 & Transit\\
\hatstara{} & FLWO 1.5\,m TRES & 2019-01-12 & 22 & 44000 & Transit\\
[1.5ex]
\hatstarb{} & FLWO 1.5\,m TRES & 2013-02-01 -- 2019-02-20 & 43 & 44000 & RV \\
\hatstarb{} & FLWO 1.5\,m TRES & 2019-02-21 & 19 & 44000 & Transit\\
\hatstarb{} & FLWO 1.5\,m TRES & 2019-03-04 & 19 & 44000 & Transit\\
\enddata 
\end{deluxetable*}

\begin{deluxetable}{rrrl}
\tablewidth{0pc}
\tabletypesize{\scriptsize}
\tablecaption{
      Relative radial velocities of \hatstara{}
    \label{tab:rv_table_stara}
}
\tablehead{ \\
    \multicolumn{1}{c}{BJD}          &
    \multicolumn{1}{c}{Relative RV}\tablenotemark{a}             &
    \multicolumn{1}{c}{$\sigma$ RV}      &
    \multicolumn{1}{c}{Instrument}      \\
    \multicolumn{1}{c}{(UTC)} &
    \multicolumn{1}{c}{$(\mathrm{km\,s}^{-1})$} &
    \multicolumn{1}{c}{$(\mathrm{km\,s}^{-1})$} 
}
\startdata
2455844.990516  &  1.437  &  0.433  & TRES \\
2455889.044893  &  0.782  &  0.159  & TRES \\
2455904.899944  &  0.763  &  0.200  & TRES \\
2456399.650041  &  0.328  &  0.178  & TRES \\
2456400.656614  &  0.462  &  0.151  & TRES \\
2456403.681071  &  0.389  &  0.121  & TRES \\
2456404.671615  &  0.138  &  0.134  & TRES \\
2456409.670360  &  0.171  &  0.142  & TRES \\
2456410.671948  &  0.293  &  0.140  & TRES \\
2457819.604459  &  0.619  &  0.175  & TRES \\
2457819.616148  &  0.473  &  0.155  & TRES \\
2457819.627797  &  0.851  &  0.150  & TRES \\
2457819.639439  &  0.482  &  0.121  & TRES \\
2457819.651291  &  0.487  &  0.172  & TRES \\
2457819.663107  &  0.587  &  0.135  & TRES \\
2457819.675045  &  0.781  &  0.118  & TRES \\
2457819.686914  &  0.753  &  0.125  & TRES \\
2457819.698568  &  0.684  &  0.119  & TRES \\
2457819.710512  &  0.728  &  0.107  & TRES \\
2457819.723092  &  0.690  &  0.140  & TRES \\
2457819.734747  &  0.487  &  0.129  & TRES \\
2457819.746511  &  0.667  &  0.098  & TRES \\
2457819.758189  &  0.733  &  0.134  & TRES \\
2457819.770017  &  0.577  &  0.161  & TRES \\
2457819.781718  &  0.593  &  0.122  & TRES \\
2457819.793343  &  0.640  &  0.120  & TRES \\
2457819.804946  &  0.364  &  0.152  & TRES \\
2457819.816560  &  0.805  &  0.178  & TRES \\
2457819.828173  &  0.847  &  0.156  & TRES \\
2457820.675448  &  0.460  &  0.110  & TRES \\
2457820.687137  &  0.732  &  0.099  & TRES \\
2457820.698797  &  0.384  &  0.148  & TRES \\
2457820.710475  &  0.553  &  0.148  & TRES \\
2457820.722152  &  0.711  &  0.123  & TRES \\
2457825.760728  &  0.264  &  0.123  & TRES \\
2457825.772683  &  0.466  &  0.162  & TRES \\
2457825.784297  &  0.350  &  0.146  & TRES \\
2457825.795980  &  0.518  &  0.147  & TRES \\
2457826.645765  &  0.480  &  0.157  & TRES \\
2457826.657442  &  0.242  &  0.165  & TRES \\
2457826.669091  &  0.062  &  0.161  & TRES \\
2457826.682487  &  0.199  &  0.176  & TRES \\
2457826.694234  &  0.194  &  0.110  & TRES \\
2457826.705923  &  0.373  &  0.136  & TRES \\
2457826.717589  &  0.292  &  0.225  & TRES \\
2457826.729296  &  0.255  &  0.151  & TRES \\
2457826.741008  &  0.199  &  0.103  & TRES \\
2458495.942762  &  0.618  &  0.228  & TRES \\
2458495.954736  &  0.642  &  0.330  & TRES \\
2458495.966877  &  0.938  &  0.285  & TRES \\
2458495.979216  &  0.286  &  0.303  & TRES \\
2458495.991201  &  0.566  &  0.356  & TRES \\
2458496.003036  &  0.810  &  0.192  & TRES \\
\enddata 
\tablenotetext{a}{
 Relative radial velocities from a multi-order cross correlation. Internal errors excluding the component of astrophysical/instrumental jitter considered in Section 3.  Velocities exclude those taken in transit. 
}
\end{deluxetable}

\begin{deluxetable}{rrrl}
\tablewidth{0pc}
\tabletypesize{\scriptsize}
\tablecaption{
      Relative radial velocities of \hatstarb{}. 
    \label{tab:rv_table_starb}
}
\tablehead{ \\
    \multicolumn{1}{c}{BJD}          &
    \multicolumn{1}{c}{RV}\tablenotemark{a}             &
    \multicolumn{1}{c}{$\sigma$ RV}      &
    \multicolumn{1}{c}{Instrument}      \\
    \multicolumn{1}{c}{(UTC)} &
    \multicolumn{1}{c}{$(\mathrm{km\,s}^{-1})$} &
    \multicolumn{1}{c}{$(\mathrm{km\,s}^{-1})$} 
}
\startdata
2456324.697671  &  24.350  &  0.642  & TRES \\
2456342.685872  &  24.867  &  0.616  & TRES \\
2457671.822407  &  25.153  &  0.708  & TRES \\
2457671.830590  &  25.270  &  0.620  & TRES \\
2457671.838751  &  26.999  &  0.532  & TRES \\
2457671.846928  &  25.516  &  0.639  & TRES \\
2457671.855187  &  25.525  &  0.727  & TRES \\
2457671.863348  &  25.804  &  1.043  & TRES \\
2457671.871531  &  25.827  &  0.599  & TRES \\
2457671.880432  &  26.936  &  0.676  & TRES \\
2457671.892204  &  24.786  &  1.114  & TRES \\
2457671.900399  &  24.601  &  0.688  & TRES \\
2457671.908554  &  24.334  &  0.934  & TRES \\
2457671.917565  &  24.760  &  0.634  & TRES \\
2457671.925743  &  25.362  &  0.773  & TRES \\
2457671.933920  &  24.981  &  0.663  & TRES \\
2457671.942121  &  25.324  &  0.659  & TRES \\
2457671.950363  &  26.011  &  0.538  & TRES \\
2457671.958645  &  25.100  &  0.825  & TRES \\
2457671.966979  &  25.709  &  0.726  & TRES \\
2457671.975226  &  24.904  &  0.592  & TRES \\
2457671.983427  &  25.817  &  0.819  & TRES \\
2457671.991639  &  24.845  &  0.677  & TRES \\
2457672.000309  &  25.261  &  0.536  & TRES \\
2457672.008573  &  24.851  &  0.584  & TRES \\
2457672.016844  &  25.661  &  0.839  & TRES \\
2457672.025149  &  25.769  &  0.434  & TRES \\
2458527.601110  &  25.803  &  1.160  & TRES \\
2458531.776169  &  25.321  &  1.741  & TRES \\
2458532.755301  &  24.806  &  0.962  & TRES \\
2458534.591655  &  24.475  &  0.630  & TRES \\
2458534.599808  &  25.870  &  0.654  & TRES \\
2458534.607915  &  25.156  &  1.032  & TRES \\
2458534.616045  &  24.988  &  0.569  & TRES \\
2458534.624158  &  25.590  &  0.992  & TRES \\
2458534.632322  &  24.981  &  0.916  & TRES \\
2458534.640470  &  26.307  &  1.199  & TRES \\
2458534.648681  &  24.680  &  0.628  & TRES \\
2458534.656811  &  25.248  &  1.155  & TRES \\
2458534.664964  &  24.315  &  0.761  & TRES \\
2458534.673094  &  25.971  &  0.920  & TRES \\
2458534.681230  &  24.672  &  0.470  & TRES \\
2458534.689354  &  24.547  &  0.848  & TRES \\
2458535.714170  &  24.300  &  0.508  & TRES \\
2458535.722375  &  23.870  &  0.994  & TRES \\
2458535.730499  &  24.905  &  2.141  & TRES \\
2458535.738629  &  23.029  &  3.309  & TRES \\
2458546.689854  &  25.441  &  0.382  & TRES \\
2458546.698076  &  26.287  &  0.618  & TRES \\
2458546.706287  &  24.819  &  0.922  & TRES \\
2458546.714516  &  26.252  &  0.741  & TRES \\
2458546.722686  &  23.979  &  0.821  & TRES \\
2458546.730914  &  25.292  &  1.133  & TRES \\
2458546.739183  &  25.804  &  1.377  & TRES \\
\enddata 
\tablenotetext{a}{
 Absolute velocities from derived from the least-squares deconvolution profiles. Internal errors excluding the component of astrophysical/instrumental jitter considered in Section 3.  Velocities exclude those taken in transit. 
}
\end{deluxetable}

\section{Analysis}
\label{sec:analysis}

\subsection{Properties of the host star}
\label{sec:hoststar}

Both \hatstara{} and \hatstarb{} are classified as
rapidly rotating A stars based on their 2MASS \citep{2006AJ....131.1163S} $J-K$ colors and the reconnaissance spectra from \emph{TRES}. Rapidly rotating stars have spectral lines that are blended and unresolved, making standard spectral classifications more difficult. In addition, the gravity darkening effect causes the derived atmospheric parameters, such as effective temperature, to be dependent on our viewing angle.
The same star would appear hotter when viewed pole-on, and cooler when viewed along the equator. We adopt the approach described in \citet{2019AJ....157...31Z} and match the spectral energy distribution of the star against a grid of synthetic magnitudes computed from the Geneva 2D rotational isochrones \citep{2012A&A...537A.146E} for a range of inclination angles. This is performed as part of the global modeling described in Section~\ref{sec:globalmodel}, as the transit light curve also contributes to constraining the inclination angle of the system. 

\begin{figure}
\centering
    \includegraphics[width=0.4\textwidth]{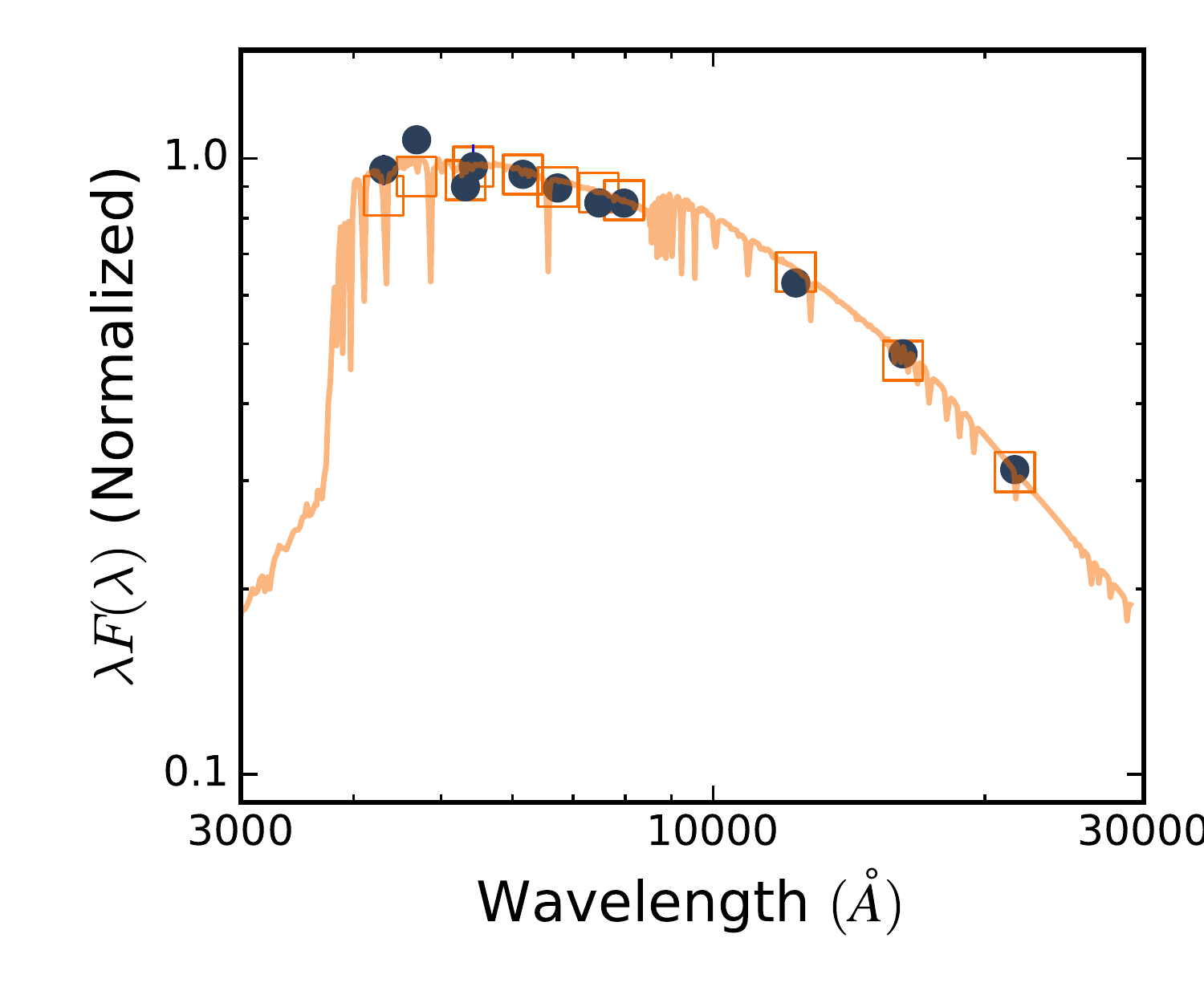}
\caption{
Spectral energy distribution of \hatstara{} with the $B$, $V$, $g'$, $r'$, and $i'$ bands from APASS \citep{2016yCat.2336....0H}, $G$, $B_P$, and $R_P$ from Gaia \citep{2018A&A...616A...1G}, and $J$, $H$, $Ks$ from 2MASS \citep{2006AJ....131.1163S}. The synthetic spectrum is generated using ATLAS9 models \citep{Castelli:2004} whilst accounting for the effect of the viewing geometry and gravity darkening of the host star. 
\label{fig:HTR413001_sed}}
\end{figure}

\begin{figure}
\centering
    \includegraphics[width=0.4\textwidth]{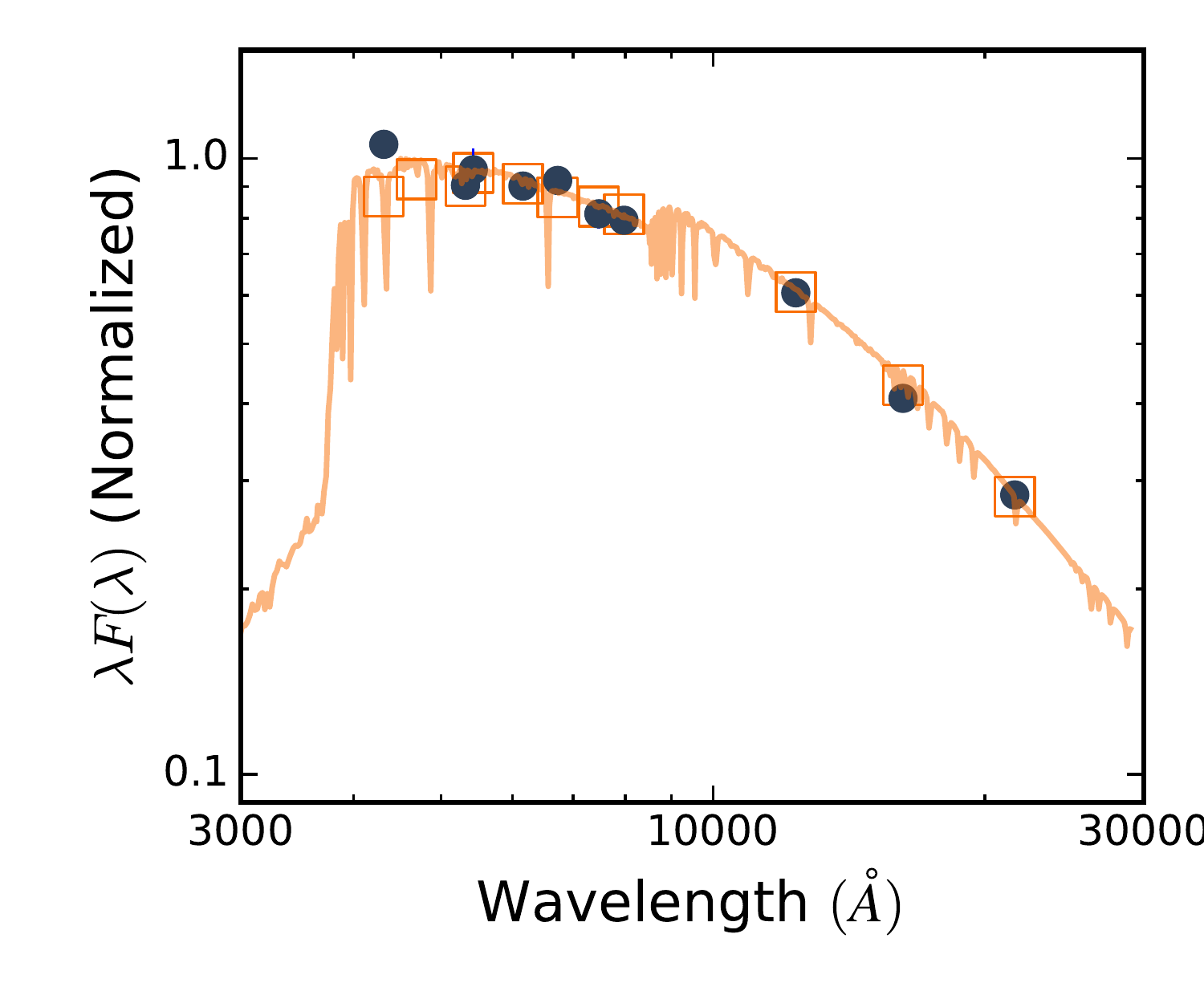}
\caption{
Spectral energy distribution of \hatstarb{}, similar to Figure~\ref{fig:HTR413001_sed}. See caption for Figure~\ref{fig:HTR413001_sed}.
\label{fig:HTR358007_sed}}
\end{figure}

The spectral energy distributions (SED) for both stars are shown in Figures~\ref{fig:HTR413001_sed} and \ref{fig:HTR358007_sed}. We find that both stars are late A dwarfs. \hatstara{} has a mass of \hatstaramass\,$M_\odot$, radius of \hatstararadius\,$R_\odot$, and effective temperature of \hatstarateff\,$\mathrm{K}$. \hatstarb{} has a mass of \hatstarbmass\,$M_\odot$, radius of \hatstarbradius\,$R_\odot$, and effective temperature of \hatstarbteff\,$\mathrm{K}$. 

We check this rotational-SED analysis with an independent fit of the SEDs to Kurucz atmosphere models of non-rotating stars \citep{1992IAUS..149..225K}. We find \hatstara{} to have $T_\mathrm{eff} = 7650 \pm 400$\,K, $R_\star = 1.88 \pm 0.19\,R_\odot$, and reddening of $A(v) = 0.01 \pm 0.01$. While \hatstarb{} has $T_\mathrm{eff} =  8400 \pm 400$\,K, $R_\star = 2.08 \pm 0.20\,R_\odot$, with reddening of $A(v) = 0.30_{-0.08}^{+0.01}$. For both stars, the non-rotational SED analysis agrees well with that from the global modeling detailed above.

As a check on the determination of the stellar parameters, we independently derived the effective temperature and metallicity of each star using the \emph{TRES} spectra and the Stellar Parameter Classification (SPC) pipeline \citep{2010ApJ...720.1118B}. We find \hatstara{} to have $T_\mathrm{eff} = 7557\pm52\,\mathrm{K}$, $\mathrm{[m/H]}=+0.05\pm0.08$ dex, while \hatstarb{} has atmospheric parameters of $T_\mathrm{eff} = 8246\pm93\,\mathrm{K}$ and $\mathrm{[m/H]} = -0.06 \pm 0.09$ dex. The spectroscopic stellar parameters agree to within $1\sigma$ with those measured from the SED, though the uncertainties are likely underestimated. The rapid rotation of the star causes difficulties in continuum normalization of the spectra, making accurate spectroscopic determination of the stellar parameters and associated uncertainties more difficult.
We incorporate the metallicity measurements from spectra as Gaussian priors in the global modeling. For a more accurate understanding of stellar properties, we simulataneously fit the SED with the transit and rotational stellar isochrones in our global modeling, instead of relying on the spectra-derived values. 

An accurate measurement of the projected stellar rotation rate is crucial for interpreting the Doppler transit data, constraining the stellar gravity darkening effect, and constraining the stellar oblateness. To measure the projected rotation velocity, we model the LSD spectral line profiles using a kernel that incorporates the effects of stellar rotation and radial-tangential macroturbulence via a numerical disk integration, and the models
the instrument line broadening as a Gaussian convolution. We find \hatstara{} to have $v\sin I_\star = 77.40\pm0.60\,\kms$ and a macroturbulent velocity of $v_\mathrm{mac} = 5.6 \pm 4.2 \,\kms$. For \hatstarb{}, the results
are $v\sin I_\star = 99.87\pm 0.65\,\kms$ and $v_\mathrm{mac} = 4.77 \pm 0.86 \,\kms$.

\subsection{Global modeling of system parameters}
\label{sec:globalmodel}

We perform a global analysis of the systems to model the large suite of observations available for \hatstara{} and \hatstarb{}. This global model simultaneously incorporates the photometric transit, radial velocities, stellar parameter constraints, the Doppler transits, and the
effect of photometric gravity darkening on the transit light curve and observed stellar properties. 

Our modeling process largely follows that described by \citet{2019AJ....157...31Z}. Rapid rotation distorts the shapes of stars; they become oblate along the equator, causing the poles to be hotter and brighter, while the equator becomes cooler and darker \citep{1924MNRAS..84..684V}. This gravity darkening effect causes both the transit light curve \citep{2009ApJ...705..683B} and the observed
spectral energy distribution of the star \citep{2015ApJ...807...58B} to
depend on the viewing direction. The photometric transit is modeled using the \emph{simuTrans} package from \citet{2018AJ....155...13H}, which accounts for both the gravity darkened non-uniform brightness distribution of the stellar disk, and the ellipsoidal nature of the rapidly rotating star. The stellar properties are inferred from the Geneva 2D rotational isochrones \citep{2012A&A...537A.146E}, which incorporates the effects of rotation on stellar evolution, and includes prescriptions for the oblateness of the stars based on their rotation rates. In the case of an oblique transiting geometry about gravity darkened stars, the resulting light curve often exhibits asymmetry due to the latidude dependence of the surface brightness distribution. This effect is detected for \hatstarbb{}, and explored in greater depth in Section~\ref{sec:asymtransit}.

The limb darkening coefficients are interpolated from the values of \citet{Claret:2011} and \citet{2017A&A...600A..30C} for the Sloan and \emph{TESS} bands. They are constrained by a Gaussian prior of width $0.02$ during the global modeling, representing the difference in the limb darkening coefficients should the stellar parameters be different by $1\sigma$. To model the transit light curves, we adopt a gravity darkening coefficient $\beta$ from interferometric observations of Vega $(\beta = 0.231 \pm 0.028)$ \citep{2012ApJ...761L...3M}. Similar interferometric gravity darkening coefficients have been measured for other rapidly rotating A stars (e.g.  $\alpha$\,Cep $\beta = 0.216\pm0.021$ \citealt{2009ApJ...701..209Z}). To account for the uncertainty in the gravity darkening coefficient, it is modeled in the global fit as a free parameter constrained about the value and uncertainty of Vega reported in \citet{2012ApJ...761L...3M}.
The model fitting procedure also includes
detrending of the ground-based follow-up light curves, via a linear combination of effects, including the pixel position of the target star, airmass, and background count values.
We account for the 30\,minute cadence of the \emph{TESS} by super-sampling and integrating the model over the exposure time.

The stellar parameters are constrained by the spectral energy distribution of the stars over the Tycho-2 \citep{2000A&A...355L..27H}, APASS \citep{2016yCat.2336....0H}, and 2MASS \citep{2006AJ....131.1163S} photometric bands, as well as the parallax from \emph{Gaia} data release 2 \citep{2018A&A...616A...1G}. Local reddening is constrained by the maximum reddening value from the dust maps of \citet{2011ApJ...737..103S}, assuming $A_v = 3.1 E(B-V)$. To account for the uncertainties in our deblending of the \emph{TESS} light curves, we also include a \emph{TESS} light curve dilution parameter, closely constrained by a Gaussian prior, with width derived from the reported uncertainties in the \emph{TESS} band magnitudes of the target and nearby stars from TIC v6. 

The Doppler transit signal is simultaneously modeled with the light curve, and provides the best constraint on the projected spin-orbit angle $\lambda$ for the orbital plane of the planets. We model variations of the stellar line profiles via a 2D integration of the rotating stellar surface being occulted by the transiting planet, incorporating the effects of differential limb darkening, radial-tangential macroturbulence, and instrument broadening. 

To derive the best fit system parameters and their associated uncertainties, we perform a Markov Chain Monte Carlo analysis using the \emph{emcee} package \citep{2013PASP..125..306F}. The resulting stellar and planetary
parameters are shown in Tables~\ref{tab:stellar} and \ref{tab:planetparam},
respectively.


\begin{deluxetable*}{lrr}
\tablewidth{0pc}
\tabletypesize{\scriptsize}
\tablecaption{
    Stellar parameters
    \label{tab:stellar}
}
\tablehead{ \\
    \multicolumn{1}{c}{~~~~~~~~Parameter~~~~~~~~}   &
    \multicolumn{1}{c}{\hatstara{}} &
    \multicolumn{1}{c}{\hatstarb{}} 
}
\startdata
\sidehead{Catalogue Information}
~~~~TIC \dotfill & 379929661 & 399870368 \\
~~~~Tycho-2 \dotfill & 0215-01594-1 & 0688-01684-1 \\
~~~~Gaia DR2 \dotfill & 3080104185367102592 & 3291455819447952768 \\
~~~~Gaia RA (2015.5) \dotfill & 08:42:01.353 & 04:58:12.560 \\
~~~~Gaia DEC (2015.5) \dotfill& +03:42:38.038 & +09:59:52.726 \\
~~~~Gaia $\mu_\alpha$ $(\mathrm{mas}\,\mathrm{yr}^{-1})$ \dotfill& $-2.856 \pm 0.074$ & $-2.657 \pm 0.096$ \\
~~~~Gaia $\mu_\delta$ $(\mathrm{mas}\,\mathrm{yr}^{-1})$ \dotfill& $0.984 \pm 0.051$  & $-4.996 \pm 0.065$\\
~~~~Gaia DR2 Parallax $(\mathrm{mas})$ \dotfill & $2.902 \pm 0.043$ & $2.996 \pm 0.061$\\
\sidehead{Stellar atmospheric properties  \tablenotemark{a}}
~~~~$\teffstar$ (K)\dotfill       &  \hatstarateff{} & \hatstarbteff{}  \\
~~~~$\feh$\dotfill                &  \hatstarafeh{} & \hatstarbfeh{}  \\
~~~~$\vsini$ (\kms)\dotfill        &  \hatstaravsini{} & \hatstarbvsini{} \\
~~~~$v_\mathrm{macro}$ (\kms)\dotfill        &  \hatstaravmac{} & \hatstarbvmac{}  \\
\sidehead{Photometric properties}
~~~~TESS $T$ (mag)\dotfill               & $9.612 \pm 0.018$ & $9.298 \pm 0.019$\\
~~~~Gaia $G$ (mag)\dotfill               & $9.77216 \pm 0.00035$ & $9.45112 \pm 0.00035$ \\
~~~~TYCHO $B$ (mag)\dotfill               &  $10.052 \pm 0.061$ & $9.621 \pm 0.045$ \\
~~~~TYCHO $V$ (mag)\dotfill               &  $9.7740 \pm 0.0050$ & $9.4700 \pm 0.0040$ \\
~~~~APASS $g'$ (mag)\dotfill               &  $9.796 \pm 0.030$ & $9.842 \pm 0.351$\\
~~~~APASS $r'$ (mag)\dotfill               & $9.855 \pm 0.041$ & $9.506 \pm 0.028$\\
~~~~APASS $i'$ (mag)\dotfill               & $9.976 \pm 0.020$ & $9.962 \pm 0.061$\\
~~~~2MASS $J$ (mag)\dotfill               & $9.373 \pm 0.024$ & $9.068 \pm 0.022$ \\
~~~~2MASS $H$ (mag)\dotfill               & $9.293 \pm 0.022$ & $9.023 \pm 0.029$ \\
~~~~2MASS $K_s$ (mag)\dotfill             & $9.280 \pm 0.023$ & $8.963 \pm 0.024$ \\
\sidehead{Stellar properties}
~~~~$\mstar$ ($\msun$)\dotfill      & \hatstaramass{} &  \hatstarbmass{} \\
~~~~$\rstar$ ($\rsun$)\dotfill      & \hatstararadius{} &  \hatstarbradius{} \\
~~~~$\loggstar$ (cgs)\dotfill       & \hatstaralogg{} & \hatstarblogg{} \\
~~~~$\lstar$ ($\lsun$)\dotfill      & \hatstaralum{} & \hatstarblum{}  \\
~~~~Stellar oblateness $R_\mathrm{pole}/R_\mathrm{eq}$ & \hatstaraoblate{} & \hatstarboblate{} \\
~~~~Line of sight inclination $I_*$\dotfill        & \hatstarairot{} & \hatstarbirot{} \\
~~~~$E(B-V)$ (mag)\tablenotemark{b} \dotfill & $0.0167_{-0.015}^{+0.011}$ &  $<0.034\,(1\sigma)$ \\
~~~~Age (Gyr)\dotfill               & $1.27_{-0.44}^{+0.28}$ & $0.60_{-0.20}^{+0.38}$ \\
~~~~Distance (pc) \dotfill           & $343.9_{-4.3}^{+4.8}$ & $329.0\pm6.5$ \\
\enddata

\tablenotetext{a}{
  Derived from the global modeling described in Section~\ref{sec:analysis}, co-constrained by spectroscopic stellar parameters and the Gaia DR2 parallax.\\
}
\tablenotetext{b}{
  Uniform prior for reddening up to the local maximum set by \citet{2011ApJ...737..103S}\\
}

\end{deluxetable*}



\begin{deluxetable*}{lrr}
\tablewidth{0pc}
\tabletypesize{\scriptsize}
\tablecaption{
    Orbital and planetary parameters 
    \label{tab:planetparam}
}
\tablehead{ \\
    \multicolumn{1}{c}{~~~~~~~~Parameter~~~~~~~~}   &
    \multicolumn{1}{c}{\hatstarab{}} &
    \multicolumn{1}{c}{\hatstarbb{}} 
}
\startdata
\sidehead{\Lc{} parameters}
~~~$P$ (days)             \dotfill    & \hatstaraperiod{} & \hatstarbperiod{} \\
~~~$T_c$ (${\rm BJD-TDB}$)    
      \tablenotemark{a}   \dotfill    & $2458495.78861_{-0.00073}^{+0.00072}$ & $2458439.57519_{-0.00037}^{+0.00045}$ \\
~~~$T_{14}$ (days)
      \tablenotemark{a}   \dotfill    & $0.2136_{-0.0014}^{+0.0014}$ & $0.1450_{-0.0020}^{+0.0028}$ \\
~~~$\arstar$              \dotfill    & $7.32_{-0.18}^{+0.16}$ & $5.45_{-0.49}^{+0.29}$ \\
~~~$\rpl/\rstar$          \dotfill    & $0.08703_{-0.00080}^{+0.00075}$ & $0.09887_{-0.00095}^{+0.00133}$ \\
~~~$b \equiv a \cos i/\rstar$
                          \dotfill    & $0.366_{-0.050}^{+0.060}$ & $-0.629_{-0.054}^{+0.081}$ \\
~~~$i$ (deg)              \dotfill    & $87.19_{-0.72}^{+0.52}$ & $96.50_{-0.91}^{+1.42}$ \\
~~~$|\lambda|$ (deg)      \dotfill    & \hatstaralam{} & \hatstarblam{} \\
\sidehead{Limb-darkening and gravity darkening coefficients \tablenotemark{b}}
~~~$a_r'$ (HAT) (linear term)        \dotfill  & 0.1194 (fixed)  & 0.1550 (fixed) \\
~~~$b_r'$ (HAT) (quadratic term)     \dotfill  & 0.3974 (fixed) & 0.3306 (fixed) \\
~~~$a_{GC}$                \dotfill   & $0.41_{-0.10}^{+0.09}$  &  $0.43_{-0.10}^{+0.10}$  \\
~~~$b_{GC}$                \dotfill  & $0.25_{-0.11}^{+0.09}$   & $0.25_{-0.11}^{+0.09}$ \\ 
~~~$a_{Rc}$                \dotfill   & $0.30_{-0.09}^{+0.10}$  &  $0.24_{-0.09}^{+0.10}$  \\
~~~$b_{Rc}$                \dotfill  & $0.21_{-0.11}^{+0.10}$   & $0.19_{-0.10}^{+0.10}$ \\
~~~$a_i'$                \dotfill   & $0.117_{-0.018}^{+0.018}$  &  $0.239_{-0.021}^{+0.018}$  \\
~~~$b_i'$                \dotfill  & $0.392_{-0.019}^{+0.020}$   & $0.338_{-0.020}^{+0.021}$  \\
~~~$a_z'$                 \dotfill  & $0.069_{-0.018}^{+0.018}$  &  \\
~~~$b_z'$                 \dotfill  & $0.389_{-0.020}^{+0.020}$   & \\
~~~$a_\mathrm{TESS}$      \dotfill  & $0.238_{-0.019}^{+0.021}$ & $0.149_{-0.021}^{+0.018}$ \\
~~~$b_\mathrm{TESS}$      \dotfill    & $0.286_{-0.019}^{+0.015}$ & $0.313_{-0.022}^{+0.019}$\\
~~~$\beta$ Gravity darkening coefficient \dotfill & $0.239_{-0.029}^{+0.026}$ & $0.242_{-0.029}^{+0.026}$ \\
\sidehead{RV parameters}
~~~$K$ (\ms)              \dotfill    & $309_{-49}^{+49}$ & $<649\,(3\sigma)$ \\
~~~$e$                    \dotfill    & 0 (fixed) & 0 (fixed)  \\
~~~RV jitter (\ms)        
                          \dotfill    & $53_{-37}^{+34}$ & $320_{-180}^{+180}$ \\
~~~Systemic RV (\ms)\tablenotemark{c}        
                          \dotfill    & $784 \pm 24$ & $25260\pm110$\\
\sidehead{Planetary parameters}
~~~$\mpl$ ($\mjup$)       \dotfill    & \hatstaraplmass{} & \hatstarbplmass{} \\
~~~$\rpl$ ($\rjup$)       \dotfill    & \hatstaraplrad{} &  \hatstarbplrad{} \\
~~~$\rhopl$ (\gcmc)       \dotfill    & $1.02_{-0.16}^{+0.18}$ & $<1.54 \, (3\sigma)$ \\
~~~$\log g_p$ (cgs)       \dotfill    & $3.521_{-0.071}^{+0.067}$ & $<3.73 \, (3\sigma)$\\
~~~$a$ (AU)               \dotfill    & $0.06555_{-0.00035}^{+0.00070}$ & $.04739_{-0.00106}^{+0.00031}$ \\
~~~$T_{\rm eq}$ (K)\tablenotemark{d}       \dotfill    & $1930_{-230}^{+80}$ & $2562_{-52}^{+43}$ \\
\enddata
\tablenotetext{a}{
    \ensuremath{T_c}: Reference epoch of mid transit that minimizes the
    correlation with the orbital period.
    \ensuremath{T_{14}}: total transit duration, time between first to
    last contact;
}
\tablenotetext{b}{
        Values for a quadratic law given separately for each of the filters with which photometric observations were obtained.  These values were adopted from the
        tabulations by \citet{Claret:2011} according to the
        spectroscopic an initial estimate of the stellar parameters. The limb darkening coefficients are constrained by strong Gaussian priors of width $0.02$ about their initial values. The Gravity darkening coefficient $\beta$ is also constrained by a Gaussian prior of width $0.028$ in the fit.   
}
\tablenotetext{c}{
    The systemic RV for the system as measured relative to the telluric lines 
}
\tablenotetext{d}{
    $T_{eq}$ calculated assuming 0 albedo and full heat redistribution
}
\end{deluxetable*}

\subsection{Blending and astrophysical false positive scenarios}
\label{sec:blend}

Many astrophysical scenarios can mimic the transit signal of a planetary system. False positive scenarios such as M-dwarf companions with similar radii as substellar counterparts are ruled out by the mass constraints imposed by our radial velocity measurements. The possibility that the transit signals are due to fainter eclipsing binaries whose eclipses are diluted by the brighter target stars are more difficult to eliminate. We adopt a number of observations, including diffraction-limited imaging, and analysis of the spectroscopic transit, to eliminate this possibility.  

\begin{figure*}
    \centering
    \begin{tabular}{cc}
        \textbf{\hatstara{}} & \textbf{\hatstarb{}} \\
        \includegraphics[width=0.5\textwidth]{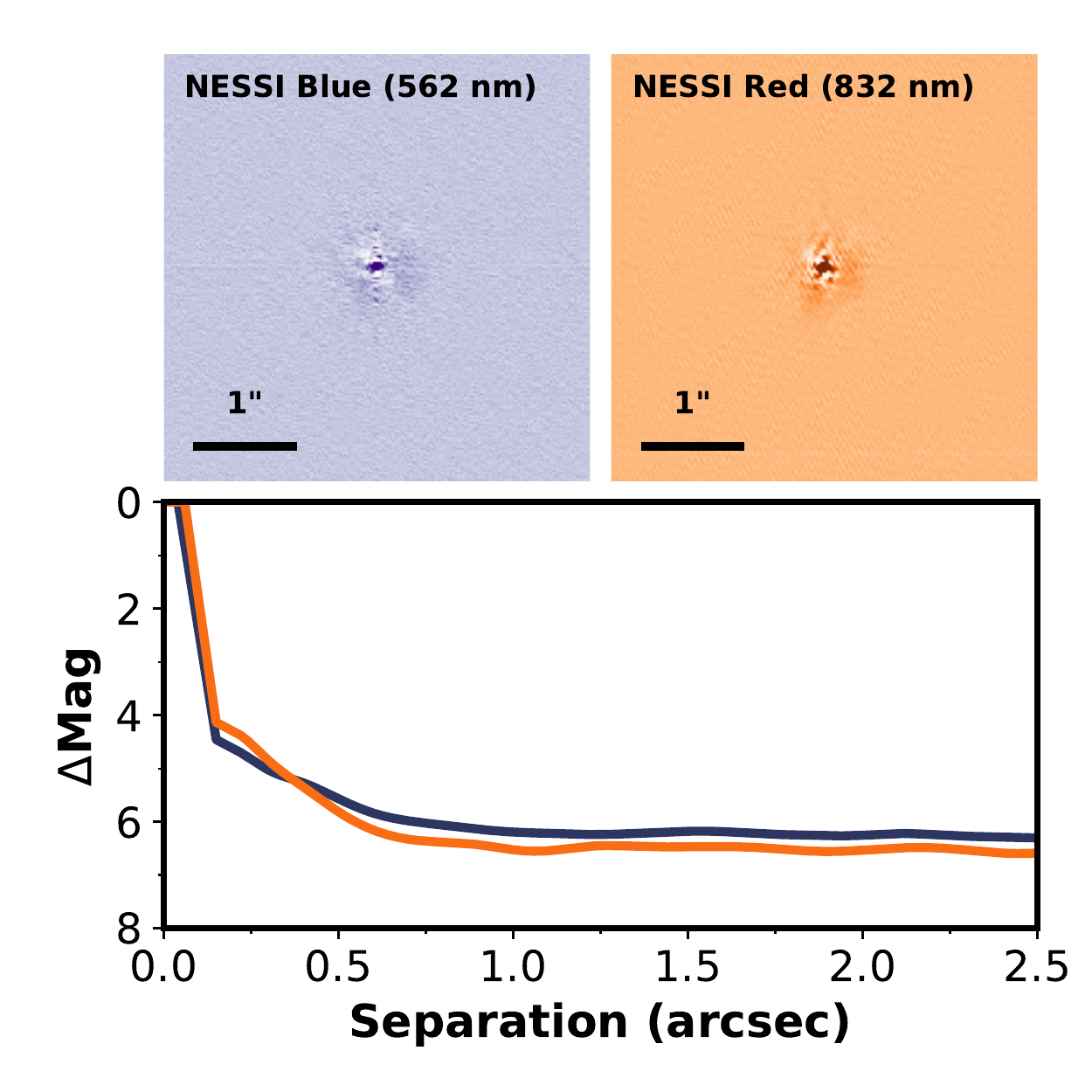} & \includegraphics[width=0.5\textwidth]{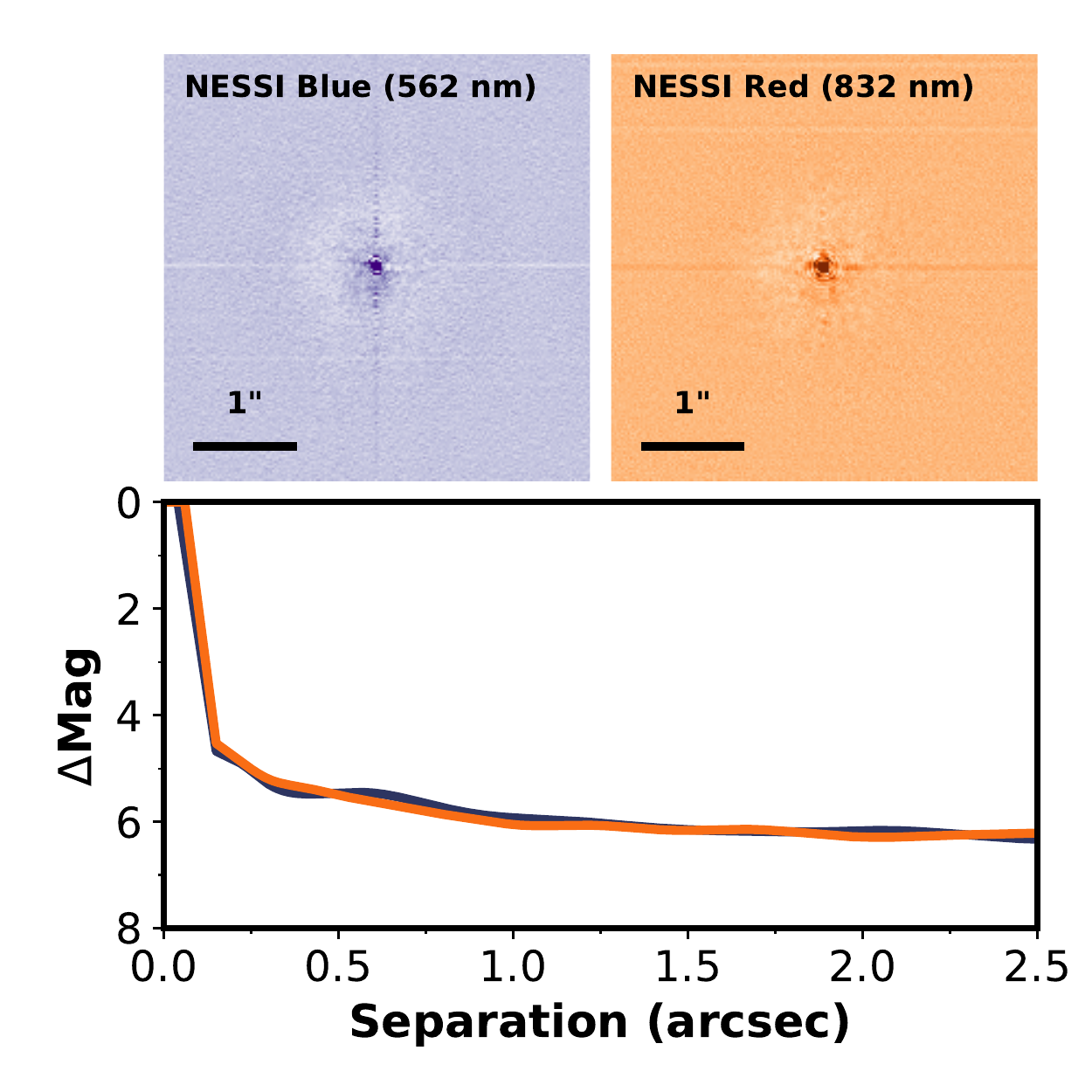}
    \end{tabular}
    \caption{Images and constraints on spatially separated stellar companions via speckle imaging for \hatstara{} and \hatstarb{} from NESSI. Companions with separations $\gtrsim 0.04\arcsec$ are ruled out. The blue and orange lines mark the $5\sigma$ limit on the detection of companions via the blue and red NESSI cameras.}
    \label{fig:speckle}
\end{figure*}

To rule out spatially nearby companions, we obtained observations with the NN-explore Exoplanet Stellar Speckle Imager \citep[NESSI,][]{2018PASP..130e4502S} on the 3.5\,m WIYN telescope at Kitt Peak National Observatory, Arizona, USA. Speckle imaging gives a resolution of $\gtrsim$0.04$\arcsec$ in both the $r$-narrow and $z$-narrow bands for both \hatstara{} and \hatstarb{}, corresponding to spatial scales as close to the stars as 14 to 22~AU (at 562\,nm and 832\,nm respectively). The corresponding constraints from NESSI are plotted in Figure~\ref{fig:speckle}. In addition, we obtained $J$ and $Ks$ band infrared seeing limited imaging \hatstara{} with the WIYN High-Resolution Infrared Camera \citep[WHIRC,][]{2011PASP..123...87S}, also finding no visual companions to the target star. 

Finally, the Doppler detection of the planetary transit confirms that the transits indeed occur around the rapidly rotating bright A star hosts, not background stars \citep[e.g.][]{2010MNRAS.407..507C}. The depth of the spectroscopic shadow agrees with the depth observed in the photometric light curves, suggesting that the dilution due to background sources is negligible.

\subsection{Detection of an asymmetric gravity darkened transit for \hatstarb{}}
\label{sec:asymtransit}

\begin{figure}
    \includegraphics[width=0.5\textwidth]{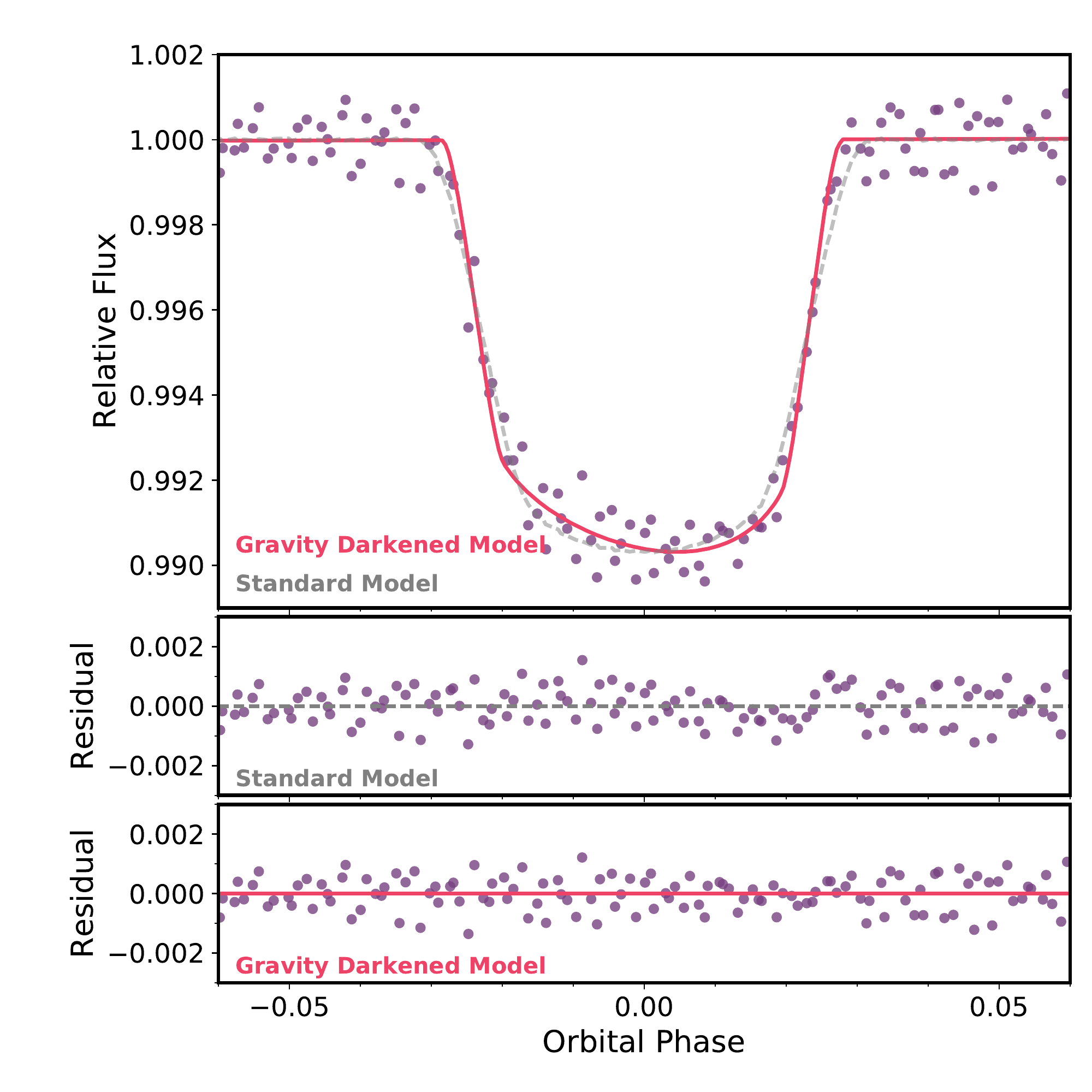} 
\caption{
The \emph{TESS} transit light curve of \hatstarb{}. Note that the transit is asymmetric, being shallower near ingress, and deeper near egress. This is due to the planet traversing from the gravity darkened equator to brighter pole during the transit. The middle panel shows the light curve residual of a standard, symmetric transit model. There are systematic variations in the residuals due to the gravity darkening effect. The bottom panel shows the residuals when the best fit gravity darkening model is subtracted. 
\label{fig:gravdark}}
\end{figure}

A transiting planet crossing a gravity darkened stellar disk may exhibit an asymmetric transit when the projected spin-orbit angle is misaligned with the stellar rotation axis.
The effects specific to gravity darkening are only visible at the parts-per-thousand level, and as such they are
difficult to detect with ground-based data. The only previous confirmed instance of asymmetric gravity darkening being observed for a planetary system is for Kepler-13. The asymmetric transit light curves of Kepler-13 were identified and modeled by \citet{2011ApJ...736L...4S}, \citet{2011ApJS..197...10B}, and \citet{2018AJ....155...13H}. Subsequent ground-based Doppler transit confirmation of the spin-orbit misalignment was performed by \citet{2014ApJ...790...30J}, and an eventual joint light curve and spectroscopic transit model developed by \citet{2015ApJ...805...28M}.

The \emph{TESS} light curves of \hatstarb{} exhibit asymmetric transits similar to those seen for Kepler-13. The transit is shallower at ingress, and deeper near egress, indicating that the planet traverses a stellar surface that is darker near ingress, and brighter near egress. Our global model reproduces such a transit, with the projected spin-orbit misaligned at \hatstaralam$^\circ$, and the stellar pole inclined to the line of sight by \hatstarairot$^\circ$ degrees. Figure~\ref{fig:gravdark} shows the \emph{TESS} transit light curve, with the best fit standard and gravity-darkened transit models over-plotted. An asymmetry  at the $500\,\mathrm{ppm}$ level can be seen in the residuals to the standard transit model, akin to that seen for Kepler-13. 

We note that we make use of the bolometric gravity darkening coefficient $\beta$ in our light curve modeling. Improvements can be made via a more careful treatment for the band-dependence of the gravity darkening effect \citep[e.g.][]{2011A&A...533A..43E}. We note though that running the global modeling whilst allowing $\beta$ to be free re-produces the same projected obliquity $\lambda$ value to within uncertainties, and as such the actual adopted gravity darkening coefficient is not critical to the modeling.

\section{The occurrence rate of hot Jupiters from \emph{TESS}}
\label{sec:occurate}

Although hot Jupiters were some of the earliest exoplanets to be discovered,  they are not intrinsically common. Radial velocity searches from the Keck, Lick, and Anglo Australia Telescope programs of 1{,}330 FGK stars revealed a hot Jupiter occurrence rate of $1.2\pm0.2\%$ \citep[$<15\,M_\mathrm{Jup}$, $<0.1$\,AU ][]{2005PThPS.158...24M}, revised to $1.20\pm0.38\%$ ($>0.1\,M_\mathrm{Jup}$, $P<10$\,days) by \citet{2012ApJ...753..160W} using the California Planet Search sample. \citet{2008PASP..120..531C} found an occurrence rate of $1.5\pm0.6\%$ ($>0.3\,M_\mathrm{Jup}$, $<0.1$\,AU) using the Keck planet search sample. Using the HARPS and CORALIE sample, \citet{2011arXiv1109.2497M} found a hot Jupiter occurrence rate of $0.89 \pm 0.36\%$ ($>0.15\,M_\mathrm{Jup}$, $<11$\,days).

These radial velocity occurrence rates are generally thought to be higher than those offered by the \emph{Kepler} survey. Studies by \citet{2012ApJS..201...15H} and \citet{2013ApJ...766...81F} of the early \emph{Kepler} data found rates of $0.4\pm0.1\%$ and $0.43\pm0.05\%$ for hot Jupiters respectively. Recent analyses with improved stellar properties from \citet{2018AJ....155...89P} found that $0.57_{-0.12}^{+0.14}\%$ of main sequence FGK stars ($5.0 > \logg > 3.9$, $4200< T_\mathrm{eff} < 6500$\,K) host hot Jupiters. The measured giant planet occurrence rate from the \emph{CoRoT} mission is higher than that from \emph{Kepler}, finding 21 giant planets $(R_p > 5\,R_\oplus)$ within 10 day period orbits, corresponding to an occurrence rate of $0.98 \pm 0.26\,\%$ \citep{2018A&A...619A..97D}. 

The stars that host hot Jupiters are more metal rich 
than random stars of the same spectral class \citep{2018AJ....155...89P, 2003A&A...398..363S,2005ApJS..159..141V,2012Natur.486..375B}. Differences between the metallicity distribution of the \emph{Kepler} stellar sample and those of the radial velocity surveys have been raised as an explanation for the differenes in the hot Jupiter occurrence rates \citep{2012ApJ...753..160W}, although \citet{2017ApJ...838...25G} showed that there is minimal difference between the \emph{Kepler} field star metallicity distribution and that of the California Planet Search sample. \citet{2015ApJ...799..229W} offered a correction for the \emph{Kepler} sample based on an improved classification of the subgiant population. They suggested that multiplicity or a lower occurrence rate of hot Jupiters around sub giants may be the cause of the disagreement. Later, \citet{2018AJ....155..244B} showed that binarity is unlikely to be responsible for any disagreements
between the Doppler and {\it Kepler} samples.

A radial velocity survey of intermediate-mass subgiants has shown that higher mass stars tend to host more gas giant planets within a few AU \citep[e.g.][]{2010PASP..122..905J,2014A&A...566A.113J,2015A&A...574A.116R,2018ApJ...860..109G}, though caveats regarding the accuracy of the mass measurements of these evolved stars should be noted \citep[e.g.][]{2013ApJ...774L...2L,2013ApJ...772..143S,2017MNRAS.472.4110S}. The giant planets around subgiants tend to be found in orbits beyond 0.1\,AU;
there appears to be a paucity of hot Jupiters around evolved stars.
These studies suggest that hot Jupiters
undergo tidal orbital decay when a star begins evolving into a subgiant
\citep{2013ApJ...772..143S}.
The planets around these ``retired A stars'' tend to be in longer period, more circular orbits than those found around main sequence stars \citep{2014A&A...566A.113J} --- although recent discoveries have unveiled numerous hot Jupiters in close-in orbits about evolved stars \citep{2018ApJ...861L...5G}. These issues inspired us to look into the hot Jupiter occurrence rate around main-sequence A stars. 

In this section, we aim to examine the hot Jupiter occurrence rate via the \emph{TESS} stellar population, with two key differences to the previous works from \emph{Kepler}. 
\begin{itemize}
    \item The \emph{TESS} stellar population encompasses bright stars covering a quarter of the sky. This sample is a significantly closer (150\,pc for a Solar-type main-sequence star) population than that from \emph{Kepler}. The \emph{TESS} sample is a closer match to the radial velocity sample of bright nearby stars, and should provide another test for any tension in the occurrence rates derived by the two techniques.
    \item The \emph{TESS} sample spans A, F, and G main sequence stars. By comparing the planet distribution around A and FG samples, we can determine if the paucity of close-in planets around ``retired A stars'' is due to post-main-sequence stellar evolution. More broadly,
    we can test whether the occurrence rates of hot Jupiters changes with stellar mass.
\end{itemize}

\subsection{Main-sequence sample} 
\label{sec:MS_sample}

We restricted our study to main-sequence stars. We did not
wish to consider evolved stars because of the problems
with selection biases, shallower transit depths, and lack of substantial follow-up observations. We do note, though, that more than half of the
{\it TESS} stars brighter than 10th magnitude
are evolved. Eventually, this will be a rich hunting ground \citep[e.g.][]{2019arXiv190101643H,2019arXiv190109950R}.

\begin{figure*}
    \includegraphics[width=0.9\textwidth]{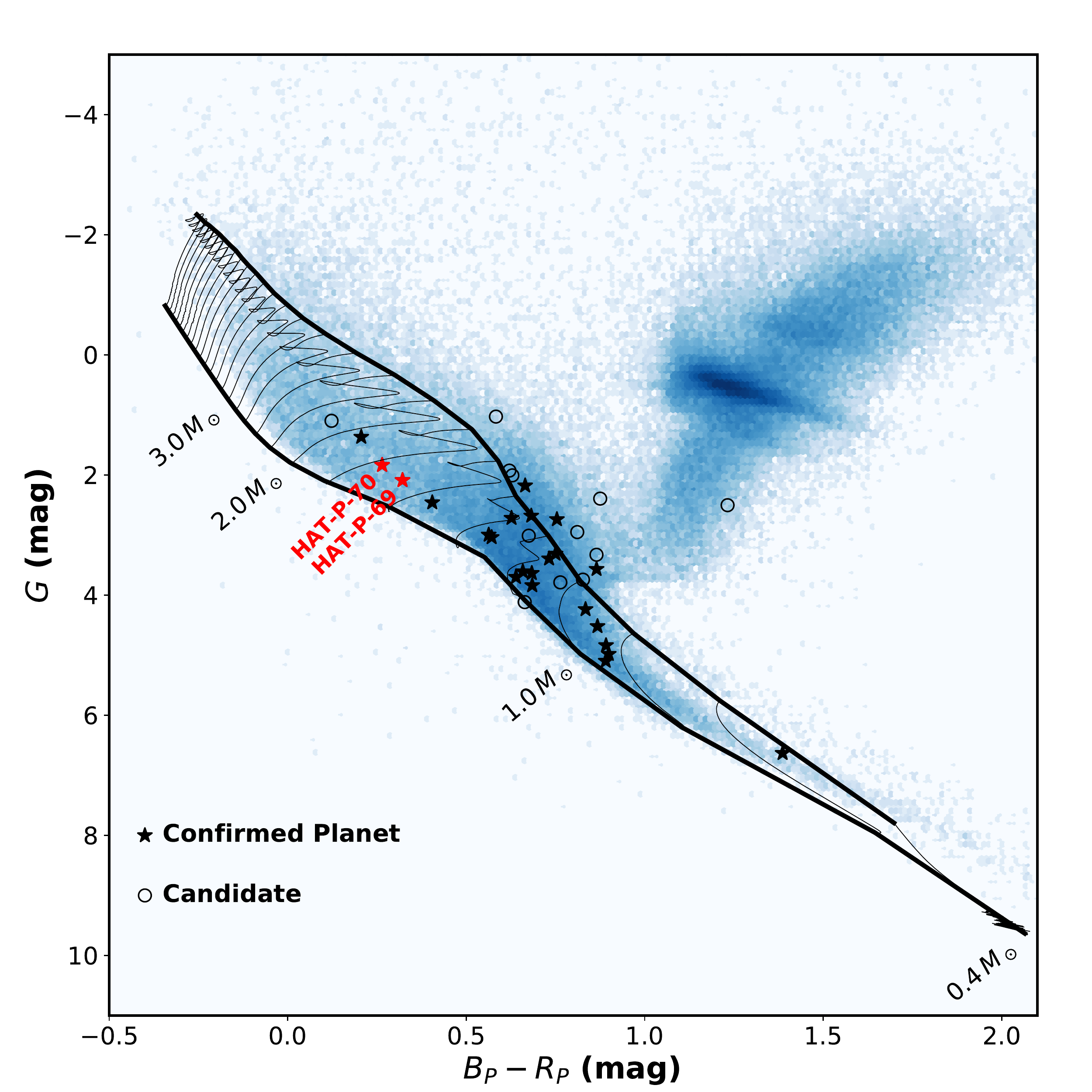} 
\caption{
The Gaia CMD for stars brighter than $T_\mathrm{mag} = 10$ observed within the first seven sectors of the \emph{TESS} mission. The Zero Age Main Sequence (lower) and Terminal Age Main Sequence (upper) boundaries are plotted to mark the main sequence. Evolution tracks from the MIST isochrones \citep{2016ApJS..222....8D} spaced at $0.2\,M_\odot$ intervals are plotted across the main sequence. Close in giant planets discovered or recovered by \emph{TESS} are plotted, marked in solid stars for confirmed planets, and open circles for planet candidates. Planet candidates off the main sequence, or around cool stars, but in the TESS Objects of Interest are are plotted on this diagram for completeness, but not included in the analysis. 
\label{fig:HR}}
\end{figure*}

Figure~\ref{fig:HR} shows the colour-magnitude diagram (CMD) of the 120{,}000 stars brighter than $T_\mathrm{mag} = 10$ that were observed by \emph{TESS}. The $B_P-R_P$ and $G$ values are taken from a cross match against the Gaia DR2 catalogue \citep{2018A&A...616A...1G}. 
To define the main sequence, we make use of the colors and magnitudes from the MESA Isochrones and Stellar Tracks (MIST) \citep{2016ApJS..222....8D}. We draw an upper and a lower boundary in the $B_P-R_P$ vs $G$ diagram based on the Zero Age Main Sequence (ZAMS) and the Terminal Age Main Sequence (TAMS) points in the solar metallicity MIST evolution tracks. As per \citet{2016ApJS..222....8D}, the ZAMS is defined by the criterion
that the core hydrogen luminosity of the star is 99.9\% that of the total core luminosity, while the TAMS is defined by the criterion that the core hydrogen fraction has fallen below $10^{-12}$. The ZAMS and TAMS boundaries are plotted in Figure~\ref{fig:HR}. Between these boundaries, we are left with 47{,}126 main sequence stars for this study. 

The restriction to stars with $T_\mathrm{mag} < 10$ allows
us to make use of the TOI catalogue available to the \emph{TESS} follow-up community, which is essentially complete for hot Jupiters.
The planet candidates around fainter stars in the FFIs
are not fully vetted. We also restrict attention to the data
from Sectors 1-7 because the candidates derived from later Sectors
have not yet received sufficient follow-up observations at the time of writing. 

\begin{figure*}
    \centering
        \includegraphics[width=0.9\textwidth]{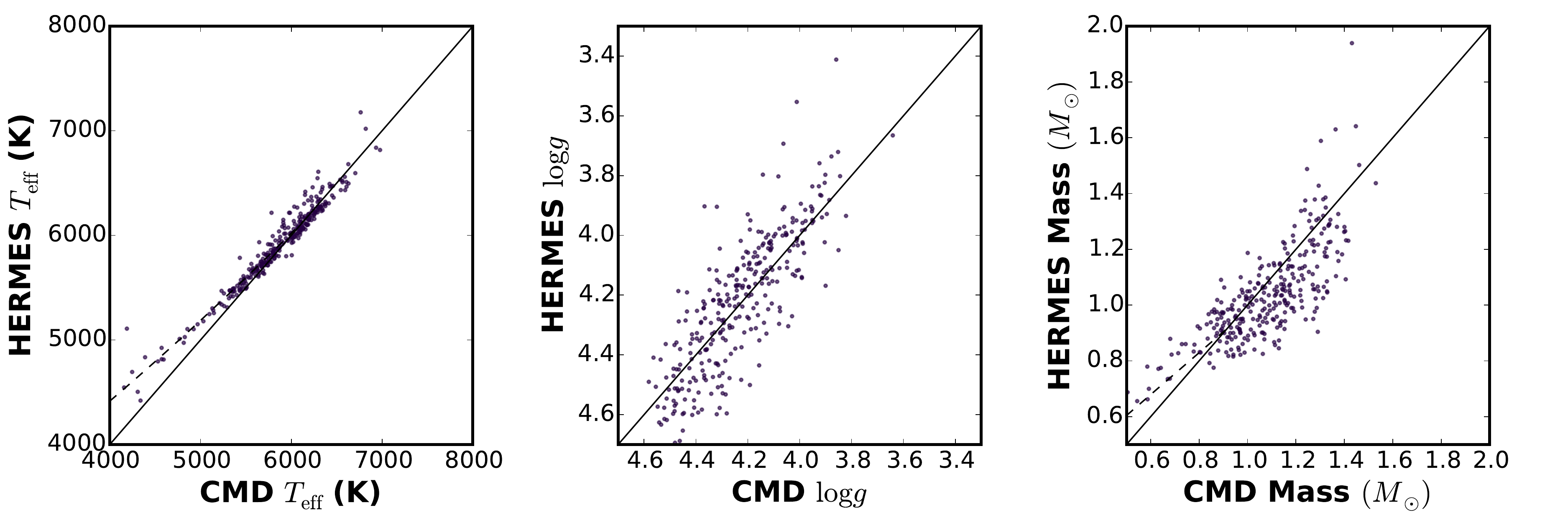} 
    \caption{Comparison between Gaia CMD derived stellar parameters and those from \emph{TESS}-HERMES. We find 301 slowly rotating stars ($v\sin I_\star < 20\,\kms$) within our sample and that have stellar parameters from \emph{TESS}-HERMES data release 1. We find a general consistency between the parameters, but apply a correction to our CMD derived $T_\mathrm{eff}$ and $M_\star$ of cool stars (marked by the dashed lines). }
    \label{fig:tess-hermes}
\end{figure*}

To check our CMD-derived stellar parameters, and to estimate the metallicity of the population, we cross-match our field stellar population against the \emph{TESS}-HERMES DR1 spectroscopic parameters for stars in the \emph{TESS} southern continuous viewing zone \citep{2018MNRAS.473.2004S}. Since the initial data release is restricted to stars within $10 < V < 13.1$, we expect a very limited number of matches. We find 491 stars to have stellar parameters from \emph{TESS}-HERMES within our sample, of which 301 have rotational broadening velocities $v\sin I_\star < 20\,\kms$. Figure~\ref{fig:tess-hermes} shows a comparison between our stellar effective temperature, surface gravity, and stellar mass against the spectroscopically measured values from \emph{TESS}-HERMES.

\begin{figure*}
    \centering
        \includegraphics[width=0.9\textwidth]{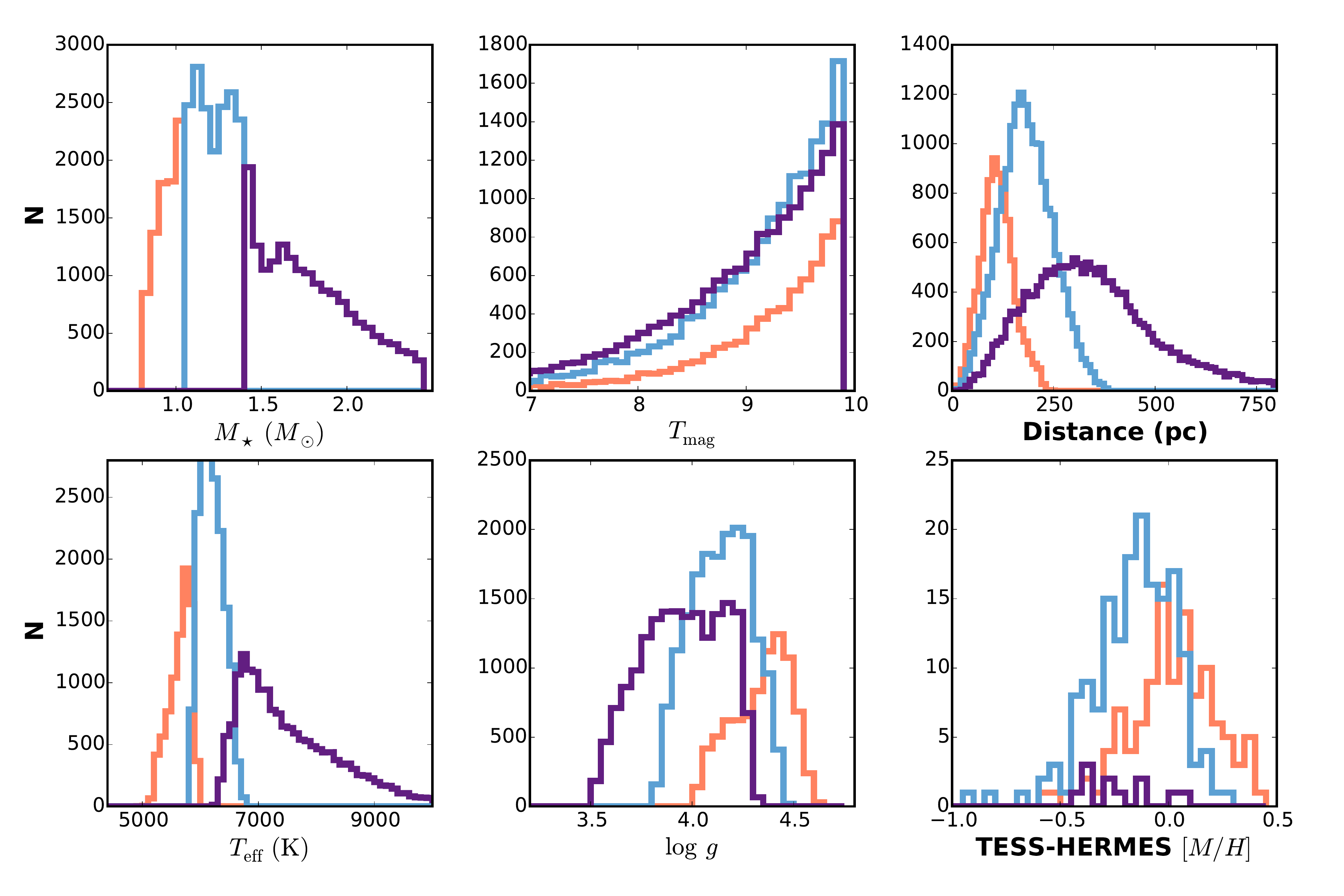} 
    \caption{Properties of the stellar population included in our sample, including mass ($M_\star$), brightness $(T_\mathrm{mag})$, distance, effective temperature $(T_\mathrm{eff})$, surface gravity $(\log g)$, and metallicity ([Fe/H]). The population is sub-divided into mass bins roughly corresponding to the A $(1.4-2.3\,M_\odot)$ (purple), F $(1.05-1.6\,M_\odot)$ (cyan), G ($0.8-1.05\,M_\odot$) (orange) spectral types. Metallicity measurements come from the 301 stars within our sample that have \emph{TESS}-HERMES measurements. }
    \label{fig:population_stats}
\end{figure*}

The median absolute deviations between CMD and spectroscopic parameters are 60\,K in $T_\mathrm{eff}$, $0.09$\,dex in $\log\,g$, and $0.09\,M_\odot$ in mass. However, we notice a systematic offset in our effective temperature and mass estimates for cool stars (dotted line in Figure~\ref{fig:tess-hermes}). We correct for this bias by fitting for a polynomial correction to our parameters as follows for temperature:
\begin{equation}
    T_\mathrm{eff} = 0.49 \, T_\mathrm{eff,CMD} + 1958 
\end{equation}
for stars with $4000 < T_\mathrm{eff,CMD} < 6120$\,K. We also apply a correction in mass:
\begin{equation}
    M_\star = 0.75 \, M_{\star,CMD}  + 0.23
\end{equation}
for $0.60 < M_{\star,CMD} < 0.92\,M_\odot$. Post correction, we find that median absolute deviations between CMD and spectroscopic parameters are 40\,K in $T_\mathrm{eff}$, and $0.08\,M_\odot$ in mass. 
Figure~\ref{fig:population_stats} shows the properties of the stellar population included in our sample. The sample is grouped into mass bins roughly corresponding to the A $(1.4-2.3\,M_\odot)$, F $(1.05-1.4\,M_\odot)$, G ($0.8-1.05\,M_\odot$) spectral types. We elaborate on the occurrence rates of planets within each mass bin in Sections~\ref{sec:completeness} and \ref{sec:occur_results}.

In particular, the metallicity distribution of the 301 stars with \emph{TESS}-HERMES measurements are plotted. We note that the population has near-solar metallicity of $\mathrm{[Fe/H]} = -0.06 \pm 0.21$. When sub-divided into the mass bins, we find the G star bin to have $\mathrm{[Fe/H]} = -0.03 \pm 0.20$, F stars to have $\mathrm{[Fe/H]} = -0.13 \pm 0.19$, and A stars to have $\mathrm{[Fe/H]} = -0.26 \pm 0.15$. We note that when sub-divided into their mass bins, the number of stars per bin become very small, and may not be representative of the population. We look forward to further fields of the \emph{TESS}-HERMES being completed, as well as similar surveys of brighter stars, for a better examination of the dependence between metallicity and the \emph{TESS} planet properties. 

\subsection{Candidate identification}
\label{sec:candidate}
Our planet sample makes use of the candidates (TOIs) released by the \emph{TESS} Science office from the first seven sectors of \emph{TESS} data around stars brighter than $T_\mathrm{mag} = 10$. The TOIs are selected from a list of threshold crossing events (TCEs) by human vetters. A threshold crossing event requires the signal to noise of the planet to be above 7.3, and that at least two transits are detected in the light curve. The human vetters reject some false positives based on standard diagnostics. For example, large secondary eclipse/phase variation detections that indicate that the eclipsing object is of stellar nature, obvious centroid offset detection that indicates the eclipsing events happened on a background object, or significant depth variation with the choice of photometric aperture. We also cross-reference the TCEs with known false positive/eclipsing binary catalogs \citep{2017A&A...608A.129T,2018AJ....156..234C}. 
Although the initial TOIs were generated from two different sources (the 2-min and the 30-min data), for uniformity we ensured that all the TOIs we used in this work are detected as TCEs through the Quick look pipeline, and that all the TCEs detected by the Quick look pipeline around stars brighter than $T_\mathrm{mag}=10$ magnitude went through the TOI process. 

We define our hot Jupiter candidates as TOIs with an orbital
period between 0.9 and 10 days,
a radius between 0.8 and 2.5\,$R_{\rm Jup}$, and a transit impact parameter smaller than 0.9. The period lower bound of 0.9\,days was adopted to incorporate WASP-18b \citep{2009Natur.460.1098H}, shortest period known hot Jupiter within \emph{TESS} sectors 1-7 \citep{2019AJ....157..178S}, into our sample. A similar minimum period cut-off was also employed by \citet{2012ApJS..201...15H} (0.7\,days) and \citet{2013ApJ...766...81F} (0.8\,days). We also note that no hot Jupiter candidates were found with periods $<0.9$\,days within our sample. 
To ensure a clean sample, we also require candidates to have Signal to Noise ratio (SNR) larger than 10 --- although, in practice, none of the giant planet candidates have a SNR between 10 and the traditional value of 7.3.
We use the stellar radii interpolated from the Gaia CMD (Section~\ref{sec:MS_sample}) to recompute the radius of the planet during the selection.

\subsection{Completeness and signal to noise estimates}
\label{sec:completeness}

\begin{figure*}
    \centering
    \begin{tabular}{ccc}
        \includegraphics[width=0.3\textwidth]{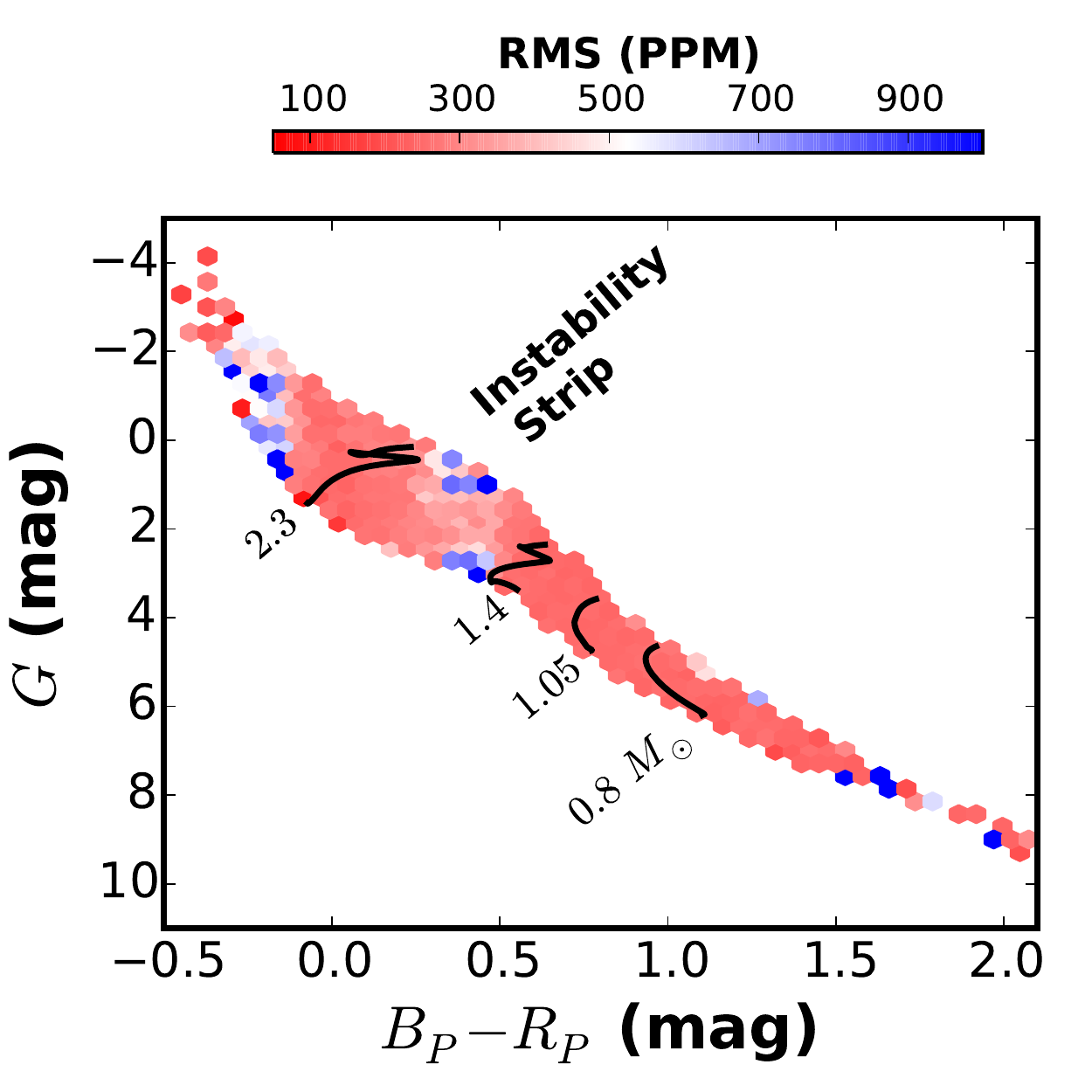} &
        \includegraphics[width=0.3\textwidth]{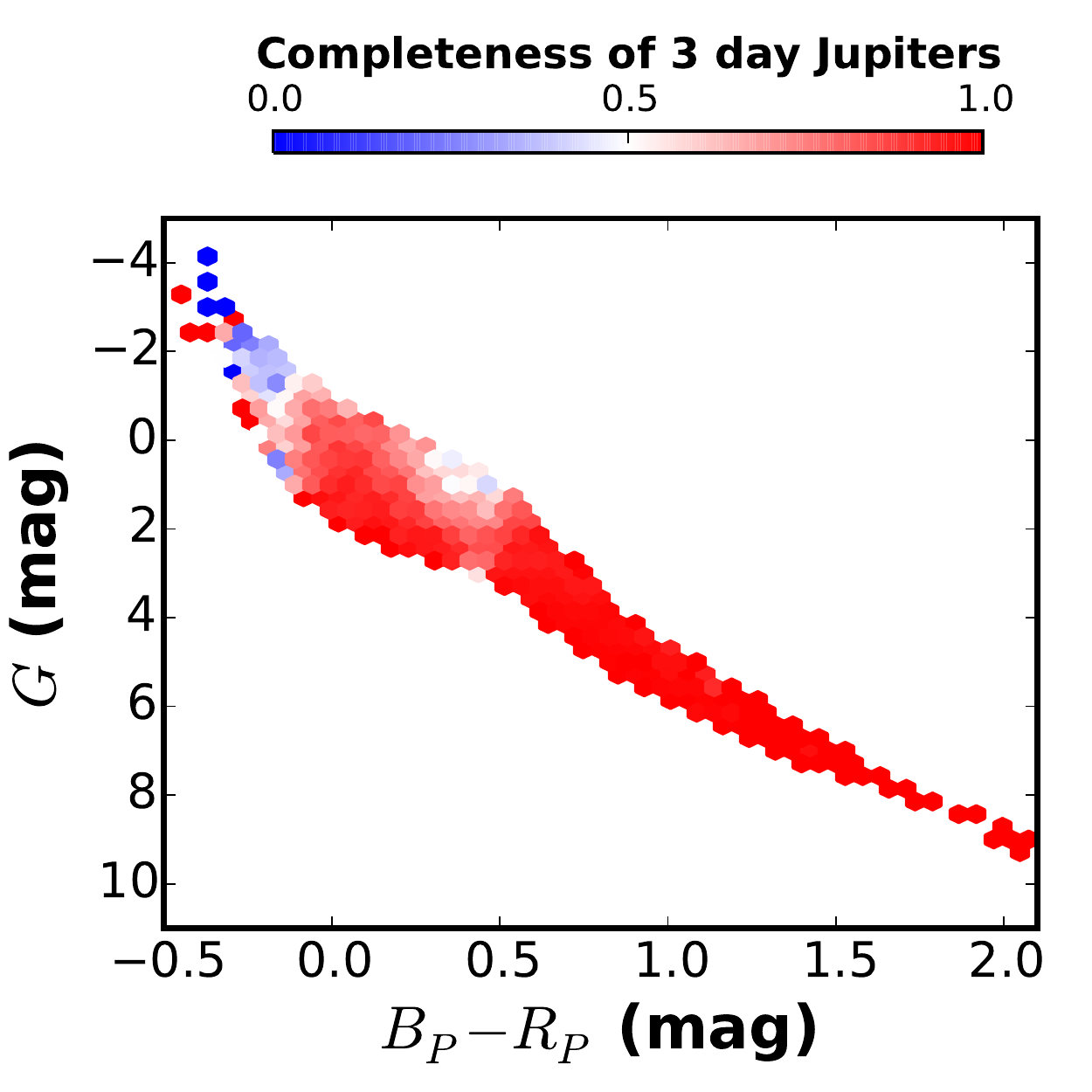} &
        \includegraphics[width=0.3\textwidth]{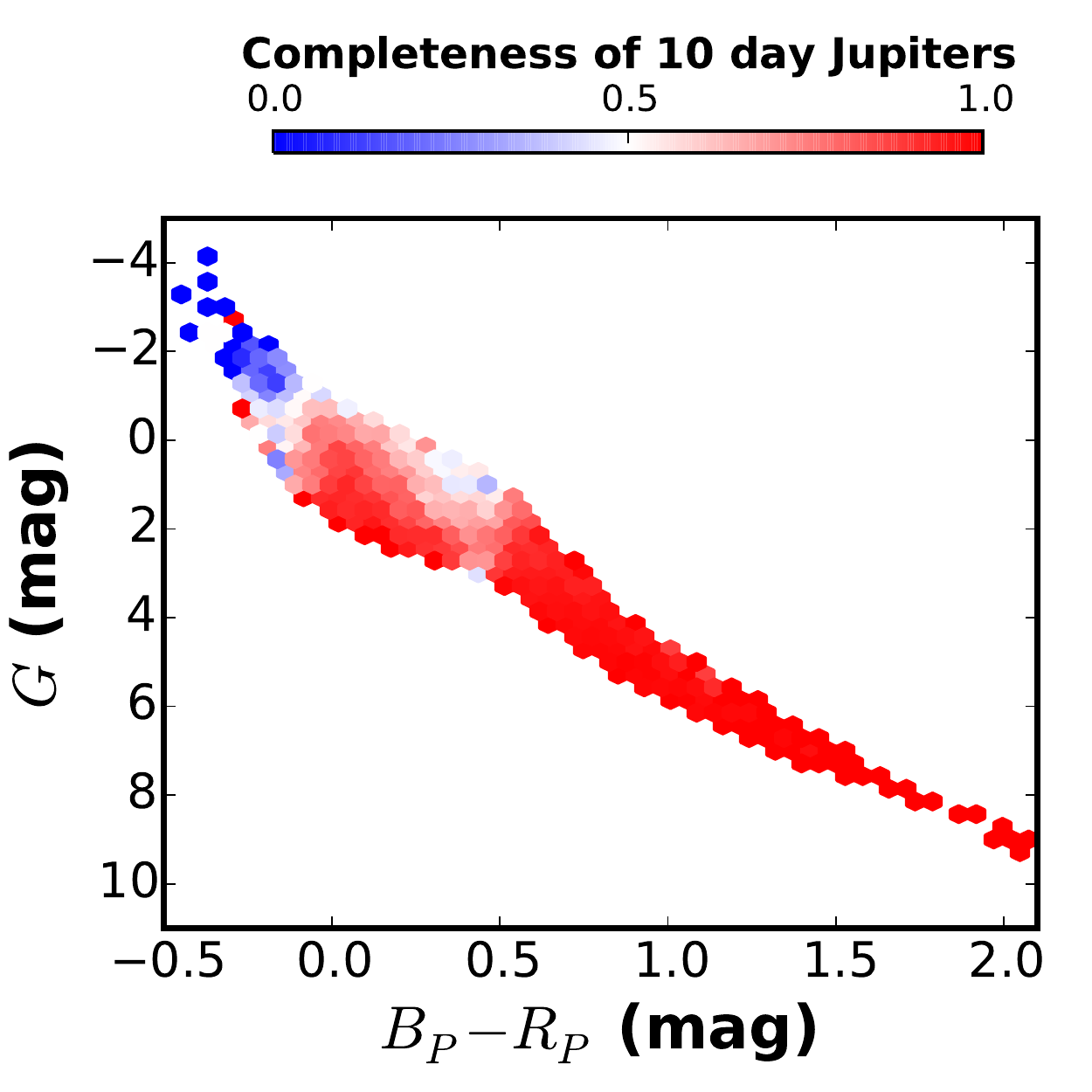} \\
    \end{tabular}
    \caption{\textbf{Left} The median light curve scatter across the main sequence. Evolution tracks for 0.8, 1.05, 1.4, and 2.3\,$M_\odot$ solar metallicity stars are plotted. The region near $1.6\,M_\odot$ exhibits higher levels of scatter than average due to stars in the instability strip. Survey completeness for a 3 day period \textbf{(Center)} and 10 day period \textbf{(Right)} Jupiter-sized planet are plotted. We find that we are 80\% complete for 10\,day period hot Jupiters across the lower main sequence ($<1.4\,M_\odot$), and 70\% complete for such planets around intermediate mass stars ($ 1.4 < M_\star < 2.3\,M_\odot$).}
    \label{fig:completeness}
\end{figure*}

Since the expected noise floor for a typical \emph{TESS} star at $T_\mathrm{mag} = 10$ per 1 hour is 200\,ppm \citep{2018ApJ...868L..39H}, any giant planet transiting a main sequence star in our sample should be detected with a high SNR. However,
some stars may exhibit large amplitude and short time scale stellar variability, 
such as stars on the instability strip of the CMD.
Strong stellar variability can reduce the sensitivity to
transit signals. To estimate our completeness rate more accurately, we measured the per point median absolute deviation (MAD) $\sigma_{mad}$ of detrended/deblended light curves for all the 47{,}126 stars used in this paper, derived from the FFIs using the Quick look pipeline. A factor of 1.48 is applied to $\sigma_{mad}$ such that it approximates the standard deviation scatter of the light curves.  The signal to noise $SNR$ of the candidates is then estimated with 
\begin{equation}
SNR = \frac{\delta}{1.48\times \sigma_{mad}}\left(\frac{T_{dur}}{0.5}N_{tr}\right)^{0.5},
\end{equation}
where $\delta$ is the approximate transit depth, $T_{dur}$ is the full transit duration in hours, and $N_{tr}$ is the number of transits that appeared
in the data from \emph{TESS} Sectors 1--7.\footnote{We have taken into account the actual duty cycles in each \emph{TESS} sector by only using the light curve available to the Box Least Search in the Quick look pipeline. This is the light curve length after accounting for bad points masking due to scattered light, pointing jitter, and data down-link gap. The number of days used in each of these seven sectors are: 21.5, 21.4, 16.5, 15.3, 21.5, 17.3, 21.5.}
We assume any planet with a calculated SNR exceeding 10 was selected as a candidate, and otherwise was not selected. We also assume that the hot Jupiters exhibit a uniform distribution in transit impact parameter between 0 and 0.9.
Figure \ref{fig:completeness} shows the survey completeness for a
Jupiter-sized planet with an impact parameter of 0.45, for both 3-day and
10-day orbits.
The transit duration is calculated under the assumption of
a circular orbit. While this assumption may not be valid
for planets with periods approaching 10 days, it has been 
shown that modestly eccentric orbits have negligible effect on survey completeness \citep{2008ApJ...679.1566B}.

\subsection{Results}
\label{sec:occur_results}

A total of 47{,}126 stars and \nTotal{} TOIs are included in the occurrence rate calculation. The TOIs are composed
of \nPlanet{} confirmed planets, \nCandidate{} planet candidates, and \nFalsePositive{} false positives. The list of planets, candidates, and false positives are given in Appendix~\ref{sec:candidates}. To summarize the previous sections, the stellar and planet population are defined within the criteria below. 
\begin{itemize}
    \item Brighter than $T_\mathrm{mag} = 10$.
    \item Lying within the solar metallicity ZAMS and TAMS boundaries on the \emph{Gaia} $B_P-R_P$ vs $G$ CMD, and thereby classified as main sequence.
    \item Planets are detected with BLS signal-to-noise ratio $>10$ and passed the vetting process.
    \item Planets with periods $0.9 \leq P \leq 10$\,days.
    \item Planets with radii $0.8 \leq R_p \leq 2.5\,R_\mathrm{Jup}$.
    \item Transits with impact parameter $b<0.9$ to avoid grazing transits.
\end{itemize}

Within this stellar sample, the population is binned by stellar mass into A $(1.4-2.3\,M_\odot)$, F $(1.05-1.4\,M_\odot)$, G ($0.8-1.05\,M_\odot$) spectral types. We estimate the occurrence rate $f$ within each stellar mass bin as the conjugate distribution of the binomial distribution (i.e. the beta distribution), 
\begin{equation}
\mathcal{P}(f) = {\mathrm{Beta}}(n_{obs}, n_{trial}-n_{obs}),
\end{equation}
in which $n_{obs}$ is the number of the transiting planets observed in the mass bin and $n_{trial}$ is the effective number of times we try to conduct the detection of those transiting planets after accounting for transit probability and completeness. Specifically, 
\begin{equation}
n_{obs} = \sum_{i =1,n_p} (1-FP_i) w_i,
\end{equation}
where $w_i$ is a weight indicating the probability that a planet/candidate falls within a particular mass bin.
The probability distribution for the mass of each planet/candidate host star is modeled as a Gaussian distribution centered on the estimated mass,
and with a dispersion equal to 10\% of the value of the estimated mass.  
The false positive rate $FP$ is estimated in each stellar mass bin using current follow-up results, and is only applied to the active candidates. For the confirmed planets, the false positive rate is set equal to zero. The false positive rate $FP$ is applied only to the active planet candidates, while $FP=0$ for confirmed planets. The false positive rate is calculated per stellar mass bin as 
\begin{equation}
FP = \frac{N_\mathrm{False positives}}{N_\mathrm{Confirmed Planets} + N_\mathrm{False Positives}}\,.
\end{equation}
Based on the photometric and spectroscopic observations that have been performed so far by the \emph{TESS} follow-up program, 
we find a false positive rate of 15\% for G stars, 41\% for F stars, and 47\%
for A stars.
Globally, the false positive rate for hot Jupiters from \emph{TESS} within our sample is 35\%. Similar false positive rates for short period giant planets (29.3\%) were reported by \citet{2013ApJ...766...81F} for the initial \emph{Kepler} candidates. The uncertainty assumes Poisson errors based on the number of planets candidates and false positives surveyed so far. 

We define $n_{trial}$ as
\begin{equation}
n_{trial} = \sum_{i=1, n_*} \int \mathcal{P}_{tran}\mathcal{P}_{det} dPdR,
\end{equation}
in which $n_*$ is the total number of observed stars fall in a particular mass bin, $\mathcal{P}_{tran}$ and $\mathcal{P}_{det}$ are the probability of a planet with period $P$ and radius $R$ transiting and being detected around star $i$, respectively. The transit probability for a planet with period $P$ around a star with radius $r_i$ and mass $m_i$ is 
\begin{equation}
\mathcal{P}_{tran, i} (P)= 0.9\,r_i\,\left(\frac{2\pi}{P}\right)^{2/3}(G\,m_i)^{-1/3} \, .
\end{equation}
The coefficient of 0.9 is present because we only consider
planets and candidates with impact parameters smaller than 0.9. The probability of detection for each star is estimated following Section~\ref{sec:completeness}, assuming any planet with SNR~$\leq 10$ has
been detected. The final integration is computed using a Monte Carlo method assuming that the intrinsic period distribution of planet is uniform within the range from 0.9 to 10 days, and the radius distribution of planet is uniform within the range from 0.8 to 1.5 $R_\mathrm{Jup}$.

\begin{figure}
\centering
    \includegraphics[width=0.4\textwidth]{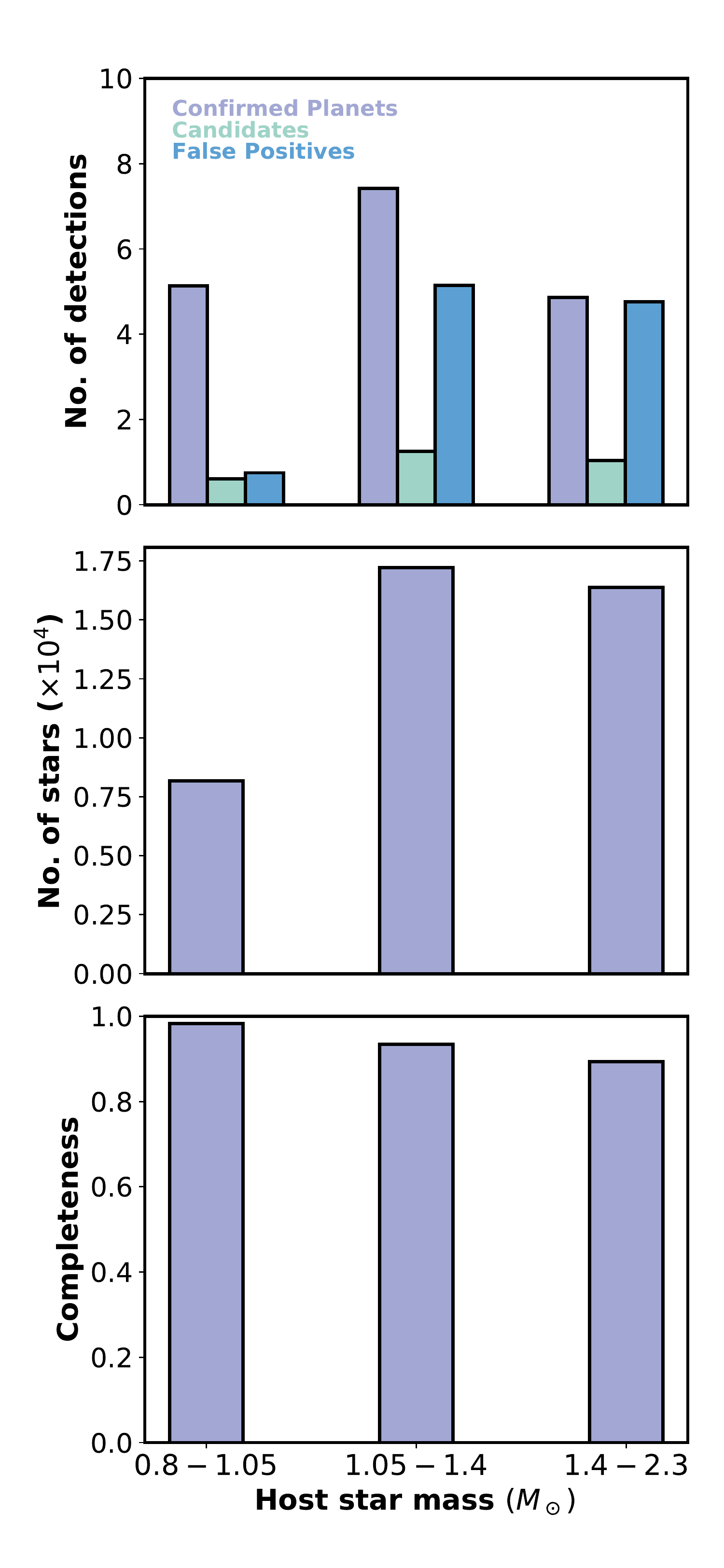}
\caption{
The sample size and completeness, for each range of stellar mass. \textbf{Top} The number of confirmed planets, candidates, and false positives for each
range of stellar mass. The non-integer number of detections within each bin is due to the mass uncertainty of each host star being taken into account.  \textbf{Center} The stellar sample size (in $10^4$ stars). \textbf{Bottom} The planet detection completeness for a 10-day Jupiter sized planet.  
\label{fig:Nplanets}}
\end{figure}

Figure~\ref{fig:Nplanets} summarizes the planet sample, search completeness, and field star population within each spectral class mass bin. 

\begin{figure*}
\centering
    \includegraphics[width=0.6\textwidth]{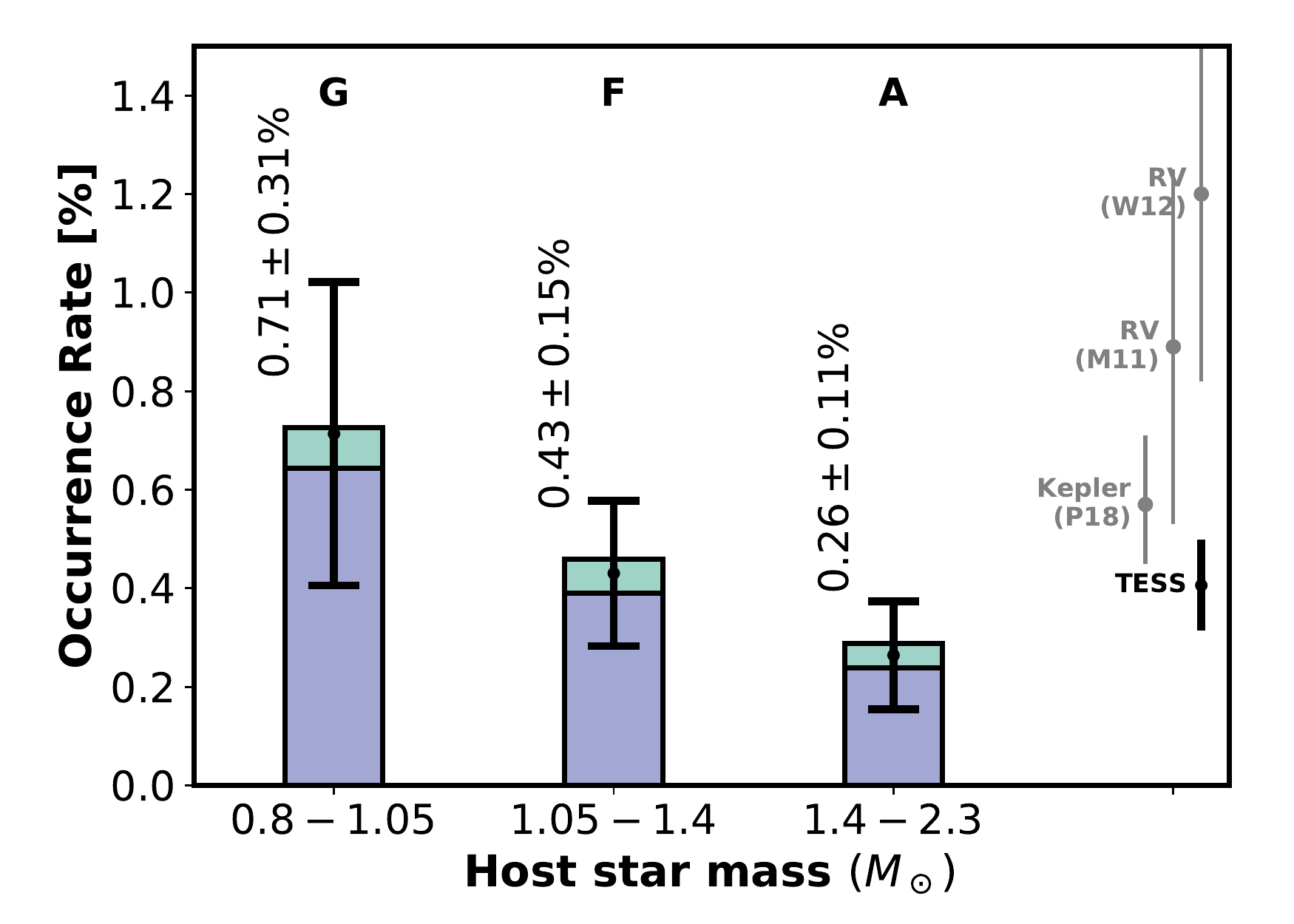} 
\caption{
The occurrence rates of hot Jupiters across the main sequence as measured by \emph{TESS}. The dark blue bars mark the occurrence rates from confirmed planets only, cyan bars mark the occurrence rates if we assume that all unconfirmed planet candidates are true planets. The center of each error bar marks the expected occurrence rate taking into account false positive rate from follow-up observations to date. The expected occurrence rate of each mass bin is labelled. We find a total occurrence rate of \ortotal{} for the entire sample. For comparison, the occurrence rate of hot Jupiters from \emph{Kepler} ($0.57_{-0.12}^{+0.14}$\% from P18 \citealt{2018AJ....155...89P}), and radial velocity surveys ($0.89 \pm 0.36$\% from M11 \citealt{2011arXiv1109.2497M} and $1.20 \pm 0.38 $\% from W12 \citealt{2012ApJ...753..160W}) are marked. 
\label{fig:occurrence_rate}}
\end{figure*}

This planet and host star sample yields a total hot Jupiter occurrence rate from \emph{TESS} of \ortotal{}. Within each mass bin, we find an occurrence rate of \orG{} for main sequence G stars, \orF{} for F stars, and \orA{} for A stars. These occurrence rates are presented in Figure~\ref{fig:occurrence_rate}. 

In this analysis, we defined the main-sequence as being bound within the solar metallicity ZAMS and TAMS lines. The actual population should exhibit a dispersion in metallicity, with the effect of stars being brighter at higher metallicity for the same evolutionary state, and vice versa for lower metallicity stars. To test the effect of a more blurred main sequence boundary, we re-performed the analysis whilst assuming a $\mathrm{[Fe/H]} = -0.27$ ZAMS boundary and a $\mathrm{[Fe/H]} = +0.15$ TAMS boundary -- encompassing the $1\sigma$ dispersion in metallicity seen in our cross-matched \emph{TESS}-HERMES stars. The resulting main-sequence sample increased to 52{,}788 stars, and included two additional confirmed planets around F stars, two new candidates about G stars, one new candidate around an F star, and one new candidate around an A star.
The net result is no significant change in the occurrence rates within each mass bin, nor any significant change for the whole sample. 

Some caution may be necessary when directly comparing our occurrence rate against that derived from \emph{Kepler} data. Our stellar sample is restricted to the main-sequence stars, whilst the \emph{Kepler} sample may contain more evolved stars \citep{2015ApJ...799..229W}. Our definition of the main sequence is also different from more traditional definitions, which are
based on surface gravity.  We do not impose a surface gravity criterion
because stars on the main sequence have different surface gravities at different masses: an intermediate-age main-sequence K star has $\log g \approx 4.5$, while A stars have $\log g \approx 3.8$ at the same evolutionary stage. Some previous works required $\log g < 3.9$ or 4.0 to define the main sequence, which may remove 10--30\% of the main sequence population between $6000 < T_\mathrm{eff} < 6500$\,K
\citep[e.g.][]{2012ApJS..201...15H,2018AJ....155...89P}.
We find that if we apply a limit of $\log g < 4.0$ to our sample, we increase the occurrence rates of hot Jupiters around F and A stars by
nearly a factor of 2. 

Although \emph{TESS} is largely complete for hot Jupiters around F and G stars, the sensitivity is poorer for more evolved early A stars, for which
the stellar radius can be as large as 4\,$R_\odot$. To check the dependence of our results on the completeness calculations, we tried drawing a boundary around smaller-radius A stars (defined by the boundary between $-0.1 < B_P-R_P < 0.5$ and $G > G_\mathrm{ZAMS} - 1.0$). For stars within this boundary,
the completeness is 80\% for hot Jupiters with a period of 10\,days.
All of the confirmed cases of hot Jupiters around 
A stars that were used in our preceding calculations also
reside within this more restricted sample. We find no significant difference ($<1\sigma$) in the occurrence rates present above and those obtained within this 'near-complete' box. 

Unrecognized binaries in the main sequence population can cause systematic errors in occurrence rate estimates. \citet{2018AJ....155..244B} found that systematic biases due to binarity may be important for small planets, but for \emph{Kepler} hot Jupiters the bias is only at the level of $\sim$5\%,
smaller than our current uncertainties. Our occurrence rates were also obtained for a main sequence defined between the ZAMS and TAMS boundaries, which has the effect of removing some binaries because they appear overluminous. In testing for the effect of metallicity on our occurrence rates, we shifted the ZAMS and TAMS boundaries, but found minimal effect on the resulting occurrence rates.

A number of caveats still exist. The number of hot Jupiters around bright stars to be identified or recovered by \emph{TESS} over the course of its mission will be at least four times that presented in this paper. We expect these occurrence rates and false positive rates to be revised over the course of the mission. In particular, the majority of new hot Jupiters from \emph{TESS} should be around intermediate mass stars; the ground-based transit surveys are least complete, and the hot Jupiter follow-up effort is most expensive within this regime. The uncertainties in our occurrence rates are currently dominated by Poisson statistics. 

\section{Conclusions}
\label{sec:discussion}

\subsection{Agreement of \emph{TESS} and \emph{Kepler} hot Jupiter occurrence rates}

We find good agreement between occurrence rates of hot Jupiters
derived from the \emph{TESS} and \emph{Kepler} surveys.
The occurrence rate from \emph{TESS} is \ortotal{}. From
\emph{Kepler}, various studies have found
occurrence rates of $0.4\pm0.1$\% \citep{2012ApJS..201...15H},
$0.43\pm0.05\%$ \citep{2013ApJ...766...81F},
$0.57_{-0.12}^{+0.14}\%$ \citep{2018AJ....155...89P},
and 
$0.43^{+0.07}_{-0.06}\%$ \citep{2017AJ....153..187M}.

The number of stars and planets within the \emph{TESS} sample is already
comparable to that from the \emph{Kepler} sample, and will soon grow.
We make use of 47{,}126 stars and \nPlanet{} planets and \nCandidate{} active candidates. Previously determined
occurrence rates of hot Jupiters were computed from 24 planet candidates around 58{,}000 stars by \citet{2012ApJS..201...15H}, and out of 14 planets around 37{,}000 stars by \citet{2018AJ....155...89P}. The light curve precision that \emph{TESS} provides for these bright stars are also comparable to that for the relatively fainter stars from the \emph{Kepler} sample. 

Our initial estimates of the sample metallicity, derived from a cross match of the bright \emph{TESS} stars against the \emph{TESS}-HERMES \citep{2018MNRAS.473.2004S} catalog suggest that our sample $(\mathrm{[Fe/H]} = -0.06 \pm 0.21)$ is similar to that of \emph{Kepler} ($-0.045 \pm 0.009$) \citep{2017ApJ...838...25G}. Future Southern spectroscopic surveys of bright stars will continue to improve our understanding of the properties of field stars surveyed by \emph{TESS}. 

The average Solar-type star from this \emph{TESS} sample is located at 150\,pc, while that observed by \emph{Kepler} would be located at 400\,pc \citep{2017ApJS..229...30M}.  Past surveys of more distant fields around galactic bulge and disk \citep{2006AcA....56....1G,2011ApJ...743..103B} also found occurrence rates of hot Jupiters to be compatible
with the rates derived from \emph{Kepler} and \emph{TESS} data,
suggesting that there is not too much variety in the occurrence
of hot Jupiters across the Galaxy.

We also remark on the near-completeness of the ground based surveys. Of the \nPlanet{} confirmed hot Jupiters within our sample, 13 were already discovered by the WASP \citep{2006PASP..118.1407P}, HATNet \citep{Bakos:2004}, and KELT \citep{2012PASP..124..230P} consortiums. Future studies of hot Jupiter properties from \emph{TESS} will continue to capitalize on the follow-up efforts already made by these surveys. 

\subsection{No evident dependence on stellar mass}

The occurrence rates of hot Jupiters within our A, F, and G mass bins agree
with each other to within $1\sigma$. Hot Jupiters are just as abundant around main-sequence A stars as they are around F and G stars. Radial-velocity surveys have reported a paucity of giant planets in close-in orbits about ``retired A stars.'' Together this
seems to support the conclusion that enhanced tidal dissipation within evolved stars accelerates the process of tidal orbital decay of hot Jupiters
\citep{2013ApJ...772..143S}. Post main-sequence tidal evolution may be strongly dependent on the mass of the planets \citep[e.g.][]{2009ApJ...705L..81V,2014ApJ...794....3V}, more stringent constraints on the distribution of these main-sequence close-in giant planets may help yield additional clues into the tidal model for hot Jupiters. 
We note, though, that sample sizes of the Doppler surveys ranged from 166
stars \citep{2014A&A...566A.113J} to 373 \citep{2015A&A...574A.116R} stars, small enough that one should only expect $\sim$1 hot Jupiter to be found even if stellar evolution has no effect on the hot Jupiter occurrence rate. 
The Doppler surveys also noted an enhanced planet fraction for longer-period gas giants about more massive stars. \citet{2018ApJ...860..109G} notes a $2\times$ increase in planet fraction about $2\,M_\odot$ stars compared to Solar mass stars, whilst \citet{2010PASP..122..905J} noted nearly $3\times$ increase in the planet fraction within the $1-2\,M_\odot$ host mass range. Curiously, the hot Jupiter occurrence rate does not reflect this trend. Hot Jupiters are no more abundant about A stars than they are about F and G stars. Since the planets around early type stars exhibit a wide distribution of obliquity angles \citep{2012ApJ...757...18A}, this may point to a lack of stellar mass preference for the dynamical migration of hot Jupiters. 

\acknowledgements  
We thank the referee for their careful reading of the manuscript, the comments significantly improved the quality of the paper. 
Work by G.Z. is provided by NASA through Hubble Fellowship grant HST-HF2-51402.001-A awarded by the Space Telescope Science Institute, which is operated by the Association of Universities for Research in Astronomy, Inc., for NASA, under contract NAS 5-26555.
HATNet operations have been
funded by NASA grants NNG04GN74G and NNX13AJ15G. Follow-up of HATNet
targets has been partially supported through NSF grant
AST-1108686. G.\'A.B., Z.C., and K.P.\ acknowledge partial support
from NASA grant NNX09AB29G. J.H.\ acknowledges support from NASA grant
NNX14AE87G. K.P. acknowledges support from NASA grants 80NSSC18K1009 and NNX17AB94G. We
acknowledge partial support also from the {\em Kepler} Mission under
NASA Cooperative Agreement NNX13AB58A (D.W.L., PI). Data presented in this paper are
based on observations obtained at the HAT station at the Submillimeter
Array of SAO, and the HAT station at the Fred Lawrence Whipple
Observatory of SAO. This research has made use of the NASA Exoplanet
Archive, which is operated by the California Institute of Technology,
under contract with the National Aeronautics and Space Administration
under the Exoplanet Exploration Program. Observations on the LCO 1m network were made through NOAO program 2017B-0039, and time on the WIYN 3.5m was awarded through NOAO programs 2016B-0078 and 2017A-0125. Observations on the SMARTS 1.5\,m CHIRON facility were  made through the NOAO program 2019A-0004. M.G. and E.J. are FNRS Senior Research Associates. The research leading to these results has received funding from the grant for Concerted Research Actions, financed by the Wallonia-Brussels Federation. Funding for the TESS mission is provided by NASA's Science Mission directorate. We acknowledge the use of public TESS Alert data from pipelines at the TESS Science Office and at the TESS Science Processing Operations Center. This research has made use of the Exoplanet Follow-up Observation Program website, which is operated by the California Institute of Technology, under contract with the National Aeronautics and Space Administration under the Exoplanet Exploration Program. This paper includes data collected by the TESS mission, which are publicly available from the Mikulski Archive for Space Telescopes (MAST). G.K. thanks the support from the National Research, Development and Innovation Office (grants K 129249). G.K. thanks the support from the National Research, Development and Innovation Office (grants K 129249). The observations by the CHIRON spectrograph have been supported by the grant of the Hungarian Scientific Research Fund (OTKA, K-81373) to G.K. WASP-South is hosted by the South African Astronomical Observatory and we are grateful for their ongoing support and assistance. Funding for WASP comes from consortium universities and from the UK’s Science and Technology Facilities Council.
Work by J.N.W.\ was supported by the Heising-Simons Foundation. This work is partly supported by JSPS KAKENHI Grant Numbers JP18H01265 and JP18H05439, and JST PRESTO Grant Number JPMJPR1775. Observations in the paper made use of the NN-EXPLORE Exoplanet and 
Stellar Speckle Imager (NESSI). NESSI was funded by the NASA Exoplanet 
Exploration Program and the NASA Ames Research Center. NESSI was built 
at the Ames Research Center by Steve B. Howell, Nic Scott, Elliott P. 
Horch, and Emmett Quigley.
I.W. is supported by a Heising-Simons \textit{51 Pegasi b} postdoctoral fellowship.
A.\,K. acknowledges support from the National Research Foundation (NRF) of South Africa.
Some of the observations reported in this paper were obtained with the Southern African Large Telescope (SALT).
\facility{HATNet, FLWO 1.5\,m, CHIRON, TESS, SALT}
\software{lightkurve \citep{lightkurve}, emcee \citep{2013PASP..125..306F}, simultrans \citep{2018AJ....155...13H}, Astropy \citep{2013A&A...558A..33A,2018AJ....156..123A}, fitsh \citep{2012MNRAS.421.1825P}}
\FloatBarrier
\appendix

\section{Planets and planet candidate}
\label{sec:candidates}

We tabulate here all \emph{TESS} Objects of Interests that made up the numerator of our occurrence rate calculation. Table~\ref{tab:confirmedplanets} presents the confirmed planets, \ref{tab:candidates} shows the planet candidates and their follow-up stats, \ref{tab:falsepositive} shows the false positives, and \ref{tab:confirmedplanetsnotincluded} shows the confirmed giant planets orbiting stars $T_\mathrm{mag} < 10$ that were not included in the sample due to the evolved states of their host stars. The planets and candidates lists are up-to-date as of 2019-06, and can be accessed via \url{tev.mit.edu}. 

\begin{deluxetable*}{rrrrrrrrrrrrrl}
\tablewidth{0pc}
\tabletypesize{\scriptsize}
\tablecaption{
    Confirmed planets with $T_\mathrm{mag} < 10$
    \label{tab:confirmedplanets}
}
\tablehead{ \\
    \multicolumn{1}{c}{TIC} &
    \multicolumn{1}{c}{TOI} &
    \multicolumn{1}{c}{Name} &
    \multicolumn{1}{c}{Status\tablenotemark{a}} &
    \multicolumn{1}{c}{Period} &
    \multicolumn{1}{c}{Depth} &
    \multicolumn{1}{c}{Gaia $G$} &
    \multicolumn{1}{c}{Gaia $B_P$} &
    \multicolumn{1}{c}{Gaia $R_P$} &
    \multicolumn{1}{c}{Distance} &
    \multicolumn{1}{c}{$T_\mathrm{eff}$}\tablenotemark{b} &
    \multicolumn{1}{c}{$M_\star$} &
    \multicolumn{1}{c}{$R_\star$} &
    \multicolumn{1}{c}{Reference} \\
    \multicolumn{1}{c}{} &
    \multicolumn{1}{c}{} &
    \multicolumn{1}{c}{} &
    \multicolumn{1}{c}{} &
    \multicolumn{1}{c}{(d)} &
    \multicolumn{1}{c}{(ppm)} &
    \multicolumn{1}{c}{(mag)} &
    \multicolumn{1}{c}{(mag)} &
    \multicolumn{1}{c}{(mag)} &
    \multicolumn{1}{c}{(pc)} &
    \multicolumn{1}{c}{(K)} &
    \multicolumn{1}{c}{$(M_\odot)$} &
    \multicolumn{1}{c}{$(R_\odot)$} &
    \multicolumn{1}{c}{} 
}
\startdata
1129033 & 398.01 & WASP-77A & P & 1.4 & 16380 & 10.1 & 10.48 & 9.59 & 105 & 5433 & 0.91 & 0.98 & {\citet{2013PASP..125...48M}}\\
 25375553 & 143.01 & WASP-111 & P & 2.3 & 6939 & 10.11 & 10.36 & 9.73 & 300 & 6305 & 1.3 & 2.03 & {\citet{2014arXiv1410.3449A}}\\
 47911178 & 471.01 & WASP-101 & P & 3.6 & 12321 & 10.14 & 10.41 & 9.75 & 202 & 6209 & 1.16 & 1.39 & {\citet{2014MNRAS.440.1982H}}\\
 65412605 & 626.01 & KELT-25 & P & 4.4 & 5812 & 9.6 & 9.68 & 9.47 & 442 & 7983 & 1.92 & 2.39 & Quinn et al. in-prep \\
 92352620 & 107.01 & WASP-94A & P & 4.0 & 12999 & 10.03 & 10.33 & 9.6 & 212 & 5949 & 1.16 & 1.67 & {\citet{2014AandA...572A..49N}}\\
 100100827 & 185.01 & WASP-18 & P & 0.9 & 10692 & 9.17 & 9.43 & 8.79 & 123 & 6291 & 1.16 & 1.3 & {\citet{2009Natur.460.1098H}} \\
 144065872 & 105.01 & WASP-95 & P & 2.2 & 11836 & 9.94 & 10.29 & 9.46 & 138 & 5627 & 0.99 & 1.28 & {\citet{2014MNRAS.440.1982H}}\\
 149603524 & 102.01 & WASP-62 & P & 4.4 & 14034 & 10.07 & 10.36 & 9.67 & 176 & 6123 & 1.13 & 1.29 & {\citet{2012MNRAS.426..739H}}\\
 166836920 & 267.01 & WASP-99 & P & 5.8 & 5386 & 9.33 & 9.64 & 8.89 & 159 & 5894 & 1.16 & 1.77 & {\citet{2014MNRAS.440.1982H}}\\
 170634116 & 413.01 & WASP-79 & P & 3.7 & 12455 & 9.97 & 10.2 & 9.63 & 248 & 6571 & 1.38 & 1.65 & {\citet{2012AandA...547A..61S}}\\
 183532609 & 191.01 & WASP-8 & P & 8.2 & 15535 & 9.61 & 10.0 & 9.11 & 90 & 5455 & 0.89 & 1.04 & {\citet{2010AandA...517L...1Q}}\\
 201248411 & 129.01 &  & P & 1.0 & 7028 & 10.59 & 11.23 & 9.85 & 61 & 4216 & 0.5 & 0.81 & Nielsen et al. in-prep\\
 230982885 & 195.01 & WASP-97 & P & 2.1 & 13510 & 10.42 & 10.79 & 9.92 & 151 & 5526 & 0.9 & 1.17 & {\citet{2014MNRAS.440.1982H}}\\
 267263253 & 135.01 & & P & 4.1 & 10068 & 9.52 & 9.75 & 9.18 & 197 & 6538 & 1.37 & 1.63 & {\citet{2019AA...625A..16J}}\\
 379929661 & 625.01 & \hatstara{} & P & 4.8 & 7627 & 9.77 & 9.9 & 9.58 & 344 & 7532 & 1.68 & 1.92 & This work\\
 399870368 & 624.01 & \hatstarb{} & P & 2.7 & 8443 & 9.45 & 9.55 & 9.29 & 333 & 7818 & 1.77 & 2.03 & This work\\
 425206121 & 508.01 & KELT-19A & P & 4.6 & 10364 & 9.86 & 10.0 & 9.6 & 302 & 7188 & 1.52 & 1.77 & {\citet{2018AJ....155...35S}}\\
 455135327 & 490.01 & HAT-P-30 & P & 2.8 & 10758 & 10.3 & 10.58 & 9.89 & 215 & 6116 & 1.12 & 1.42 & {\citet{2011ApJ...735...24J}}\\
\enddata
\tablenotetext{a}{
    P: Confirmed Planet
}
\tablenotetext{b}{
    Parameters $T_\mathrm{eff}$, $M_\star$, $R_\star$ from isochrone interpolation of the Gaia colour-magnitudes. These can deviate from literature values, but are consistent with the remainder of the analysis of the field star population. 
}
\end{deluxetable*}

\begin{deluxetable*}{rrrrrrrrrrrrl}
\tablewidth{0pc}
\tabletypesize{\scriptsize}
\tablecaption{
    Planet Candidates
    \label{tab:candidates}
}
\tablehead{ \\
    \multicolumn{1}{c}{TIC} &
    \multicolumn{1}{c}{TOI} &
    \multicolumn{1}{c}{Status\tablenotemark{a}} &
    \multicolumn{1}{c}{Period} &
    \multicolumn{1}{c}{Depth} &
    \multicolumn{1}{c}{Gaia $G$} &
    \multicolumn{1}{c}{Gaia $B_P$} &
    \multicolumn{1}{c}{Gaia $R_P$} &
    \multicolumn{1}{c}{Distance} &
    \multicolumn{1}{c}{$T_\mathrm{eff}$} &
    \multicolumn{1}{c}{$M_\star$} &
    \multicolumn{1}{c}{$R_\star$} &
    \multicolumn{1}{c}{Follow-up } \\
    \multicolumn{1}{c}{} &
    \multicolumn{1}{c}{} &
    \multicolumn{1}{c}{} &
    \multicolumn{1}{c}{(d)} &
    \multicolumn{1}{c}{(ppm)} &
    \multicolumn{1}{c}{(mag)} &
    \multicolumn{1}{c}{(mag)} &
    \multicolumn{1}{c}{(mag)} &
    \multicolumn{1}{c}{(pc)} &
    \multicolumn{1}{c}{(K)} &
    \multicolumn{1}{c}{$(M_\odot)$} &
    \multicolumn{1}{c}{$(R_\odot)$} &
    \multicolumn{1}{c}{Status} 
}
\startdata
 129637892 & 155.01 & PC & 5.4 & 8232 & 9.38 & 9.66 & 8.98 & 188 & 6144 & 1.26 & 1.87& Passed spectroscopic vetting \\
 281408474 & 628.01 & PC & 3.4 & 6353 & 10.06 & 10.38 & 9.61 & 179 & 5850 & 1.05 & 1.45& Undergoing spectroscopic vetting\\
 293853437 & 629.01 & PC & 8.7 & 2075 & 8.73 & 8.79 & 8.66 & 336 & 8400 & 2.06 & 2.49& Undergoing spectroscopic vetting \\
\enddata
\tablenotetext{a}{
    PC: Active Planet Candidate
}\end{deluxetable*}

\begin{deluxetable*}{rrrrrrrrrrrrl}
\tablewidth{0pc}
\tabletypesize{\scriptsize}
\tablecaption{
    Candidates determined to be false positives
    \label{tab:falsepositive}
}
\tablehead{ \\
    \multicolumn{1}{c}{TIC} &
    \multicolumn{1}{c}{TOI} &
    \multicolumn{1}{c}{Status\tablenotemark{a}} &
    \multicolumn{1}{c}{Period} &
    \multicolumn{1}{c}{Depth} &
    \multicolumn{1}{c}{Gaia $G$} &
    \multicolumn{1}{c}{Gaia $B_P$} &
    \multicolumn{1}{c}{Gaia $R_P$} &
    \multicolumn{1}{c}{Distance} &
    \multicolumn{1}{c}{$T_\mathrm{eff}$} &
    \multicolumn{1}{c}{$M_\star$} &
    \multicolumn{1}{c}{$R_\star$} \\
    \multicolumn{1}{c}{} &
    \multicolumn{1}{c}{} &
    \multicolumn{1}{c}{} &
    \multicolumn{1}{c}{(d)} &
    \multicolumn{1}{c}{(ppm)} &
    \multicolumn{1}{c}{(mag)} &
    \multicolumn{1}{c}{(mag)} &
    \multicolumn{1}{c}{(mag)} &
    \multicolumn{1}{c}{(pc)} &
    \multicolumn{1}{c}{(K)} &
    \multicolumn{1}{c}{$(M_\odot)$} &
    \multicolumn{1}{c}{$(R_\odot)$} 
}
\startdata
7624182 & 412.01 & NEB & 1.1 & 859 & 8.83 & 8.87 & 8.77 & 466 & 8099 & 2.2 & 3.28 \\
 14091633 & 447.01 & SB1 & 5.5 & 20670 & 9.2 & 9.46 & 8.83 & 126 & 6316 & 1.17 & 1.3 \\
 49899799 & 416.01 & SB1 & 7.0 & 7442 & 8.65 & 8.94 & 8.24 & 132 & 6065 & 1.23 & 1.89 \\
 55452495 & 336.01 & NEB & 8.9 & 4962 & 10.1 & 10.22 & 9.92 & 610 & 7287 & 1.88 & 2.95 \\
 92359850 & 387.01 & NPC & 4.2 & 3880 & 10.01 & 10.28 & 9.61 & 219 & 6165 & 1.26 & 1.61 \\
 123898871 & 630.01 & BEB & 4.9 & 11111 & 10.08 & 10.26 & 9.81 & 311 & 6980 & 1.45 & 1.74 \\
 156987351 & 476.01 & BEB & 3.1 & 6560 & 9.09 & 9.26 & 8.84 & 246 & 7104 & 1.62 & 2.11 \\
 175482273 & 369.01 & BEB & 5.5 & 3229 & 9.52 & 9.58 & 8.92 & 172 & 6192 & 1.26 & 1.58 \\
 350743714 & 165.01 & SB1 & 7.8 & 4399 & 10.01 & 10.25 & 9.61 & 211 & 6260 & 1.27 & 1.51 \\
 365781372 & 627.01 & BEB & 1.1 & 5368 & 10.17 & 10.3 & 9.95 & 640 & 7127 & 1.87 & 3.11 \\
\enddata
\tablenotetext{a}{
    NEB: Nearby Eclipsing Binary; BEB: Blended Eclipsing Binary; SB1: Single lined spectroscopic binary; NPC: Transit caused by nearby source that may still be planetary in origin. 
}
\end{deluxetable*}

\begin{deluxetable*}{rrrrrrrrrrrrrl}
\tablewidth{0pc}
\tabletypesize{\scriptsize}
\tablecaption{
    Confirmed planets around evolved stars with $T_\mathrm{mag} < 10$ not included in occurrence rate calculation
    \label{tab:confirmedplanetsnotincluded}
}
\tablehead{ \\
    \multicolumn{1}{c}{TIC} &
    \multicolumn{1}{c}{TOI} &
    \multicolumn{1}{c}{Name} &
    \multicolumn{1}{c}{Status} &
    \multicolumn{1}{c}{Period} &
    \multicolumn{1}{c}{Depth} &
    \multicolumn{1}{c}{Gaia $G$} &
    \multicolumn{1}{c}{Gaia $B_P$} &
    \multicolumn{1}{c}{Gaia $R_P$} &
    \multicolumn{1}{c}{Distance} &
    \multicolumn{1}{c}{$T_\mathrm{eff}$} &
    \multicolumn{1}{c}{$M_\star$} &
    \multicolumn{1}{c}{$R_\star$} &
    \multicolumn{1}{c}{Reference} \\
    \multicolumn{1}{c}{} &
    \multicolumn{1}{c}{} &
    \multicolumn{1}{c}{} &
    \multicolumn{1}{c}{} &
    \multicolumn{1}{c}{(d)} &
    \multicolumn{1}{c}{(ppm)} &
    \multicolumn{1}{c}{(mag)} &
    \multicolumn{1}{c}{(mag)} &
    \multicolumn{1}{c}{(mag)} &
    \multicolumn{1}{c}{(pc)} &
    \multicolumn{1}{c}{(K)} &
    \multicolumn{1}{c}{$(M_\odot)$} &
    \multicolumn{1}{c}{$(R_\odot)$} &
    \multicolumn{1}{c}{} 
}
\startdata
 290131778 & 123.01 &  & P & 3.3 & 3177 & 8.23 & 8.43 & 7.77 & 162 & 6234 & 1.46 & 2.72 & {\citep{2019AJ....157...51W}} \\
 231670397 & 104.01 & WASP-73 & P & 4.1 & 3586 & 10.26 & 10.57 & 9.82 & 319 & 5950 & 1.21 & 2.33 & {\citep{2014AandA...563A.143D}} \\
 339672028 & 481.01 & & P & 10.3 & 4590 & 9.85 & 10.22 & 9.35 & 180 & 5661 & 0.98 & 1.8 & Rojas et al. in prep \\
 410214986 & 200.01 & DS Tuc & P & 8.1 & 3576 & 8.32 & 8.7 & 7.81 & 44 & 5466 & 0.93 & 0.92 & Newton et al. submitted\\
 452808876 & 453.01 & WASP-82 & P & 2.7 & 6400 & 9.9 & 10.18 & 9.49 & 277 & 6126 & 1.27 & 2.21 & {\citep{2016AandA...585A.126W}} \\
\enddata
\end{deluxetable*}

\bibliographystyle{apj}
\bibliography{hatbib}

\end{document}